\documentclass[12pt,twocolumn,apj]{openjournal}

\usepackage{changepage}
\usepackage{booktabs}
\usepackage{aas_macros}
\usepackage{epsfig}
\usepackage{amsmath}
\usepackage{amssymb}
\usepackage{comment}
\usepackage{leftidx}
\usepackage[rgb]{xcolor}
\definecolor{MyGreen}{rgb}{0.0,0.6,0.3}
\definecolor{MyPurple}{rgb}{0.6,0,0.3}
\usepackage{fancyref}
\usepackage{hyperref}
\hypersetup{
    colorlinks=true,
    linkcolor=blue,
    filecolor=blue,      
    urlcolor=MyPurple,
    citecolor=MyGreen,
}
\usepackage{color,colortbl}
\usepackage{tensind}
\tensordelimiter{?}
\DeclareGraphicsExtensions{.bmp,.png,.jpg,.pdf}
\usepackage{verbatim}
\usepackage[normalem]{ulem}
\usepackage{orcidlink}
\usepackage{soul}
\begin{document}
\title[Red Supergiant Boil-off] {Boil-off of red supergiants: mass loss and type II-P supernovae}

\author{Jim Fuller\orcidlink{0000-0002-4544-0750}}
\email{jfuller@caltech.edu}
\affiliation{TAPIR, Mailcode 350-17, California Institute of Technology, Pasadena, CA 91125, USA}

\author{Daichi Tsuna\orcidlink{0000-0002-6347-3089}}
\affiliation{TAPIR, Mailcode 350-17, California Institute of Technology, Pasadena, CA 91125, USA}
\affiliation{Research Center for the Early Universe (RESCEU), School of Science, The University of Tokyo, Bunkyo, Tokyo 113-0033, Japan}

\begin{abstract}

The mass loss mechanism of red supergiant stars is not well understood, even though it has crucial consequences for their stellar evolution and the appearance of supernovae that occur upon core-collapse. We argue that outgoing shock waves launched near the photosphere can support a dense chromosphere between the star's surface and the dust formation radius at several stellar radii. We derive analytic expressions for the time-averaged density profile of the chromosphere, and we use these to estimate mass loss rates due to winds launched by radiation pressure at the dust formation radius. These mass loss rates are similar to recent observations, possibly explaining the upward kink in mass loss rates of luminous red supergiants. Our models predict that low-mass red supergiants lose less mass than commonly assumed, while high-mass red supergiants lose more. The chromospheric mass of our models is $\sim$0.01 solar masses, most of which lies within a few stellar radii. This can help explain the early light curves and spectra of type-II P supernovae without requiring extreme pre-supernova mass loss. We discuss implications for stellar evolution, type II-P supernovae, SN 2023ixf, and Betelgeuse.

\end{abstract}


\maketitle

\section{Introduction}
\label{sec:intro}

It has long been known that red giants and red supergiants (RSGs) lose mass through dense, slow-moving winds (e.g., \citealt{reimers:75}). These winds can carry mass away at high rates $10^{-8} M_\odot/{\rm yr} \lesssim \dot{M} \lesssim 10^{-4} M_\odot/{\rm yr}$ (e.g., \citealt{antoniadis:24}), partially or fully stripping the hydrogen envelopes of RSGs before their ultimate death via a supernova explosion or collapse to a black hole. RSGs are formed from massive stars ($10 \, M_\odot \lesssim M \lesssim 40 \, M_\odot$) during their post-main sequence evolution, with large radii of $300 \, R_\odot \lesssim R \lesssim 1000 \, R_\odot$, high luminosities of $10^4 \, L_\odot \lesssim L \lesssim 3 \times 10^5 \, L_\odot$, and cool surface temperatures of $3500 \, {\rm K} \lesssim T_{\rm eff} \lesssim 4500 \, {\rm K}$ (e.g., \citealt{levesque:05}). Their low escape speeds $v_{\rm esc} \sim 100 \, {\rm km}/{\rm s}$ and high luminosities probably contribute to their high mass loss rates, yet the physical mechanisms underlying mass loss are not well understood.

\begin{figure}
\centering
\includegraphics[scale=0.38]{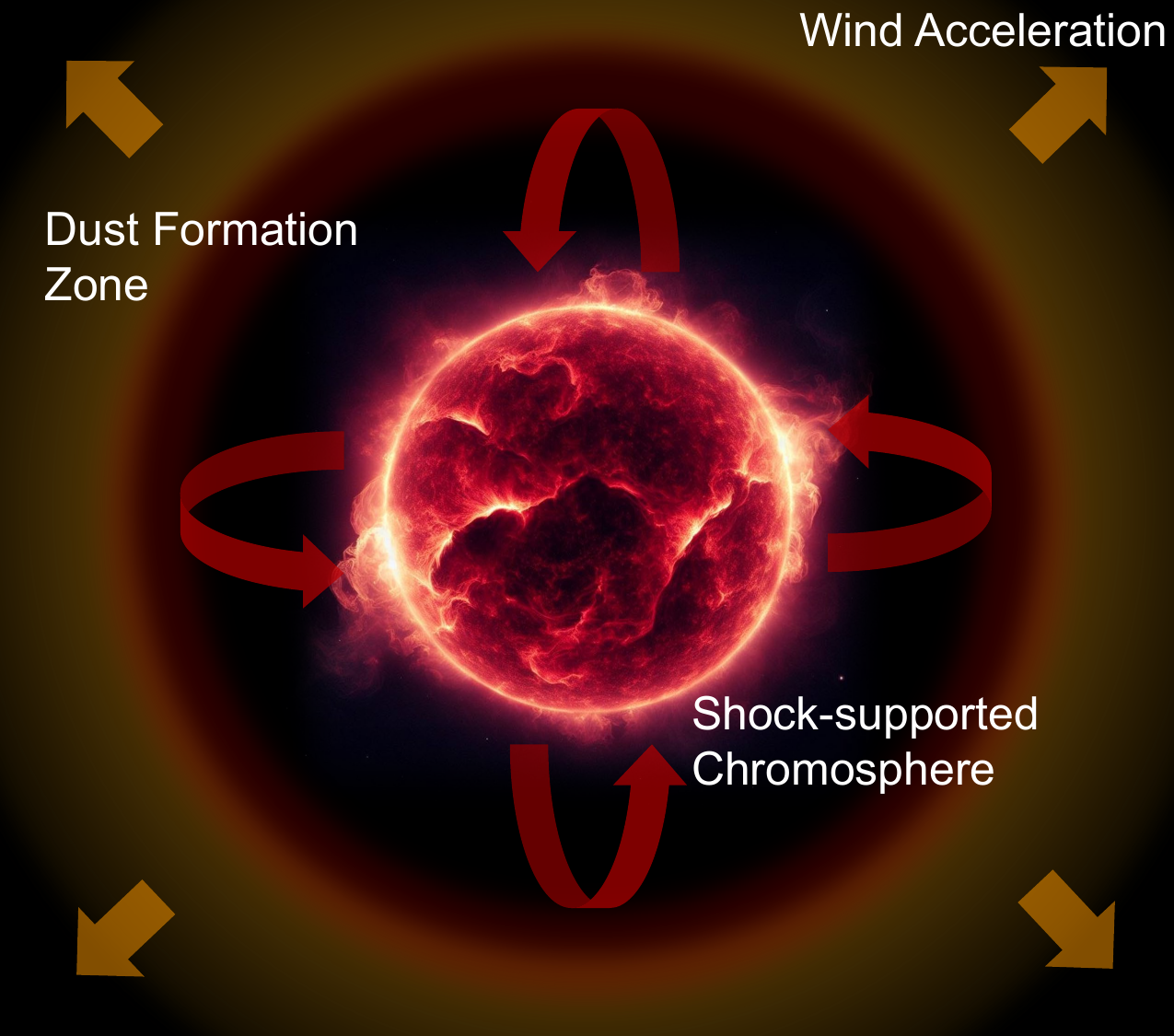}
\caption{\label{fig:cartoon} 
Cartoon illustrating the chromospheric and wind dynamics of red supergiants (not to scale). Outgoing shock waves launched by vigorous convective motions support optically thin material above the photosphere, forming an extended and dynamically varying chromosphere. Material that rises above $\sim$5 stellar radii forms dust, increasing its opacity and creating a radiation-driven wind.   
}
\end{figure}

It is widely accepted that these winds are accelerated by radiation pressure on dust (e.g., \citealt{hoyle:62}), which forms well above their photospheres where the temperature drops to $T \lesssim 1500 \, {\rm K}$ (\citealt{field:74,hofner:18}) such that dust can form (Figure \ref{fig:cartoon}). The high opacity of dust means that dust-laden gas has a large enough opacity to exceed the Eddington limit, such that it will be accelerated outwards by the momentum of photons emitted from the star. However, the dust can only form at radii several times larger than the photosphere of the star, $R_{\rm d} \gtrsim 5 \, R$, and the mechanisms by which material is lofted up to the dust formation radius are highly uncertain. This wind-launching problem for RSGs is crucial for determining their mass loss rates, and it is the main focus of this paper. 

Perhaps the most detailed investigations of mass loss from cool stars have been performed for asymptotic giant branch (AGB) stars (see \citealt{hofner:18} for a review). Although these stars have lower mass and smaller luminosities than RSGs, they have similar surface temperatures and escape speeds, and the mass loss mechanism is believed to be similar to that described above. For AGB stars, some combination of turbulent pressure and shock waves driven by radial pulsations are thought to propel gas above the photosphere \citep{hofner:03}. Because this chromospheric material is dense, it can cool (unlike the Sun's corona, \citealt{cranmer:11}) so that its temperature maintains radiative equilibrium with the star and decreases approximately as $T \propto r^{-1/2}$ in the chromosphere. Evermore sophisticated radiation hydrodynamics simulations of this process \citep{freytag:08,freytag:17,freytag:23} support this general picture and also give detailed predictions for atomic, molecular, and dust spectral features in the spectra of these stars.

For the last decade, transient surveys have been detecting type II-P supernovae at increasingly early times (sometimes just hours after shock-breakout) when the appearance of the supernova is very sensitive to the properties of the exploding RSG and its immediate surroundings (e.g., \citealt{yaron:17,hiramatsu:21,bruch:21,li:23,mezaretamal:24}). With typical ejecta velocities of $\sim \! 10^9 \, {\rm cm}/{\rm s}$, the outgoing SN ejecta takes $\sim$5 days to expand from the star's photosphere out to the dust formation radius of $R_{\rm d} \! \sim 10 \, R \sim \! 5 \times 10^{14} \, {\rm cm}$. Early SN observations thus probe the SN as it is sweeping up material between the star's surface and dust formation radius. The results of these observations have been clear: a large fraction of type II-P SNe show indications of interaction between the SN ejecta and dense circumstellar material (CSM) within the dust formation radius 
\citep{khazov:16,bruch:21,bruch:23}.

This CSM is orders of magnitude denser than expected from stellar evolution models of RSGs, which predict sharp drops in density above the photosphere. The CSM is also a few orders of magnitude denser than expected if the CSM is formed by a wind with mass loss rate and velocity typical for RSGs, $\dot{M} \sim 10^{-6} \, M_\odot/{\rm yr}$, $v_{\rm wind} \sim 30 \, {\rm km}/{\rm s}$, \citep{dessart:17,moriya:17,morozova:17,moriya:18,morozova:18}. The frequent conclusion of SN observers and modelers has thus been that RSGs lose mass at much higher rates during the final years of their lives. This idea has been supported by some RSG models that have predicted pre-supernova outbursts that eject mass in the weeks-months before explosion \citep{woosleyheger:15,fuller:17}.

In this paper, we argue that the RSG wind-launching problem and the type-II SN interaction problem are inextricably linked. We show that CSM surrounding exploding RSGs is naturally produced by a dense, radially extended chromosphere between the stellar surface and dust formation radius. This chromosphere is approximately in hydrostatic equilibrium in a time-averaged sense, and it is supported by the momentum flux of upward-traveling weak shock waves (analogous to turbulent pressure support) excited by vigorous convection near the stellar photosphere. By computing the density of the chromospheric material at the dust formation radius, we also predict RSG mass loss rates. We show that these models can explain several features of RSG mass loss rates as a function of RSG luminosity. Finally, we show that the extended chromospheres of these RSGs helps to explain the unexpected early light curves and flash-ionized emission lines observed in many type II-P SNe.

\section{Chromospheric Dynamics}
\label{sec:chromosphere}

\subsection{Equations of Motion}

In red supergiants, both 1D stellar models with convective mixing length theory (MLT) and 3D stellar models predict typical convective speeds near the photosphere of $v_{\rm con} \sim 10 \, {\rm km}/{\rm s}$ \citep{goldberg:22}. In comparison, the escape speeds from the are 
\begin{align}
v_{\rm esc} &= \sqrt{2 G M/R} \nonumber \\
&\simeq 100 \, {\rm km/s} \bigg(\frac{M}{14 M_\odot}\bigg)^{\!1/2} \bigg(\frac{R}{500 R_\odot}\bigg)^{\!-1/2} \, ,
\end{align}
where $M$ and $R$ are the mass and photospheric radius of the star. Hence, typical convective elements move slower than the escape speed and only rise a small height above the photosphere, $\Delta R \sim R (v_{\rm con}/v_{\rm esc})^2 \sim R/100$, so the density must fall sharply.

Simulations show that turbulent pressure support dominates over thermal pressure support for material above the photosphere. For a turbulent pressure $P_{\rm turb} \sim \rho v_{\rm turb}^2$, where $\rho$ is the density and $v_{\rm turb}$ is the RMS turbulent velocity, we naively expect a chromospheric scale height of $H \sim P_{\rm turb}/(\rho g) \sim v_{\rm con}^2/g \sim 2 r (v_{\rm turb}/v_{\rm esc})^2$. This is a factor of $\sim$50 smaller than the stellar radius for typical RSGs, and so we might expect the density to fall very steeply if it is supported by turbulent pressure. 

However, an important factor to consider is the intermittency of convection, which causes large fluctuations in the convective velocities near the surface of the star. Some regions of the star, or some epochs in time, have larger convective velocities and will have larger chromospheric scale heights. We show this can greatly increase the chromospheric density and associated mass loss rate. 

To build a model for the chromospheric structure, we consider hydrostatic equilibrium between gravity and momentum depostion from outgoing shock waves.  
The momentum equation can be written in the form
\begin{equation}
\label{eq:hydro}
    \frac{\partial}{\partial t} \big( \rho {\vec v} \big) + \nabla \cdot \big( \rho {\vec v}{\vec v} \big) = - \nabla P + \rho {\vec g} \, .
\end{equation}
The $\rho {\vec v}$ term is proportional to the mass flux, which may be non-zero but is assumed to be constant in space/time. Hence, averaging over time, this term vanishes. For simplicity, we will neglect thermal pressure since it is small for moderate or strong shocks (see \citealt{bertschinger:85}) for a model including thermal pressure). We also neglect radiative forces because the low opacity and density of the chromosphere causes it to be optically thin, but later we will account for radiative acceleration of dust grains in the dust formation region. 

Rather than assuming isotropic turbulent motion, we consider purely radial motion, which is a better model to capture the behavior of shocks propagating nearly radially out into the chromosphere. Shocks in RSG chromospheres are nearly isothermal due to the short radiative relaxation time in the post-shock gas \citep{schirrmacher:03}. Hence, to good approximation we expect an adiabatic index of $\gamma\simeq 1$. For a strong nearly isothermal shock, the post-shock and post-cooling velocity $v$ of a fluid element is nearly equal to the shock velocity $v_{\rm s}$ (e.g., \citealt{shu:92}). Hence, the radial momentum flux of an outward going shock wave is $\rho v_{\rm s}^2$, where $v_s$ is the shock speed.

Applying these simplifications to equation \ref{eq:hydro}, hydrostatic equilbrium requires
\begin{equation}
\label{eq:hydroeq}
    \frac{1}{r^2} \frac{d}{dr} \big( \rho r^2 v_{\rm s}^2 \big) = - \frac{G M  \rho}{r^2} \, 
\end{equation}
where $r$ is the radial coordinate. Here, $\rho$ should be regarded as the time-averaged density, whereas the actual density and fluid velocity fluctuate in time as shock waves pass by. We approximate $M$ as constant above the photosphere because the chromospheric mass is a very small fraction of the star's total mass.

Simulations (e.g., \citealt{freytag:17,goldberg:22}) show that the turbulent velocity is nearly constant in the region above the photosphere of a AGB or RSG star. This may be explained by the steepening of turbulent motions into moderate shocks (Mach numbers $\mathcal{M} \sim 3$) in the chromosphere. Such moderate shocks propagating down a density gradient typically steepen until their Mach numbers are of order unity, and then maintain nearly constant Mach number \citep{coughlin:18} and hence nearly constant speed in the nearly isothermal chromosphere. 

Another justification for a nearly constant shock speed above the photosphere is the isothermal nature of the shocks produced. Following \cite{matzner:99}, a shock wave propagating through a star accelerates the material to a velocity of 
\begin{equation}
\label{eq:vs}
    v_s \sim v_0 \bigg(\frac{m}{m_0} \bigg)^{\! \beta-1/2} \bigg(\frac{\rho}{\rho_0} \bigg)^{\! -\beta} \bigg(\frac{r}{r_0} \bigg)^{\! -3\beta} \, .
\end{equation}
Here, $v_0$, $m_0$, $\rho_0$, and $r_0$ are the shock velocity, mass coordinate, density, and radius at an arbitrary reference location. The value of $\beta$ depends on the equation of state of the gas \citep{sakurai:60}, but is approximately $\beta \approx [2 + \sqrt{2 \gamma/(\gamma-1)}]^{-1}$ for strong shocks \citep{whitham:74}, which gives $\mu \simeq 0.2$ for $\gamma=4/3$ as appropriate for adiabatic and radiation-pressure dominated post-shock gas within stellar interiors. But for nearly isothermal shocks appropriate for RSG chromospheres, $\gamma \simeq 1$ and $\beta \simeq 0$. The mass coordinate $m$ is nearly constant for a chromosphere with small total mass. In this case, from equation \ref{eq:vs}, the post-shock velocity $v_s \sim v_0$ is nearly constant as the shock propagates outward.

A third justification for a constant shock speed as a function of radius are the models of \cite{bertschinger:85}, who examined atmospheres of Mira variables supported by a periodic train of outgoing shocks. In the plane-parallel limit, they showed that periodic solutions have shock velocities that are constant as a function of height, while the density falls exponentially (when averaging over time), as shown by the calculation below. The plane-parallel limit applies when the separation between subsequent shocks is less than the stellar radius, which is only marginally true in our case. Simulations (e.g., \citealt{goldberg:22}) show that shocks are launched approximately on the convective turnover time $t_{\rm con} \sim 0.5 R/v_{\rm con}$ such that the separation between shocks is $\sim 0.5 R$.

The shock support term can also be derived heuristically. For strong shocks passing through a medium, the rate of momentum deposition per unit radius is $\dot{p} = d(4 \pi r^2 \rho v_{\rm s})/dt$. For a time-steady state, the time derivative is $d/dt \rightarrow v_s d/dr$. If $v_{\rm s}$ is constant as a function of radius, it can be moved inside the derivative. The momentum deposition per unit volume is thus $\dot{p}/(4 \pi r^2) = (1/r^2) d (\rho r^2 v_{\rm s}^2)/dr$. Remarkably, in the steady-state model, the momentum deposition rate is independent of the frequency at which shocks pass by a given point (i.e., number of shocks per unit time), and is only dependent on the shock amplitude $v_{\rm s}$.

\subsection{Chromospheric Structure}

Proceeding with the assumption that $v_{\rm s}$ is constant as a function of radius, equation \ref{eq:hydroeq} can be solved to yield the density profile
\begin{equation}
\label{eq:rho}
    \rho = \rho(R) \bigg(\frac{R}{r}\bigg)^2 e^{- \frac{v_{\rm esc}^2}{2 v_{\rm s}^2}\big(1-\frac{R}{r}\big)} \, .
\end{equation}
This is similar to the structure of an isothermal atmosphere, but with the sound speed replaced by $v_{\rm s}$, and with the extra geometric factor of $(R/r)^2$ due to the use of the divergence of the momentum flux rather than a turbulent pressure. Just above the stellar surface, equation \ref{eq:rho} is identical to the density profile found by \cite{bertschinger:85} for a periodic train of shocks in a plane-parallel atmosphere. Importantly, the density profile depends only on the amplitude of shocks, and not on their temporal frequency.

As mentioned above, shocks are generated by turbulent convection so that $v_{\rm s}$ is not actually constant, but fluctuates stochastically in time. Since this term enters into the exponential of equation \ref{eq:rho}, this could cause huge changes in the chromospheric density profile. We assume the convective velocity distribution follows a Gaussian probability distribution
\begin{equation}
\label{eq:pv}
    p(v_{\rm s}) = \sqrt{\frac{2}{\pi}} \frac{1}{v_{\rm con}} e^{-\frac{v_{\rm s}^2}{2 v_{\rm con}^2}} \, ,
\end{equation}
where $v_{\rm con}$ is the typical convective velocity. From this definition, the mean turbulent velocity is $v_{\rm con}$. We use a Gaussian rather than Maxwellian distrubtion because we consider only the radial component of the convective velocities.

\begin{figure}
\centering
\includegraphics[scale=0.36]{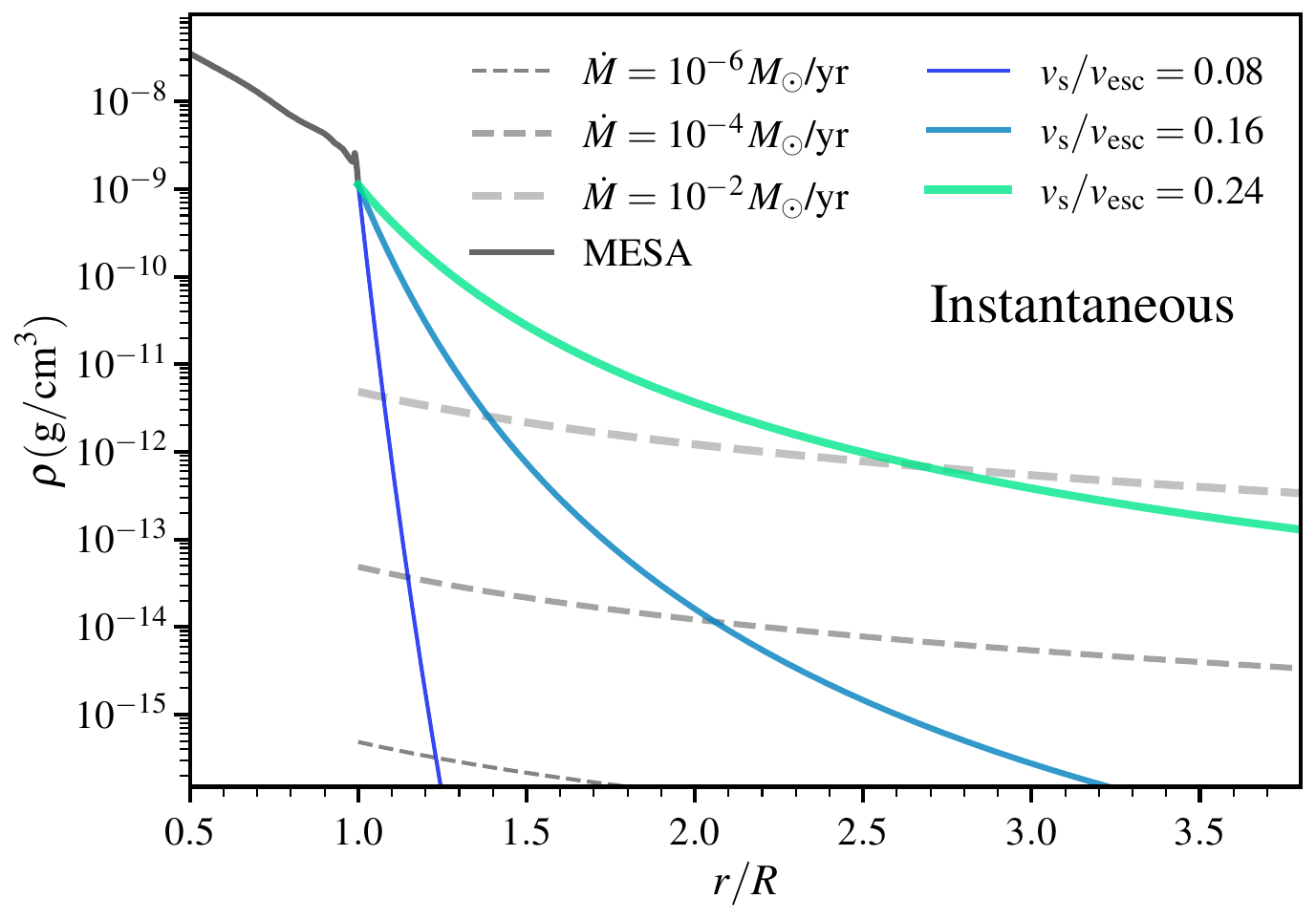}
\includegraphics[scale=0.36]{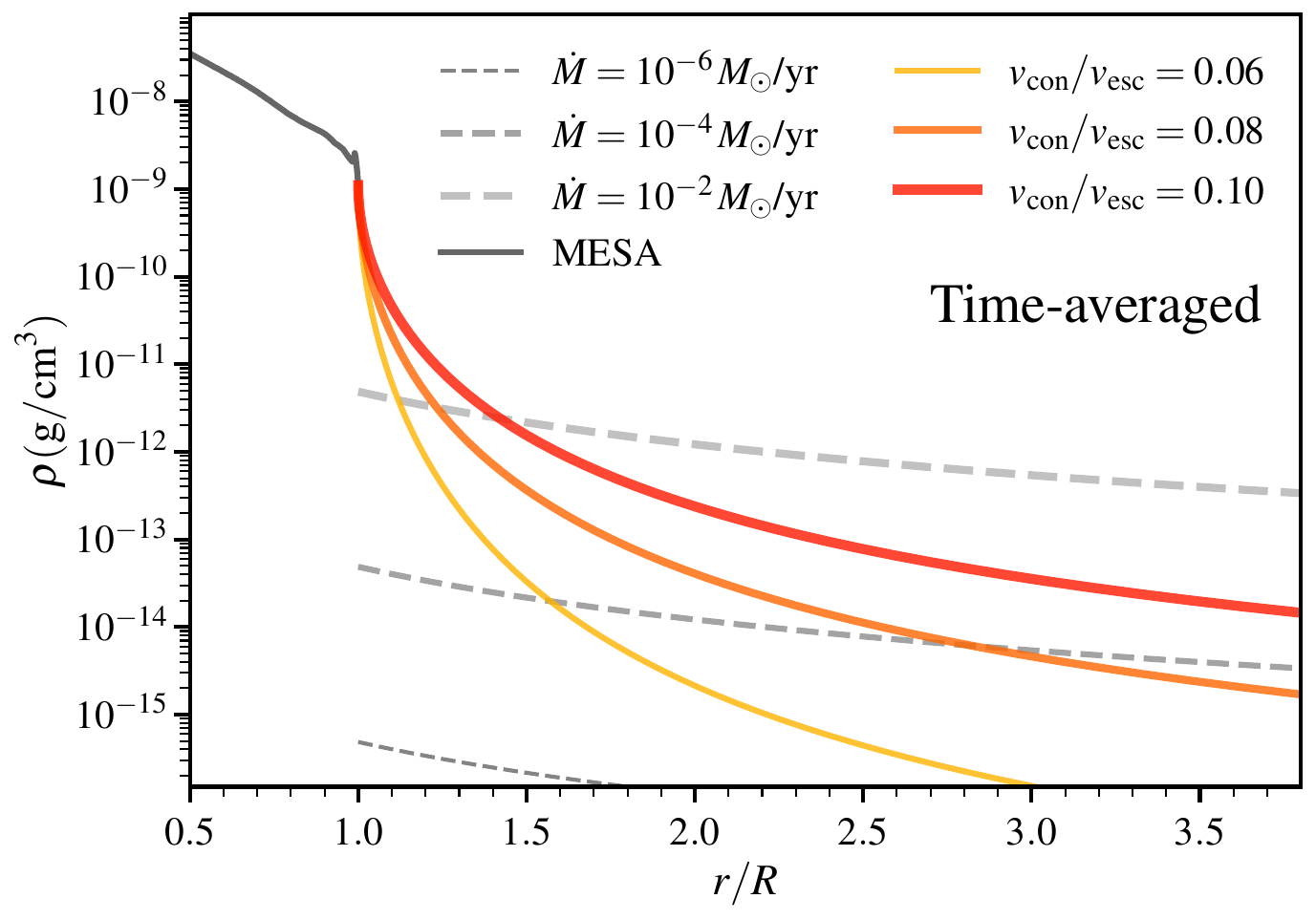}
\caption{\label{fig:DensProf} 
{\bf Top:} Density profiles of a chromosphere model (solid lines) for representative values of $v_{\rm s}/v_{\rm esc}$ for a $15 \, M_\odot$ MESA model of a red supergiant just before core-collapse. This model has an RMS convective velocity of $v_{\rm con}/v_{\rm esc} \simeq 0.08$, and a mass loss rate of $\dot{M} \approx 10^{-5} \, M_\odot/{\rm yr}$. Dashed lines show density profiles for winds of the indicated mass loss rates and a constant wind speed of $v_{\rm wind} = 30 \, {\rm km}/{\rm s}$.
{\bf Bottom:} Time-averaged density profiles of the same model, for different values of the RMS convective velocity that are representative of red supergiants.
}
\end{figure}

\begin{figure}
\centering
\includegraphics[scale=0.36]{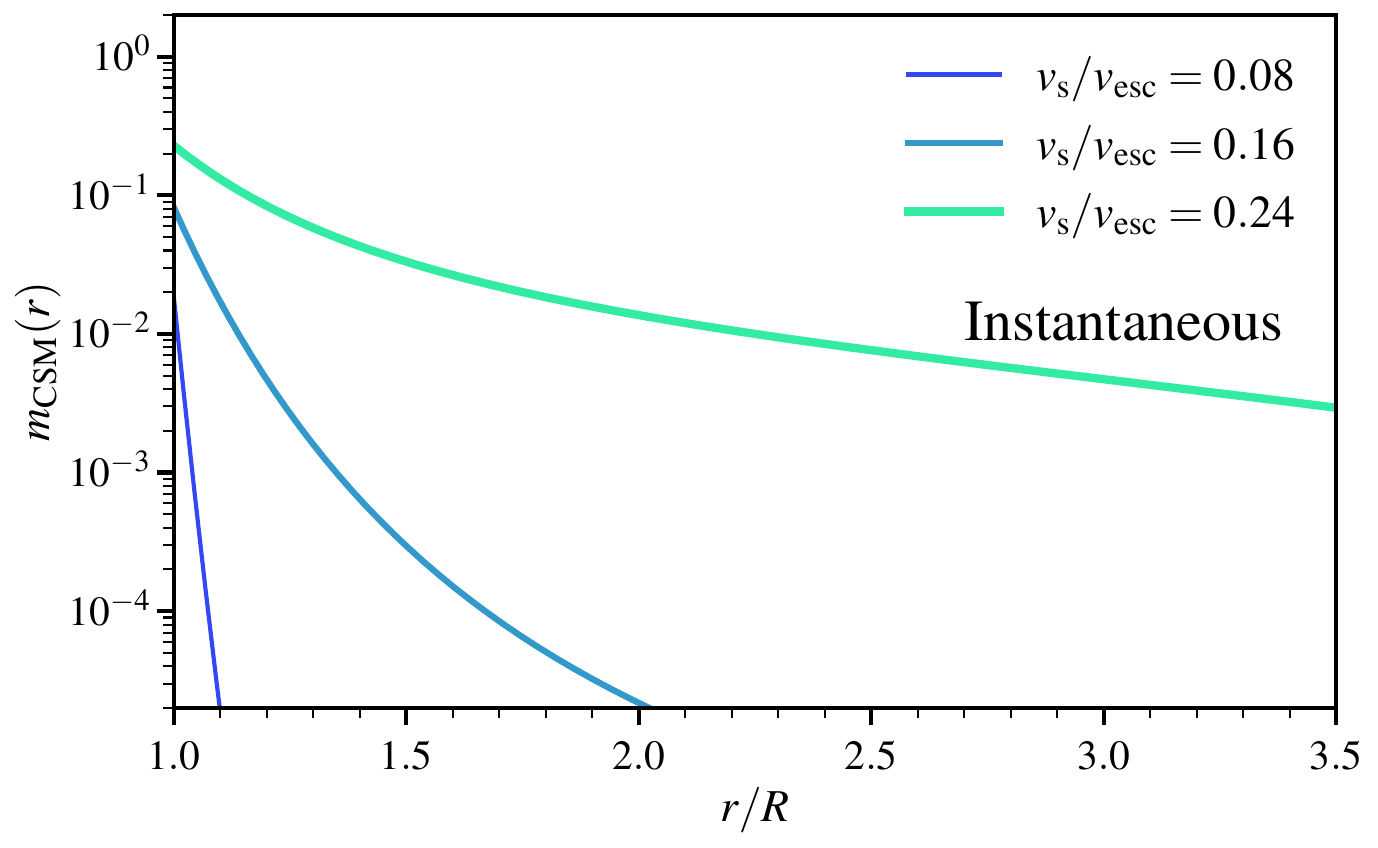}
\includegraphics[scale=0.36]{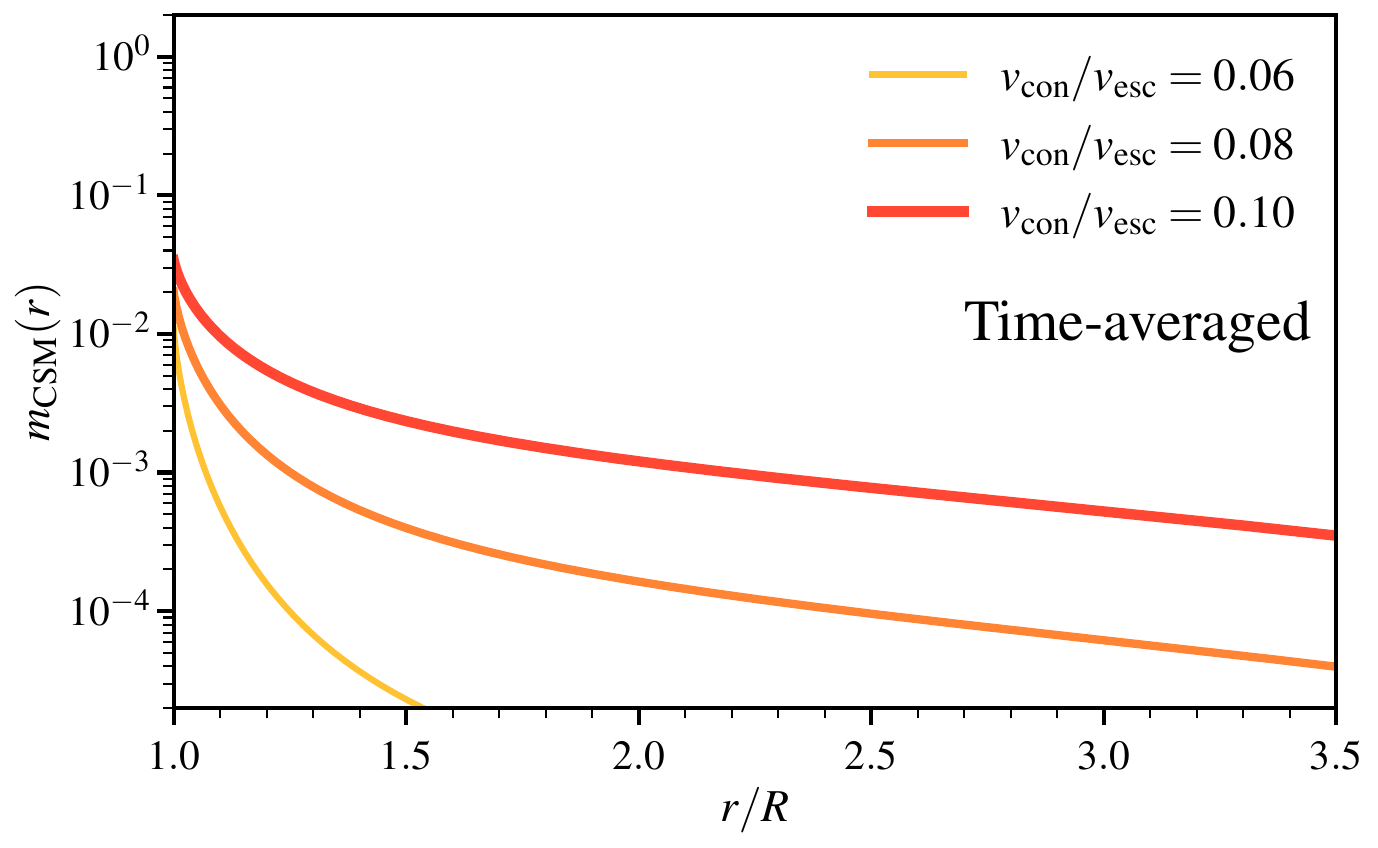}
\caption{\label{fig:MassProf} 
{\bf Top:} Profile of overlying chromospheric mass as a function of radius corresponding to the models in Figure \ref{fig:DensProf}. We expect chromospheres of $\sim \! 10^{-2}-10^{-1} \, M_\odot$, mostly within $\sim$2 stellar radii.
{\bf Bottom:} Time-averaged mass profiles of the same model.
}
\end{figure}

If we assume that $v_{\rm s}$ is constant throughout the chromosphere, the time-averaged density profile is
\begin{align}
\label{eq:rhoav}
    \rho_{\rm av}(r) &= \int^\infty_0 p(v_{\rm s}) \, \rho(r) \, dv_{\rm s} \nonumber \\
    &= \sqrt{\frac{2}{\pi}} \rho(R) \bigg(\frac{R}{r}\bigg)^2 \int^\infty_0 e^{-\frac{v_{\rm s}^2}{2 v_{\rm con}^2} - \frac{v_{\rm esc}^2}{2 v_{\rm s}^2}\big(1-\frac{R}{r}\big) } \frac{dv_{\rm s}}{v_{\rm con}}  \nonumber \\
    &= \rho(R) \bigg(\frac{R}{r}\bigg)^2 e^{-\frac{v_{\rm esc} \sqrt{1-R/r}}{v_{\rm con}} } \, .
\end{align}
Remarkably, accounting for fluctuating convective velocities transforms the isothermal atmosphere density profile of equation \ref{eq:rho} into an exponential distribution. This is important because the ratio of $v_{\rm esc}/v_{\rm con} \sim 10$ is a large number, so the time-averaged density profile of equation \ref{eq:rhoav} falls off much more slowly than that of equation \ref{eq:rho}. The integrand of equation \ref{eq:rhoav} is maximized where $v_{\rm s} \sim \sqrt{v_{\rm esc} v_{\rm con}} \sim 3 v_{\rm con}$, hence the time-averaged density profile at large radii is dominated by uncommon moments of more intense turbulent motion where the convective velocities are $\sim$2 sigma larger than their usual values.

Figure \ref{fig:DensProf} shows the density profiles of equation \ref{eq:rho} and equation \ref{eq:rhoav} above a RSG photosphere. The underlying model is a $13.8 \, M_\odot$ RSG with $R=844 R_\odot$ generated using MESA \citep{paxton:11,paxton:13,paxton:15}. This model had a ZAMS mass of $15 \, M_\odot$ and has been evolved to core-collapse. We use MLT with $\alpha_{\rm MLT}=2.5$ to measure the convective velocity, which we evaluate where the optical depth is $\tau = P_{\rm rad} c/(P v_{\rm con})$, where $P_{\rm rad}$ and $P$ are the radiation and total pressures. This yields approximately correct RMS convective velocities (see \citealt{goldberg:22}), which for RSGs are quite similar to the sound speed $c_s$. This model has $v_{\rm con} \simeq 0.08 v_{\rm esc}$ near the star's surface.

We plot the density profiles of equation \ref{eq:rho} for $v_{\rm s}/v_{\rm esc} = $0.08, 0.16, and 0.24 to demonstrate how moments of large $v_{\rm s}$ greatly increase the chromospheric density. We also plot the time-averaged density profile of equation \ref{eq:rhoav} for values of $v_{\rm con}/v_{\rm esc} = $0.06, 0.08, and 0.10. These values are representative of our RSG models, with low-mass RSGS at the lower end of this range and high-mass RSGS or pre-SN RSGs at the upper end. For this $15 \, M_\odot$ model, the corresponding range in mass loss rate is $\dot{M} \sim 10^{-7}-10^{-4} \, M_\odot/{\rm yr}$ (see Section \ref{sec:massloss}), demonstrating its senstivity to $v_{\rm con}/v_{\rm esc}$.

Figure \ref{fig:DensProf} shows that the chromospheric density falls off steeply within one stellar radius above the photosphere, but then approaches a $r^{-2}$ profile. Nonetheless, within a few stellar radii, the chromospheric density is typically orders of magnitude larger than that implied by simple constant-velocity wind models. For comparison, a constant-velocity wind at mass loss rate $\dot{M}_{\rm wind}$ has a density
\begin{equation}
    \rho_{\rm wind} = \frac{\dot{M}_{\rm wind}}{4 \pi r^2 v_{\rm wind}} \, .
\end{equation}
We plot $\rho_{\rm wind}$ for $v_{\rm wind} = 30$ km/s and three different mass loss rates. The chromospheric density profiles of our models can be much larger than those of simple wind models below a few stellar radii, even for large mass loss rates of $\dot{M}_{\rm wind} = 10^{-4} \, M_\odot$/yr. Between 1-2 stellar radii, our chromospheres may have densities comparable to an extreme wind of $\dot{M}_{\rm wind} = 10^{-2} \, M_\odot$/yr, even though the true mass loss rates of our models are orders of magnitude smaller (Section \ref{sec:massloss}).

Correspondingly, the circumstellar mass is much larger in our models than naive expectations. We define the overlying CSM mass as
\begin{equation}
    \label{eq:mcsm}
    M_{\rm CSM}(r) = \int^{R_{\rm d}}_r 4 \pi \rho r^2 dr \,
\end{equation}
where $R_{\rm d}$ is the dust formation radius. Figure \ref{fig:MassProf} shows the overlying CSM mass for both instantaneous and time-averaged cases. We find time-averaged chromospheric masses out to $R_{\rm d}$ (see next section) of $\sim \! 10^{-2}\, M_\odot$, with large variation for different values of $v_{\rm s}/v_{\rm esc}$ or $v_{\rm con}/v_{\rm esc}$. Most of this mass is near the photosphere, but the time-averaged mass above two stellar radii can be $\sim \! 10^{-3} \, M_\odot$. For comparison, a constant wind model with a similar mass loss rate of $\dot{M} = 10^{-5} \, M_\odot$/yr contains only $\sim \! 10^{-5} \, M_\odot$ within two stellar radii.

\section{Mass Loss}
\label{sec:massloss}

\subsection{Dust Formation and Wind Acceleration}
\label{sec:dust}

\begin{figure}
\centering
\includegraphics[scale=0.36]{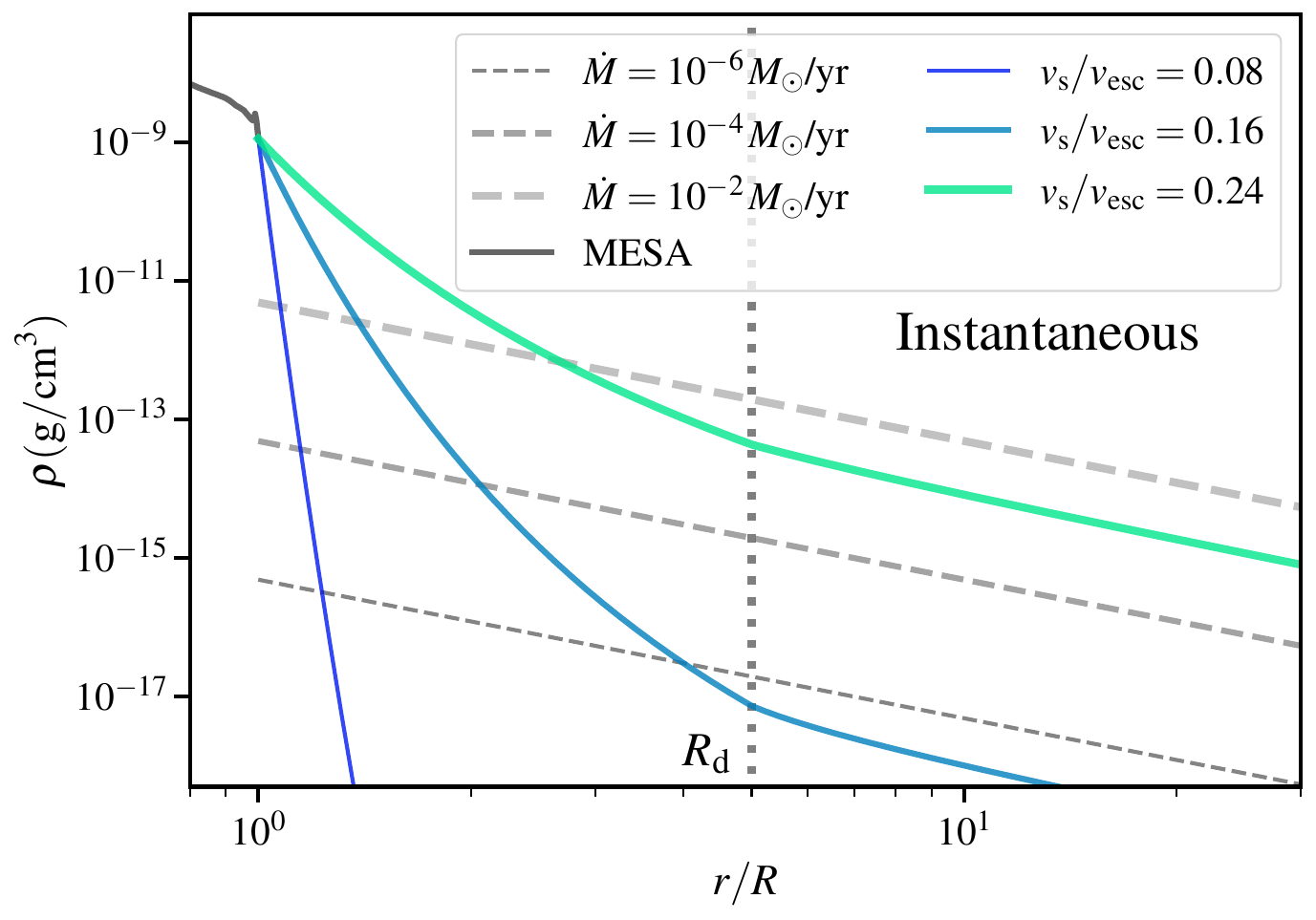}
\includegraphics[scale=0.36]{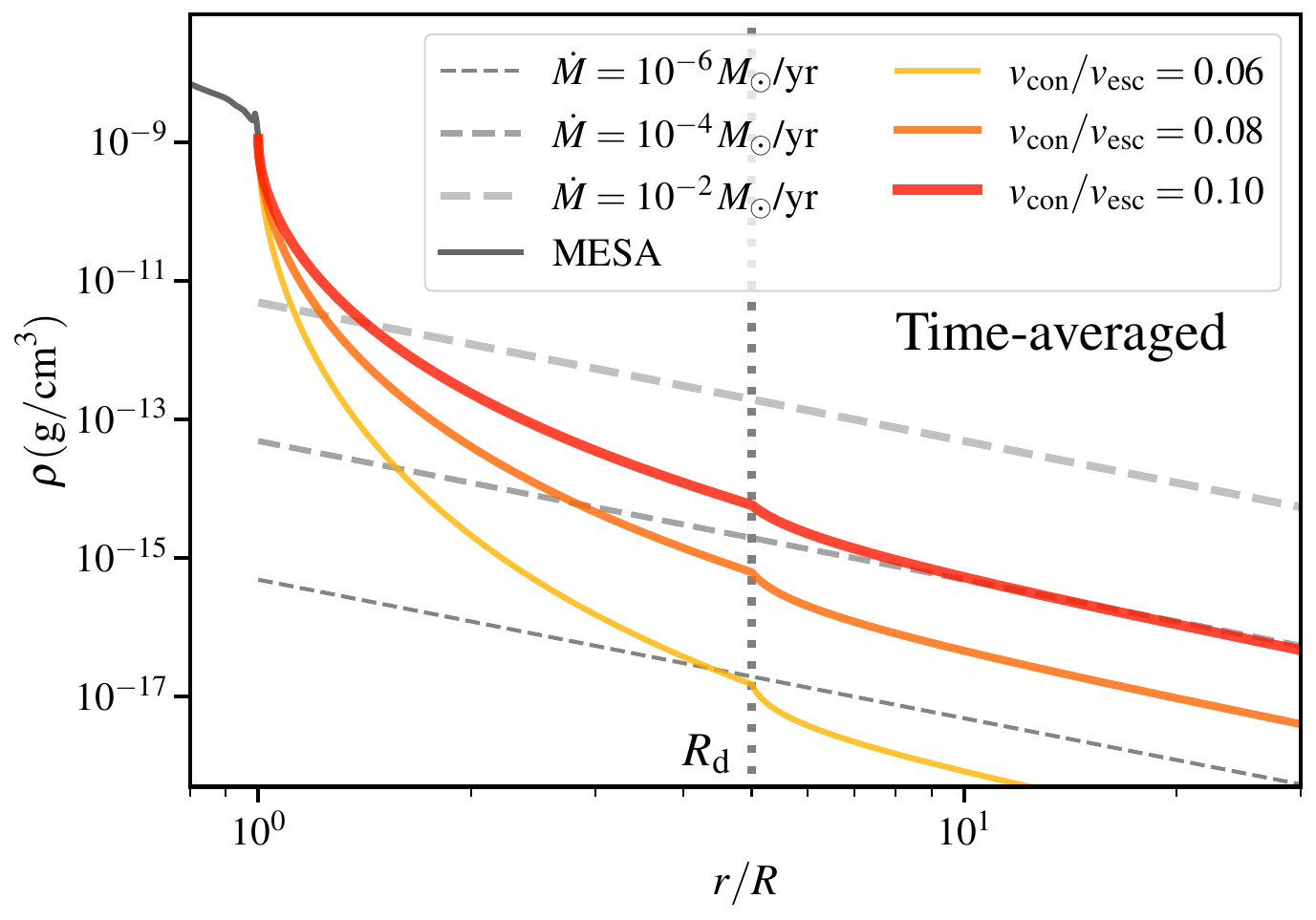}
\caption{\label{fig:DensProfLarge} 
{\bf Top:} Same as Figure \ref{fig:DensProf}, but now showing the transition from chromosphere to wind at a dust formation radius of $R_{\rm d} = 5 R$ (dashed black line). Near the stellar surface, the density falls roughly exponentially, and it falls roughly as $r^{-2}$ at larger radii. {\bf Bottom:} Same as top panel, but showing the time-averaged density profiles.
}
\end{figure}

\begin{figure}
\centering
\includegraphics[scale=0.36]{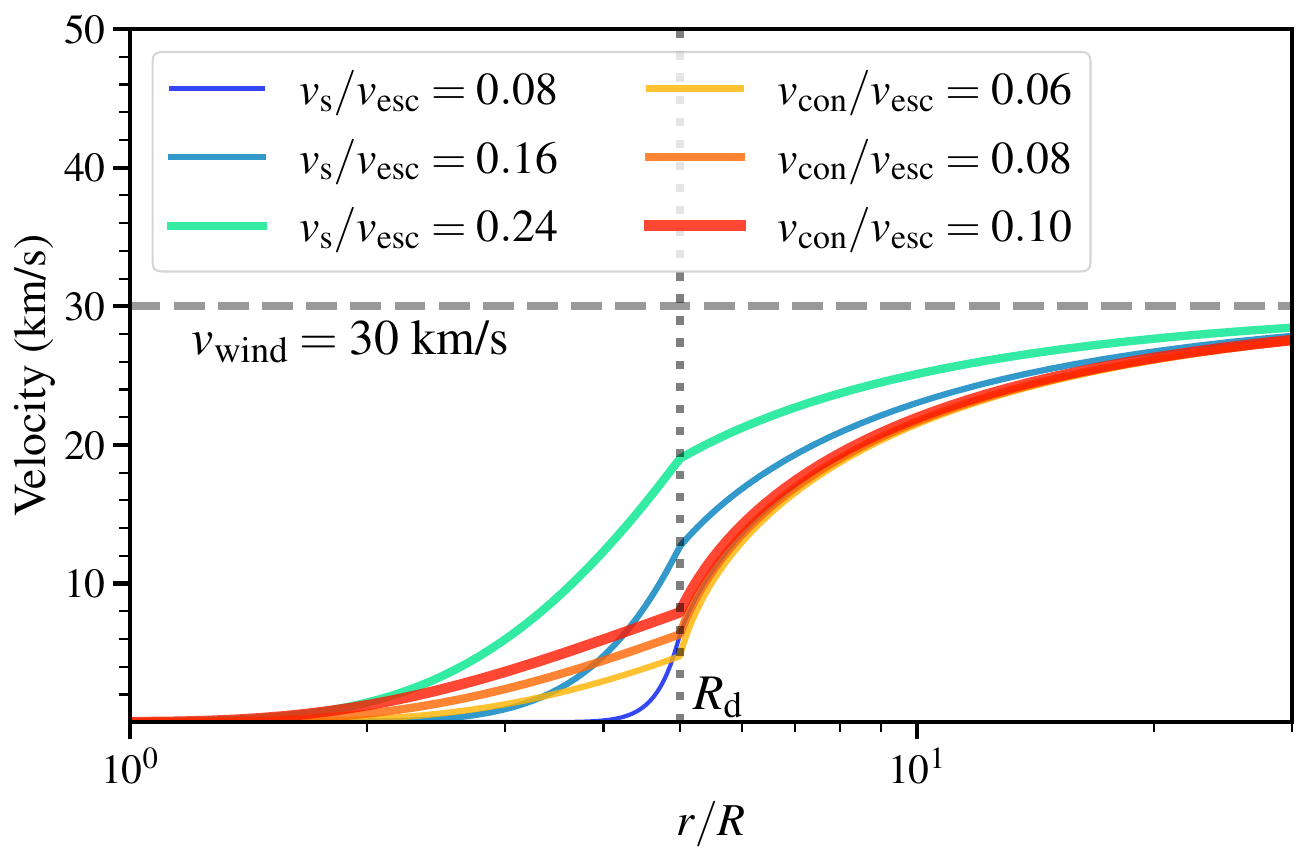}
\caption{\label{fig:VelProf} 
The velocity profiles of the chromosphere/wind for the same models as Figure \ref{fig:DensProfLarge}, showing how the wind speeds are small below the dust formation radius, but approach the terminal wind speed at large distances.
}
\end{figure}

If chromospheric material extends to large radii, it can cool enough to form large amounts of dust. Radiation pressure on this dust will then accelerate this material outwards, driving mass loss from the RSG. Except just below shocks, the temperature profile of the chromosphere is approximately in radiative equilibrium with the star due to molecular and dust absorption/emission processes \citep{schirrmacher:03}. The grey radiative equilibrium temperature for an optically thin chromosphere in the Eddington approximation is (e.g., \citealt{chandrasekhar:34,bowen:88,winters:97})
\begin{align}
\label{eq:temp}
    T(r) &= T_* \bigg[ \frac{1 - \big(1 - (R/r)^2 \big)^{1/2}}{2} \bigg]^{\!1/4} \nonumber \\
    &\approx \frac{T_*}{2^{1/2}} \bigg(\frac{R}{r}\bigg)^{\!1/2} \qquad {\rm for} \, r\gtrsim 2R \, .
\end{align}
where $T_*$ is the photospheric temperature. For a given dust formation temperature $T_{\rm d}$, the corresponding dust formation radius is 
\begin{equation}
\label{eq:rd}
    R_{\rm d} \approx \frac{R}{2} \bigg(\frac{T_*}{T_{\rm d}}\bigg)^{\!2} \, .
\end{equation}
For RSG surface temperatures of $T_* \approx 3600 \, {\rm K}$ and dust formation temperatures of $T_{\rm d} \sim 1200 \, {\rm K}$, dust condenses at a dust formation radius of $R_{\rm d} \sim 4.5 \,R$.

To estimate the wind velocity profile due to radiative driving of dust in our model, we approximate the radiative acceleration as zero below $R_{\rm d}$ and falling off as $1/r^2$ above the dust formation radius. We also neglect turbulent pressure support above $R_{\rm d}$ so that the momentum equation in this region is
\begin{equation}
    v \frac{dv}{dr} = -\frac{GM}{r^2} + \frac{f_{\rm Ed} GM}{r^2} \qquad {\rm for} \,\, r > R_{\rm d} \, .
\end{equation}
Here, the constant $f_{\rm Ed} = L/L_{\rm Ed}$ determines the magnitude of the radiation force on the dust-gas mixture and will be computed below. For an initial velocity $v(R_{\rm d})=v_{\rm d}$ at the dust formation radius, the solution to this equation is
\begin{equation}
\label{eq:vr}
    v_{\rm out} = \bigg[ v_{\rm d}^2 + x_{\rm d} G M \bigg(\frac{1}{R_{\rm d}} -\frac{1}{r} \bigg)  \bigg]^{1/2} \qquad {\rm for} \,\, r > R_{\rm d} 
\end{equation}
where $x_{\rm d} = 2(f_{\rm Ed}-1)$. The velocity at $r\rightarrow\infty$ is then 
\begin{equation}
\label{eq:vinf}
v_\infty = \sqrt{ v_{\rm d}^2 + x_{\rm d} GM/R_{\rm d}} \, .
\end{equation}

To match our solution above and below the dust formation radius, we choose an outflow velocity $v_{\rm d} = v_{\rm con}$ at the dust formation radius. This is motivated by the fact that the dust formation radius is where the flow transitions from an approximately hydrostatic to an unbound outflow, such that the outflow velocity begins to exceed $v_{\rm con}$, and the advection term becomes important in the momentum equation. The mass loss rate is then
\begin{align}
\label{eq:mdot}
    \dot{M} &\approx 4 \pi R_d^2 \rho(R_d) \, v_{\rm con} \nonumber \\
    &\approx 4 \pi R^2 \rho(R) v_{\rm con} \, e^{-\frac{v_{\rm esc} \sqrt{1-R/R_{\rm d}}}{v_{\rm con}} } \, .
\end{align}
This procedure is similar to computing the mass loss of a Parker wind at the sonic point. However, in our problem the sonic point is not important because the dust formation radius is at much smaller radius than the sonic radius (i.e., the effective Bondi radius), which lies at $R_{\rm s} = GM/v_{\rm con}^2 \sim 50 R$. The mass loss rate is determined primarily by the chromospheric density at the dust formation radius, and it is only weakly sensitive to the radiative acceleration of the wind above this point.

We choose to use observed wind velocities $v_\infty$ to constrain $x_{\rm d}$, but one could use dust opacities and mass fractions to calculate $x_{\rm d}$ and then predict the corresponding value of $v_{\infty}$. From equation \ref{eq:vinf}, we have $x_{\rm d} = (R_{\rm d}/G M) ( v_\infty^2 - v_{\rm d}^2)$. In this work, we use $v_\infty = 30$ km/s, in line with observational estimates \citep{sjouwerman:98,vanloon:01,marshall:04}, although this velocity likely depends weakly on the stellar luminosity (e.g., \citealt{mauron:11}). This requires $x_{\rm d} \sim 2$, approximately what we expect for realistic dust-laden gas opacities of $\kappa_{\rm d} \sim 5 \, {\rm cm^2}/{\rm g}$ \citep{hofner:03}.

To compute the time-averaged outflow speed below the dust formation radius, we simply require $\dot{M}$ to be constant with radius such that $v_{\rm out} = \dot{M}/(4 \pi \rho r^2)$.
Finally, to compute the time-average density above the dust formation radius, we again use the constant mass loss rate so that the density is
\begin{equation}
\label{eq:rhod}
    \rho(r) = \frac{\dot{M}}{4 \pi r^2} \bigg[ v_{\rm d}^2 + x_{\rm d} G M \bigg(\frac{1}{R_{\rm d}} -\frac{1}{r} \bigg)  \bigg]^{-1/2} \qquad {\rm for} \,\, r > R_{\rm d} \, .
\end{equation}

As mentioned above, dust formation begins at temperatures ranging from $T\simeq 1000-2000 \, {\rm K}$, with different species condensing at different temperatures. Here we set the dust formation radius to $R_{\rm d} = 5 R$, a reasonable value for RSGs. Different choices for $R_{\rm d}$ would result in slightly different density/velocity profiles above $R_{\rm d}$. The mass loss rate of our model is only weakly sensitive to the value of $R_{\rm d}$ because the factor of $R_{\rm d}^{-2}$ in the density profile (equation \ref{eq:rho}) cancels out the factor of $R_{\rm d}^2$ in the mass loss rate (equation \ref{eq:mdot}). In reality, dust starts to condense over a range of temperatures/radii, and there is no single dust formation radius.

Figures \ref{fig:DensProfLarge} and \ref{fig:VelProf} shows our computed time-averaged density and velocity profiles of the star's CSM for different values of $v_{\rm con}/v_{\rm esc}$. The density profiles are the same as Figure \ref{fig:DensProf} below the dust formation radius, but we now extend the plot above $R_{\rm d}$. The velocity profile has low outflow velocities below $R_{\rm d}$, before the transition into a wind above the dust formation radius. The density is much larger than that of a constant-velocity wind below $R_{\rm d}$, and then it approaches the constant-velocity solution above $R_{\rm d}$ where the wind speed approaches is asymptotic value.

For the time-averaged models, there is a downward kink in the density profiles at the dust formation radius $R_{\rm d}$. This is caused by the upward kink in the outflow velocity at $R_{\rm d}$ (Figure \ref{fig:VelProf}) due to radiative acceleration. While the true velocity and density profiles are unlikely to contain sharp kinks, we do expect a small dip in CSM density above dust formation radius due to the acceleration of the wind.

\subsection{Red Supergiant Mass Loss Rates}
\label{sec:RSGs}

\begin{figure}
\centering
\includegraphics[scale=0.36]{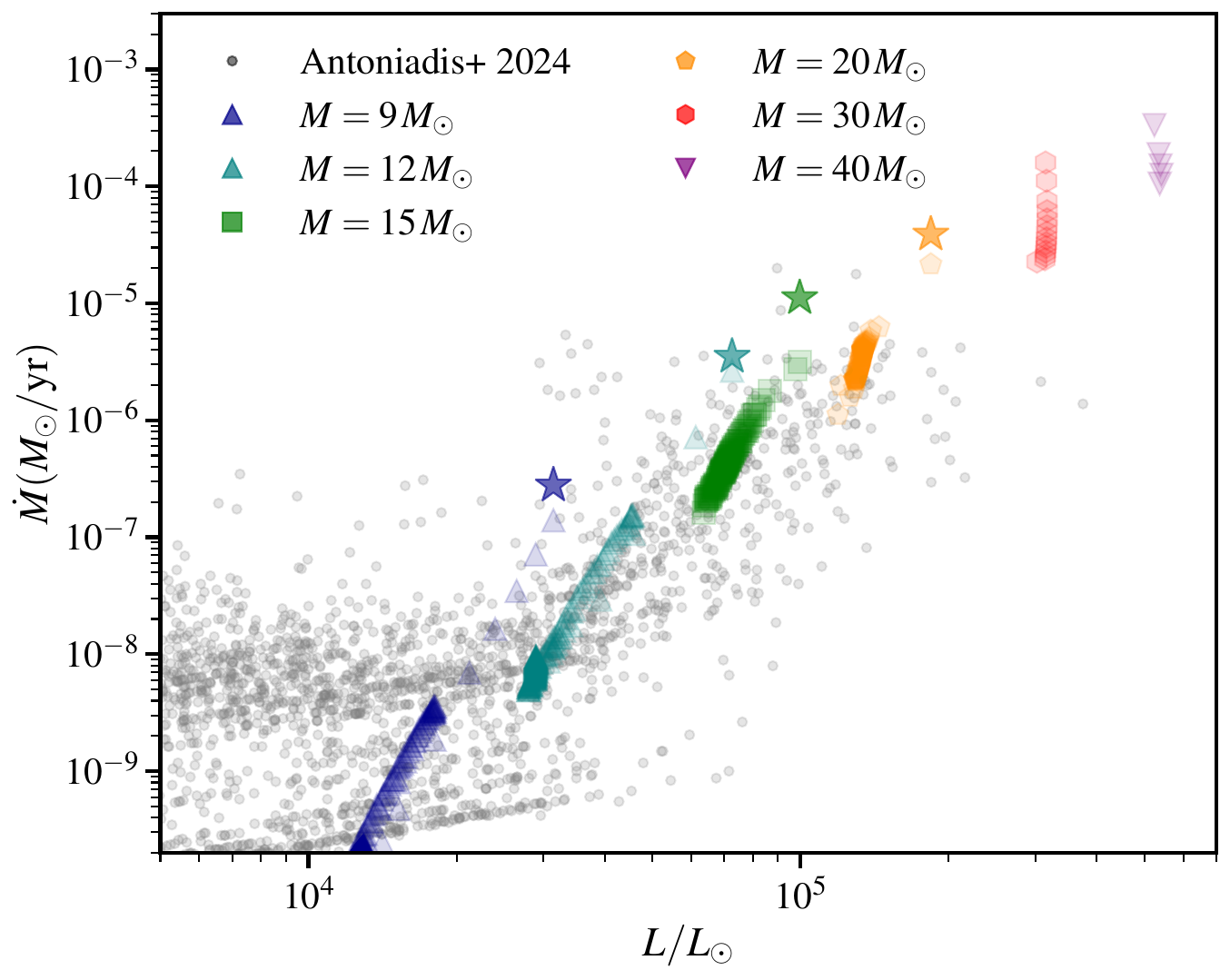}
\caption{\label{fig:MassLossAnt} 
Mass loss rates of stellar models as a function of their luminosity. The small gray symbols are measured mass loss rates of red supergiants in the LMC from \citealt{antoniadis:24}, while colored symbols are our models. We only show the red supergiant phase of evolution, when $T_{\rm eff} < 4200 \, {\rm K}$. We plot one symbol for every $10^4$ years of evolution, and use a large star symbol to denote the pre-supernova mass loss rate. The full evolution of $30\, M_\odot$ and $40 \, M_\odot$ models is not shown because they turn into yellow supergiants after heavy mass loss. An alternative mass loss mechanism is required to explain stars with $L \! \lesssim 2 \times 10^{4} \, L_\odot$.
}
\end{figure}

\begin{figure}
\centering
\includegraphics[scale=0.36]{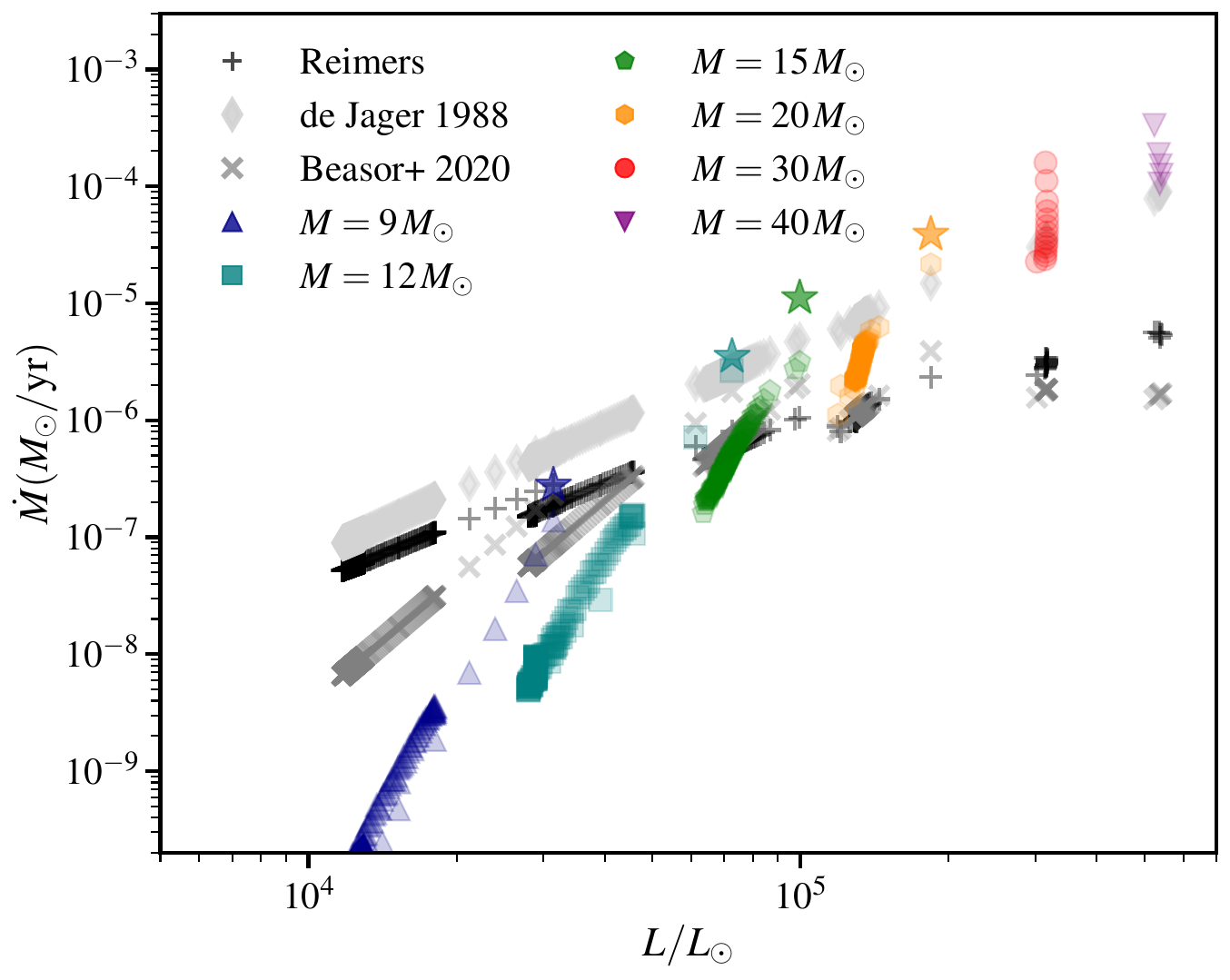}
\caption{\label{fig:MassLoss} 
Mass loss rates for the same models as Figure \ref{fig:MassLoss}. The black symbols are the Reimer's mass loss rates of the same models for $\eta=0.5$, while the dark gray symbols are for the \citealt{dejager:98} prescription, and dark gray symbols are the empirical prescription of \citealt{beasor:20}. Our model predicts a steeper scaling of mass loss with luminosity than any of these prescriptions.  
}
\end{figure}


We implement the mass loss rate of the previous section (equation \ref{eq:mdot}) into stellar evolution models generated with MESA as described above. At surface temperatures above $4500 \, {\rm K}$ we use the ``Dutch" wind prescription with scaling factor $\eta = 0.5$. At cooler surface temperatures, we use our wind prescription. Our MESA inlists and \verb|run_star_extras| files are available at \href{https://zenodo.org/records/11591641}{this link}.

Figure \ref{fig:MassLossAnt} shows the predicted mass loss rates for RSGs with masses of $10-40 \, M_\odot$. We only plot the RSG phase where $T_{\rm eff} < 4200 \, {\rm K}$, where the surface convection zone is deep. Usually the mass loss rate from boil-off is negligible at higher temperatures when the star is more compact. However, those models can have narrow super-Eddington sub-surface convective zones and we do not trust the MLT prediction for $v_{\rm con}$ in those models, so we do not investigate those cases here.

Our mass loss rates are similar for a given luminosity to those recently measured for RSGs in the LMC \citep{antoniadis:24,wen:24}. These mass loss rates are substantially smaller than those measured in the SMC \citep{yang:23}, but those may be systematically overestimated according to \citealt{antoniadis:24}. Most importantly, the predicted scaling of mass loss with luminosity above $L > 3 \times 10^4 \, L_\odot$ is quite similar between our models and all of these measurments. Interestingly, the LMC/SMC measurements indicate a kink at $L \sim 3 \times 10^4 \, L_\odot$, with the mass loss rate flattening at smaller luminosities, which does not happen in our models. The observed mass loss rates at low luminosities are closer to the predictions of a Reimer's mass loss prescription. It is likely that a different mass loss mechanism (e.g., mass loss driven by coronal heating due to Alfv\'en waves, \citealt{cranmer:11}) dominates at low luminosities.

Figure \ref{fig:MassLoss} compares our models to the Reimers \citep{reimers:75} mass loss rate
\begin{equation}
\label{eq:reimers}
    \dot{M}_{\rm Reim} = \eta \, 4 \times 10^{-13} \frac{(L/L_\odot) (R/R_\odot)}{M/M_\odot} M_\odot/{\rm yr} \, ,
\end{equation}
and we use $\eta=0.5$.
Additionally, we show the commonly used empirical mass loss prescription of \cite{dejager:98}, which can be written
\begin{equation}
    \dot{M}_{\rm de Ja} = 4.5 \times 10^{-6} \bigg(\frac{L}{10^5 L_\odot}\bigg)^{\!1.769} \bigg(\frac{T_{\rm eff}}{4000 \, {\rm K}}\bigg)^{\!-1.676} \, M_\odot/{\rm yr}
\end{equation}
Finally, we compare to the recent empirical mass loss formula of \citep{beasor:20}, which can be expressed
\begin{equation}
\label{eq:beasor}
    \dot{M}_{\rm Beas} = 10^{-3.5} 10^{-3 M/(20 M_\odot)} \bigg(\frac{L}{L_\odot}\bigg)^{\!3.6} \, M_\odot/{\rm yr}  \, .
\end{equation}
We do not plot the mass loss prescription of \cite{schroder:05}, which lies between the Reimers and de Jager prescriptions for our models.

We can see that the predicted mass loss rates are usually in the range expected for RSGs, $10^{-8} \, M_\odot/{\rm yr} \lesssim \dot{M} \lesssim 10^{-4} \, M_\odot/{\rm yr}$, with canonical mass loss rates of $\dot{M} \sim 10^{-7}-10^{-5} \, M_\odot/{\rm yr}$ obtained for models with $M \! \sim \! 15-20 \, M_\odot$. We predict strong sensitivity to both luminosity and mass during the RSG phase: low-luminosity $\sim \! 12 \, M_\odot$ stars are predicted to have very small mass loss rates, $\dot{M} \sim 10^{-8}-10^{-7} \, M_\odot/{\rm yr}$, during core helium-burning, below the Reimer's and Beasor rates. However, when these models expand and brighten during the helium-shell and carbon-burning phases, their mass loss rates increase by more than an order of magnitude to $\dot{M} > 10^{-6} \, M_\odot/{\rm yr}$.

The strong luminosity dependence of our predicted mass loss rates results from their sensitivity to the ratio of $v_{\rm con}/v_{\rm esc}$. Brighter RSGS typically have larger luminosities (with slightly larger values of $v_{\rm con}$) and larger radii (with slightly smaller values of $v_{\rm esc}$), which produces much larger mass loss rates. When RSGs expand and brighten after core helium depletion, their mass loss rates increase substantially, and we predict typical pre-SN mass loss rates of $\sim 10^{-5} \, M_\odot/{\rm yr}$.
For RSGs at the same luminosity/radius, more massive RSGs have much smaller mass loss rates because they have larger values of $v_{\rm esc}$. This is consistent with the empirical mass loss prescription of \cite{beasor:20}.

Our $15\, M_\odot$ and $20 \, M_\odot$ models have similar mass loss rates to both the Reimer's and Beasor prescriptions, with typical values of $\dot{M} \sim 10^{-6} \, M_\odot/{\rm yr}$. The models with $M \leq 15 \, M_\odot$ lose a small fraction of their hydrogen envelope before core-collapse, while our $20 \, M_\odot$ model loses about $3 \, M_\odot$ during the RSG phase. The pre-supernova mass loss rate of these models is about ten times larger than during helium burning due to their expansion after helium depletion.

Massive RSGs of $M \! \gtrsim \! 30\, M_\odot$ are predicted to have very large mass loss rates of $\dot{M} \gtrsim 10^{-5} \, M_\odot/{\rm yr}$ during core helium-burning, much larger than predicted by the Reimers or Beasor mass loss rates. The $30 \, M_\odot$ model makes it less than halfway through helium burning, living as an RSG for about $1.3 \times 10^5 \, {\rm yr}$ before losing most of its hydrogen envelope. The $40\, M_\odot$ model loses its envelope very quickly, living as an RSG for about $5 \times 10^4 \, {\rm yr}$. In both cases, the models start to evolve into yellow supergiants as they lose mass, where we become less confident in the MLT velocities and our predicted mass loss rates. Those models also require very short time steps, so we terminate their evolution before the end of core helium-burning. It is likely that such stars will lose all of their hydrogen envelopes and turn into Wolf-Rayet stars, but more work will be needed to develop models that reliably follow the stars through that transition. 

\subsection{Implications of mass loss}

Our massive models can exhibit exponentially increasing mass loss rate with time. As these stars lose mass, their luminosities/radii remain nearly constant but their masses decrease, decreasing $v_{\rm esc}$ and increasing the mass loss rate, creating a runaway process.
Combined with the rarity of such stars due to the IMF, this implies a correspondingly small number of observed very luminous RSGs with large mass loss rates, perhaps consistent with their low observed numbers \citep{beasor:22}. In contrast, the Reimers prescription predicts much lower mass loss rates, longer RSG durations, and a larger number of observed very luminous RSGs. The boil-off phenomenon of our models can thus help explain the absence of RSGs observed with luminosities larger than $L_{\rm max} \approx 10^{5.5} \, L_\odot$ \citep{davies:18}, while simultaneously preserving low-mass loss rates of less massive RSGs.

\cite{vink:23} suggested that the upward kink in mass loss rates at high luminosity ($L \gtrsim 10^{4.5} \, L_\odot$) is caused by a transition from an optically thin to optically thick wind. This is motivated by an analogy with line-driven winds from hot stars, which show a similar feature as photons begin to scatter multiple times in optically thick winds, allowing them to drive higher mass loss rates. However, as far as we are aware, there is no model of radiatively driven winds in cool stars, e.g., due to photon pressure on molecular absorption lines. While photon pressure on dust helps accelerate RSG winds, the mass loss rate may instead be determined by how much mass can be uplifted to the dust formation radius, so it is not clear that optically thick dust absorption would actually drive higher mass loss rates. Our model can explain the dependence of mass loss on luminosity above the kink, but it cannot explain the dependence below the kink. We speculate that a different mechanism with weaker luminosity dependence dominates mass loss at luminosities below $10^{4.5} \, L_\odot$. Future work will be necessary to distinguish between our suggestion and that of \cite{vink:23}.

In this work, we do not investigate the effects of metallicity. While our prescription does not explicitly dependent on metallicity, RSG radii (and hence $v_{\rm esc}$) do depend on metallicity, which will affect mass loss rates. Preliminary models predict slightly smaller mass loss rates by a factor of less than $\sim$2 due to smaller radii for LMC metallicity, and very similar to the measurements of \citep{antoniadis:24}. A competing effect is that low-metallicity RSGs have lower mass at the same luminosity, increasing mass loss rates. Preliminary SMC metallicity models have similar mass loss rates to the solar metallicity models, given the same luminosity, but lower than the measurements of \citep{yang:23}. 

An apparent prediction of our model is that mass loss does not scale strongly with metallicity. However, there must be some dependence, because very low metallicity stars would not be able to form enough dust to launch the winds at all. A possible prediction of our model is that higher metallicity stars have similar mass loss rates but higher terminal wind speeds due to the larger opacity of dust in their winds. More realistic modeling of dust formation and radiative acceleration will be needed to confirm these hypotheses.

\begin{figure}
\centering
\includegraphics[scale=0.36]{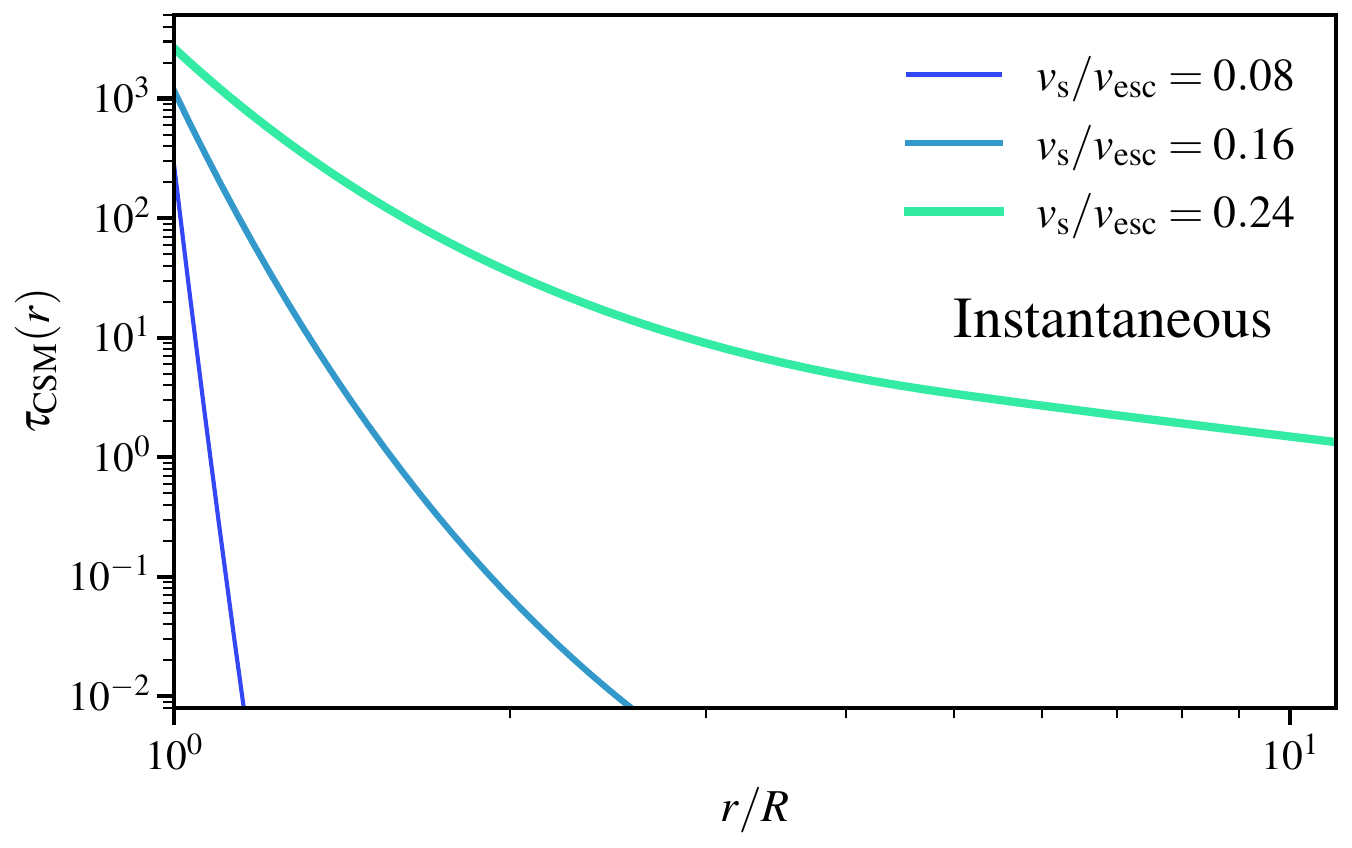}
\includegraphics[scale=0.36]{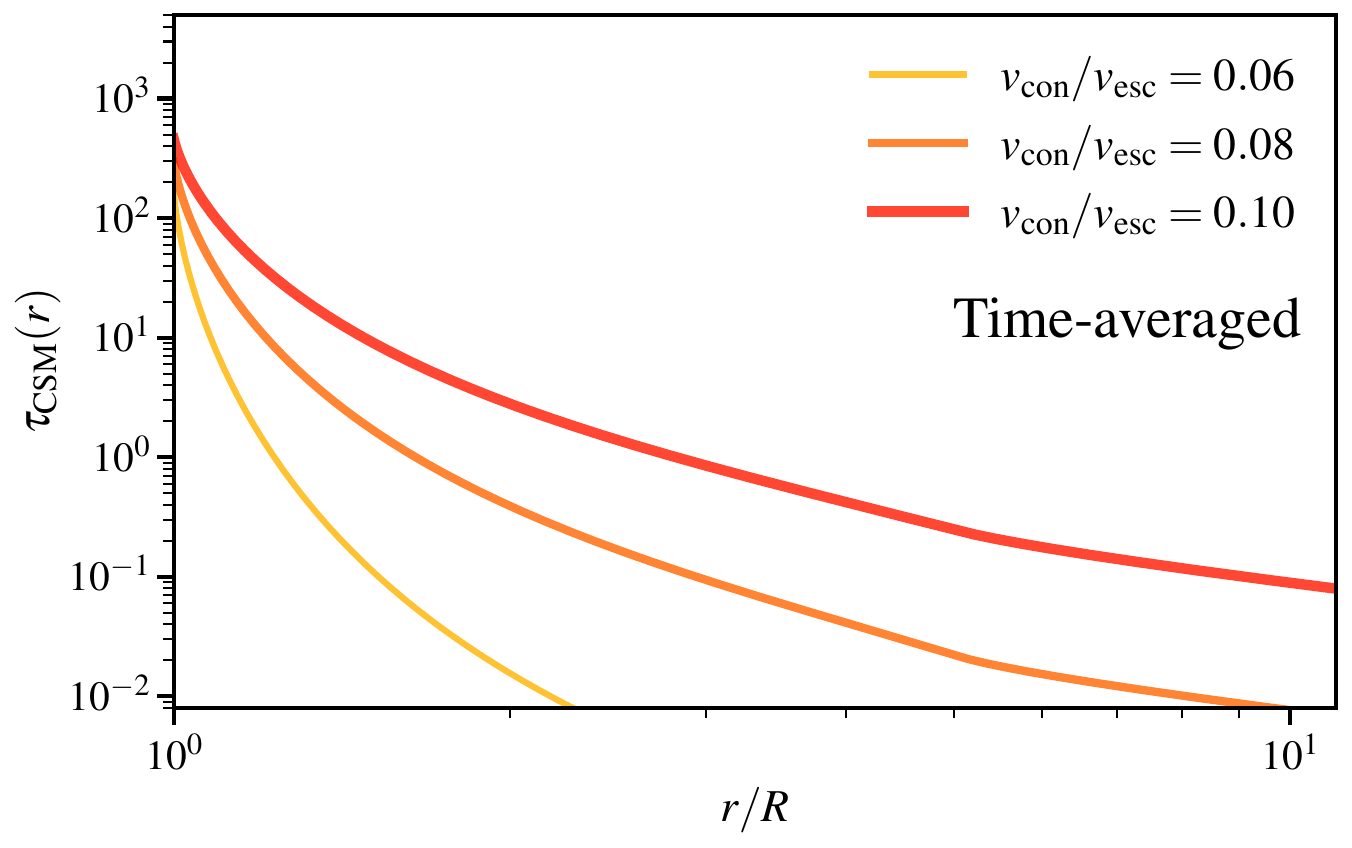}
\caption{\label{fig:TauProf} 
{\bf Top:} The overlying optical depth profile of the chromosphere after it is fully ionized by supernova shock breakout such that its opacity is $\kappa \simeq 0.35 \, {\rm cm}^2/{\rm g}$. {\bf Bottom:} Same as top panel, but for time-averaged models. The chromosphere can be optically thick to electron scattering out to more than 2 stellar radii, and its shock breakout radius (where $\tau = c/v_{\rm schock} \sim 30$) can be greater than 1.2 stellar radii.
}
\end{figure}

\section{Effect on Type II-P Supernovae}
\label{sec:supernova}

Our model predicts typical densities of $\rho \gtrsim 10^{-14} \, {\rm g}/{\rm cm}^3$ within three stellar radii of RSGs. This is a few orders of magnitude larger than expected for constant velocity winds with similar mass loss rates (Figure \ref{fig:DensProf}). Therefore, we expect signs of CSM interaction in type-II SNe, even for ``ordinary" pre-SN mass loss rates of $\sim \! 10^{-6}-10^{-5} \, M_{\odot}/{\rm yr}$. As mentioned above, the total CSM mass of our models is in the range of $10^{-3}-10^{-1} \, M_\odot$, slightly smaller than estimates based on light-curve modeling of type-IIP SNe exhibiting signs of CSM interaction \citep{morozova:17,dessart:17,moriya:18}.  Our chromospheres qualitatively resemble the less massive but more extended CSM structure needed to match both the light curve and narrow emission lines of type II-P \citep{dessart:23}, but detailed spectral modeling will be needed to determine whether they can reproduce observations. Here we model the impact of the chromosphere on the light curves of type II-P SNe.

\begin{figure*}
\centering
\includegraphics[scale=0.7]{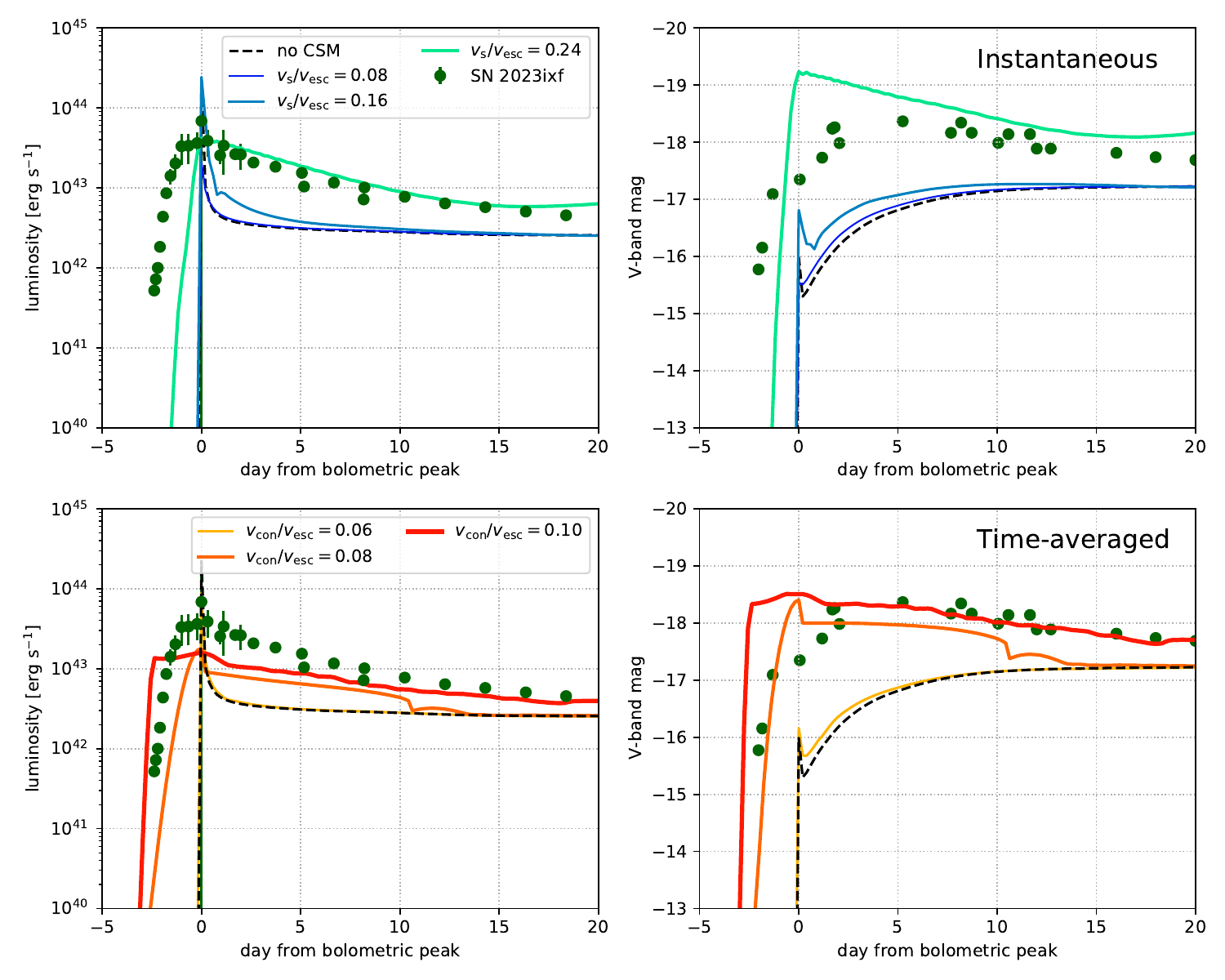}
\caption{\label{fig:lightcurves} 
Bolometric and V-band light curves of supernovae from our stellar models with extended chromospheres, given an explosion energy of $E=10^{51} \, {\rm erg}$. For comparison, we also plot the observed light curve of SN 2023ixf \citep{zimmerman:23}, which showed signs of interaction with circumstellar material, similar to many other type II-P supernovae.
}
\end{figure*}

An important property of the CSM is its optical depth. While the star is a RSG, the opacity of this neutral hydrogen/helium is very small ($\kappa \sim 10^{-3} \, {\rm cm}^2/{\rm g})$, and the CSM is optically thin except for large values of $v_{\rm con}/v_{\rm esc}$. The CSM not greatly obscure the RSG in most cases, unless dust absorption in the wind can redistribute most of the optical/NIR emission to long wavelengths. However, when the star explodes, the shock breakout emission ionizes the CSM, causing its opacity to increase approximately to the Thompson scattering value of $\kappa \approx 0.35 \, {\rm cm^2}/{\rm g}$. As the CSM recombines (but before being overtaken by the expanding SN ejecta), it may produce narrow flash-ionized emission lines as observed in early spectra of many type II-P SNe.

Figure \ref{fig:TauProf} shows the optical depth of overlying CSM for our models. Assuming Thomson scattering, we predict the CSM to have an optical depth of $\tau \sim 10^2-10^3$ at the star's surface. The shock breakout radius where $\tau = c/v_{\rm ej} \sim 30$ ranges from $\sim \! 1.1-1.5 \, R$ for our models, which would greatly alter the shock breakout emission and early SN light curve. We do not easily produce CSM with breakout radii larger than $10^{14} \, {\rm cm}$, as observed in $\sim$1/3 of supernovae according to \citep{irani:23}. However, the optical depth of the CSM can be larger than unity out to a few stellar radii (beyond $10^{14} \, {\rm cm}$), which could strongly affect the SN appearance during its first week or two.

We do not perform spectral modeling here, but we note that non-LTE modeling of flash-ionized emission lines has constrained outflow velocities to low values of $v_{\rm wind} \lesssim 30 \, {\rm km/s}$ in the case of SN 2013cu (e.g., \citealt{grafener:16}). That event was a hydrogen-poor type-IIb SN for which our models are not directly applicable. In any case, our quasi-hydrostatic chromosphere models predict low outflow/shock speeds within the dust formation radius of exploding type II-P SN progenitors, corresponding to narrow widths of $v \lesssim 30 \, {\rm km/s}$ of flash-ionized emission lines, although the widths may be increased by acceleration due to radiation pressure from the subsequent SN \citep{tsuna:23}. 

\subsection{Light curve modeling}

To estimate the effect of the CSM on type-IIP SNe, we perform light curve modeling with the SNEC code \citep{morozova:15}. The initial models are constructed with 
our CSM models for $v_{\rm s}/v_{\rm esc}=0.08$—$0.24$, and our time-averaged models with $v_{\rm con}/v_{\rm esc}=0.06$—$0.10$. These are attached to a 15 $M_\odot$ MESA model at core-collapse, as shown in Figures \ref{fig:DensProf} and \ref{fig:DensProfLarge}. We excise the innermost $1.62\ M_\odot$ of the star corresponding to the oxygen-silicon interface, as suggested by \cite{morozova:18}. The explosions are all simulated with 3000 grid cells, with energy deposited in the innermost $0.1\ M_\odot$ as a thermal bomb to obtain a final explosion energy of $10^{51}$ erg. The mass of $^{56}$Ni is fixed to $0.05\ M_\odot$, but this value does not affect the early part of the light curves, which is our focus.

When the SN shock propagates an extended dense CSM (e.g., the case for $v_{\rm s}/v_{\rm esc}=0.24$), the shock downstream is compressed into a thin dense shell due to radiative cooling. By default, SNEC obtains the ``observed luminosity” by measuring the local luminosity at the photospheric radius where $\tau=2/3$. However, due to the thin-shell nature of the shocked region, we find numerical instabilities in the luminosity after the photosphere enters the shocked region \citep[e.g., Fig 4 of][]{tsuna:23}. To avoid this, we instead measure the bolometric luminosity at the outermost part of the computational region ($\approx 3\times 10^{16}$ cm) where photons are free-streaming ($\tau\ll 2/3$). We use this luminosity and the photospheric radius, and adopt the same bolometric corrections as done in SNEC to obtain the V-band light curves.

Figure \ref{fig:lightcurves} shows the bolometric and V-band SN light curves of our models. As the ratio of $v_{\rm s}/v_{\rm esc}$ or $v_{\rm con}/v_{\rm esc}$ increases, the CSM mass increases, and the initial peak in the light curve becomes brighter and longer lasting. This peak arises primarily from shock-cooling emission of the extended CSM. For comparison, we also plot the observed light curve of SN 2023ixf \citep{zimmerman:23}, which shows prominent signs of interaction with a dense and confined CSM. Models without CSM or with low-density CSM (e.g., $v_{\rm s}/v_{\rm esc}=0.08-0.16$) do not come close to matching the observed light curve. Models with denser CSM (e.g., $v_{\rm s}/v_{\rm esc}=0.24$ or $v_{\rm con} = 0.08-0.10$) come closer to matching the data, exhibiting broad early peaks in their light curves that can last for more than a week. The agreement is not perfect, but illustrates the point that realistic chromospheres can significantly alter the shape of the early light curve in a manner resembling observations.


Important effects not included in these models are asphericity and intermittency expected for realistic RSG chromospheres. Our light curve models use angle-averaged or time-averaged chrmospheres, whereas we expect the CSM to be very aspherical in real stars, and to be fluctuating greatly in time. This will increase the effect of the CSM in a fraction of events where an energetic convective plume is on the observable side of the star at the time of explosion. In other cases where there is not an energetic convective plume at the time of the explosion, the effect of the CSM will be minor. Additionally, the CSM will be clumpy, which will shorten the diffusion time for some photons and lengthen in for others, broadening the early light curve peak due \citep{goldberg:22b}. Future multi-dimensional light curve modeling of SNe with clumpy CSM will be needed for detailed comparison with observations.

We note that the model V-band light curves can be affected by SNEC's assumption of gas-radiation equilibrium and a fully thermalized radiation field. We may expect departures from thermal spectra for the luminosity generated around shock breakout, and that powered by prolonged CSM interaction \citep[e.g.,][]{nakar:2010,svirski:2012,morozova:18,haynie:2021,tsuna:21}. 
On the other hand, for the shock cooling phase after breakout the thermalization is expected to be more complete \citep{morozova:18}. Additionally, the morphology of the light curve around the peak is possibly affected by the relatively low resolution within the breakout layer ($\tau\approx$ several tens), most notably for the ones with densest CSM. 


\section{Discussion}
\label{sec:disc}

\subsection{Comparison with red supergiant observations}

Our model predicts that chromospheric material is clumpy, stochastically varying with time, and moving both inwards and outwards with typical velocities of $\sim$10 of km/s. These speeds are slightly smaller than observed for gas within 2 stellar radii of Antares and Betelgeuse, moving at velocities of $\sim \! 20-50$ km/s \citep{ohnaka:17,lopez:22}. We also expect that RSGs with larger atmospheric motions should have larger mass loss rates, as tentatively observed \citep{josselin:07}. Such observations are typically made though near-infrared emission lines from molecules, which can extend out to a few stellar radii \citep{arroyo-torres:15}. It is not clear if our models can account for molecular emission at such large radii, and this should be investigated in future work.

Observations of AGB stars may also be useful, since the same processes are likely to be at work in those stars. For instance, CO emission lines measured with ALMA can be used to infer the density, temperature, and velocity profiles of material in the chromosphere/wind.  \citep{khouri:24} finds density profiles for R Dor that are very similar to that shown in Figure \ref{fig:DensProf}, characterized by a steep drop in density by a factor of $\sim \! 10^3$ within $1.6 \, R$, followed by a more gradual decline in density at larger radii. However, the temperature distribution they infer is somewhat steeper than that of radiative equilibrium (equation \ref{eq:temp}). More sophisticated thermodynamic modeling and an extension of our model to AGB stars (which are significantly cooler than RSGs) will help refine its predictions. 

Our model predicts that most of RSG mass loss occurs through episodic ejection events driven by shocks. These mass ejection events would be accompanied by dust formation events, which may have caused the great dimming of Betelgeuse in 2020 \citep{montarges:21,taniguchi:22}. That event was preceded by radial velocity and molecular absorption disturbances in the photosphere several months before the dust formation \citep{dupree:22,jadlovsky:23}. Additional tomography of Betelgeuse showed two strong outgoing shock waves around the time of the great dimming event \citep{kravchenko:21,jadlovsky:24}. These observations are qualitatively consistent with our picture of the chromosphere expanding and forming dust during mass ejection events, on a time scale of $\sim \! R_{\rm d}/v_{\rm esc} \sim 1 \, {\rm yr}$. We discuss this further in Section \ref{sec:predictions}.

\subsection{Comparison with prior models}

Ours is not the first paper to investigate many of these ideas. \cite{soker:21} proposed RSGs contain extended ``effervescent zones" where some material falls back towards the star while other material is accelerated outwards by photon pressure, arguing this region will be denser than expected from simple wind models. Given the low opacity of chromospheric material, we believe that hydrodynamic propulsion will be more important than photon momentum, at least for material below the dust formation radius.

\cite{moriya:18} examined how the acceleration of RSG winds can affect the CSM density near the stellar surface. They used parameterized $\beta$ velocity profiles for the wind speed, and then calculated the corresponding wind density profile. If the wind accelerates slowly, the CSM density near the star is correspondingly larger, helping to explain SNe observations requiring dense CSM, in a manner similar to our work. While we find qualitatively similar density profiles to those of a $\beta$-law model, the density profiles from our model are likely to be more realistic because they are based on a physical mechanism rather than an arbitrarily chosen velocity profile. These differences may be important when predicting the early evolution of light curves and spectra due to CSM interaction.

It is instructive to compare our model to the recent work of \cite{kee:21}. They consider turbulent pressure support of RSG atmospheres, assuming a constant turbulent speed $v_{\rm turb}$ within the chromosphere. However, they do not consider time time-variable nature of turbulent motions, so their predictions are similar to isothermal density profile of equation \ref{eq:rho}. In their models, $v_{\rm turb}$ is a free parameter that must be set to $\approx 20 \, {\rm km}/{\rm s}$ in order to achieve sensible mass loss rates. The problem is that these mass loss rates are incredibly sensitive to the chosen value of $v_{\rm turb}$: changing $v_{\rm turb}$ from 10 km/s to 20 km/s increases their mass loss rate by about six orders of magnitude (their Figure 3). It seems unlikely that stars naturally fine-tune their turbulent velocities to such an extent. Our model builds on theirs, showing that time-averaging reduces the sensitivity to $v_{\rm turb}$, and we link this value to the near-surface convective velocities rather than letting it be a free parameter.

Although sometimes forgotten by the massive star community, models for shock-supported stellar chromospheres have been developed for decades (e.g., \citealt{hill:1979,bertschinger:85,bowen:88}), primarily in the context of radially pulsating AGB stars and Mira variables. Our model is largely a reapplication of previous models to RSGs, but here the shocks are launched by convective motions rather than large-scale pulsation. This allows for analytic estimates of chromospheric densities and stellar mass loss rates, which can be readily compared with data or implemented into stellar evolution codes. The most important new aspect of our model is the inclusion of a spectrum of shock speeds $v_s$ rather than assuming a single value. The high-velocity tail of this spectrum greatly increases the time-averaged chromospheric density and mass loss rate.

\cite{hill:1979} found approximately periodic and hydrostatic (in a time-averaged sense) chromospheric solutions as long as the dimensionless ratio $P_0/P$ (their equation 2) is small. In the relevant limit $v_{\rm con} \ll v_{\rm esc}$, this requires
\begin{equation}
    \frac{P_0}{P} \simeq \frac{4 r v_{\rm con}}{v_{\rm esc}^2 P_{\rm s}} \, .
\end{equation}
As discussed in Section \ref{sec:chromosphere}, for a a typical period between shocks of $P_{\rm s} \sim 0.5 R/v_{\rm con}$, we obtain $P_0/P \sim 8 (r/R) (v_{\rm con}/v_{\rm esc})^2$, which is less than unity below the dust formation radius for all our RSG models. Hence we expect our assumption of an approximately hydrostatic to be valid except when fairly strong shocks propagate through the chromosphere, which should be investigated more thoroughly with numerical simulations.

\subsection{Comparison with red supergiant simulations}

It is useful to compare our model with simulations of RSGs, such as the 3D radiation hydrodynamical simulations of \cite{goldberg:22} (see also \citealt{chinavassa:11,chinavassa:11b,ma:23,chiavassa:24}). These simulations show RMS near-surface convective velocities of $\sim$10 km/s, similar to those predicted by mixing length theory, for values of $\alpha_{\rm MLT} \sim 2-4$. These help justify our use MLT to compute the values of $v_{\rm con}$ in our models. However, our predicted mass loss rates are very sensitive to the convective velocities (see Section \ref{sec:uncertainties}), so it will be necessary to directly compute mass loss in simulations in order to test and calibrate our model.

However existing simulations (e.g., those of \citealt{goldberg:22}) may have difficulty correctly predicting the chromospheric density structure. For numerical reasons, the simulations have a density floor, so this material surrounding the star falls onto it. There is also a numerical transient at the start of the simulations that ejects some mass, and some of this mass falls back onto the star over the course of the simulations. These effects cause an accretion shock to form between the stellar surface and the infalling mass (visible in their Figures 6, 10, and 12), which compresses the chromosphere. A similar phenomenon may be present in the CO5BOLD simulations of \cite{wedemeyer:17} (see their Figure 2). This may compress the chromospheric structure, decreasing the density of material at $r \! \gtrsim 1.5 R$. An additional issue is that \cite{goldberg:22} assume fully ionized gas in their energy equation and mean molecular weight, whereas real RSGs have primarily neutral atomic H/He in their photospheric and chromosopheric layers. Including recombination energy may be important for the dynamics of material near and above the photosphere. Future simulations that address these issues can better determine the chromospheric structure of RSGs. 

Some very useful existing simulations are those of AGB stars and their dust-driven winds \citep{freytag:08,freytag:17,freytag:23}. These simulations indeed exhibit RMS velocities of material in and above the photosphere that are nearly constant radius, with RMS radial velocities of $\sim$5-10 km/s and Mach numbers of order $\mathcal{M} \sim 2-3$. The density profiles of those models also resemble ours in Figure \ref{fig:DensProf}, with a sharp drop in density by a few orders of magnitude just above the surface of the star, followed by a more gradual decline at larger radii. Their AGB models have higher values of $v_{\rm con}/v_{\rm esc}$, cooler surface temperatures, and smaller dust formation radii than we expect for RSGs, so a detailed comparison should await simulations in the RSG regime.

\subsection{Red supergiant mass loss and variability}
\label{sec:predictions}

Our model predicts that both the chromospheric structure and the mass loss rate of RSGs are dynamic, varying stochastically on a convective turnover time of $\sim$months. While the plots in this paper are angle-averaged and time-averaged quantities, the actual chromospheric density is expected to exhibit large fluctuations around this mean. Variations of dust spectral features from the wind acceleration region (which are used for mass loss measurements) are likely slower, varying on a time scale of $\sim$months-years. This could cause inferred mass loss rates to change greatly with time, which could account for differences between different groups' measurements, such as the different mass loss rates of \cite{beasor:22} compared to those of \cite{vanloon:05}. Future observations should investigate time variations in RSG spectra and inferred mass loss rates. 

We can also estimate the likelihood of observing a star with an enhanced CSM density and mass loss rate. As discussed above, the mass loss is dominated by uncommon convective episodes with $v \sim \sqrt{v_{\rm con} v_{\rm esc}} \sim 3 v_{\rm con}$. The probability of a given convective element reaching such velocities is $p \sim e^{-v_{\rm esc}/2v_{\rm con}} \sim 10^{-3}$. For a star with $\sim$10 surface convective elements and a convective turnover time of $\sim$100-300 days, we expect to wait $\sim$30-100 years between major mass ejection events.

Each mass loss event would last for a convective turnover time of $\sim$100 days, ejecting mass at a rate $\sim \! 10^{-4}-10^{-3} \, M_\odot$/yr (see green line in Figure \ref{fig:DensProfLarge}), ejecting a total mass of $M_{\rm ej} \! \sim \! 3\times 10^{-5}-3\times 10^{-4} \, M_\odot$ and about $3 \times 10^{-7}- 3\times 10^{-6} \, M_\odot$ of dust. For a dust-laden gas opacity of $\kappa_{\rm d} \sim 3 \, {\rm cm}^2/{\rm g}$, this mass can have an optical depth $\tau \sim 1-10$ at the dust formation radius, temporarily obscuring the star and creating a dimming event similar to that of Betelgeuse in 2020. Such stars would be expected to exhibit great dimming events every century or so, possibly in line with historical evolution of Betelgeuse. Stars with large mass loss rates of $10^{-4} \, M_\odot/{\rm yr}$ (as inferred for the progenitor of SN 2023ixf) would suffer a mass ejection event every few years. Since the gas would remain near the star for a time scale of $R_{\rm d}/v_{\rm wind} \sim 2 \, {\rm yr}$, the probability of observing such a star with a very dense CSM at the moment of a SN explosion is near unity.

\subsection{Comparison with supernovae and SN 2023ixf}
\label{sec:2023ixf}

Observations of type-IIP supernovae have frequently indicated the presence of dense CSM within $\sim \! 10^{15}$ {\rm cm} ($\sim$20 stellar radii) of exploding RSGs. This material is revealed by flash-ionized emission lines (e.g., \citealt{khazov:16}) from recombination following ionization by shock breakout and continued irradation from the shocked CSM. The emission lines typically fade within $\sim$5 days after shock-breakout \citep{bruch:23} as this CSM is swept up by the SN ejecta that travels outwards at $\sim \! 10^{9}$ cm/s ($\sim \! 10^{14}$ cm/day). Early light curves of type-IIP SNe often cannot be explained by bare RSGs, but can be explained by interaction with dense CSM. This material has frequently been interpreted and modeled (e.g., \citealt{fuller:17,morozova:20,bruch:23}) as arising from intense pre-supernova mass loss in the final decade of the star's life.

Our model naturally predicts dense chromospheric material within $\sim$10 stellar radii ($\sim 5 \times 10^{14} \, {\rm cm}$) of exploding RSGs. This material will be swept up by the SN ejecta in $\sim$5 days, in line with the disappearance of flash-ionized emission lines. We believe such material is essentially the chromosphere of the RSG, rather than being mass ejected from the star just before explosion. Pre-SN outbursts likely occur in rarer cases, e.g., SN 2020tlf \citep{jacobson-galan:22}, and in type-IIn SNe \citep{ofek:14,strotjohann:21}. Importantly, our model produces dense CSM with only moderate pre-SN mass loss rates of $\dot{M} \sim 10^{-5} M_\odot/{\rm yr}$. This is not enough to completely obscure the star (see Section \ref{sec:supernova}), avoiding the problem of progenitor obscuration by superwinds \citep{davies:22}.

Observations of SN 2023ixf and other type II-P SNe are qualitatively consistent with our model. In SN 2023ixf, modeling of flash-ionized lines \citep{zhang:23} suggests CSM densities of $\sim \! 10^{-14} \, {\rm g}/{\rm cm}^3$ at $\sim$5 stellar radii, only slightly above the prediction of our time-averaged model in Figure \ref{fig:DensProfLarge}. The outer ``edge" of this dense CSM has been inferred to be at $\sim$10 stellar radii, near the dust formation radius where we expect the CSM density to approach that of a constant-velocity wind model \citep{bostroem:23}. The average CSM density inferred from that work, $\rho \sim \! 5 \times 10^{-14} \, {\rm g}/{\rm cm}^3$, is slightly larger than our time-averaged model, which has a volume-averaged CSM density of $\rho \sim \! 5 \times 10^{-15} \, {\rm g}/{\rm cm}^3$ between 1.5 stellar radii and the dust formation radius.

\cite{zimmerman:23} infers an effective shock breakout radius for SN 2023ixf of $\sim \! 2 \times 10^{14} \, {\rm cm}$ (about 4 stellar radii), substantially larger than our time-averaged model (Figure \ref{fig:TauProf}). X-ray observations and SED modeling also suggest large mass loss rates of $\dot{M} \sim 10^{-4} \, M_\odot/{\rm yr}$ \citep{grefenstette:23,jencson:23}. We cannot explain this with our models unless the progenitor's mass loss rate at core-collapse rate was larger than average. This is certainly possible since our model predicts stochastic variations in time.  Moreover, we expect the CSM to be clumpy such that it is denser and more optically thick on one side of the star, increasing the inferred shock-breakout radius for many viewing angles.

The CSM density at $\sim \! 10^{15} \, {\rm cm}$ is inferred to have a density of $\sim \! 4 \times 10^{-16} \, {\rm g}/{\rm cm}^3$ \citep{zimmerman:23}, which is slightly larger than predicted by our time-averaged model, and again could be reconciled by a model with a larger instantaneous mass loss rate. \cite{zimmerman:23} interprets this as evidence for a sharp drop in CSM density somewhere above the shock-breakout radius. However, it is not clear from the data that the CSM has a sharp edge: what is required is that the CSM density fall steeper than $r^{-2}$ from $\sim \! 10^{14}-10^{15}$ cm, but it could be a smooth drop as predicted by our models. The CSM densities measured from X-ray emission of SN 2023ixf \citep{grefenstette:23} are similar to those above, and once again a factor of several larger than our model in Figure \ref{fig:DensProfLarge}.

Multiple efforts to model the light curve of SN 2023ixf \citep{jacobson-galan:23,hiramatsu:23,li:23} have concluded that its early light curve requires dense CSM. For instance, \citep{jacobson-galan:22} infers a density of $\rho \sim 10^{-12} \, {\rm cm}$ at $\sim$2 stellar radii, slightly larger than our time-averaged model in Figure \ref{fig:DensProf}. These high densities are often interpreted as evidence for pre-supernova mass loss rates of $\sim \! 10^{-2}-10^{-1} \, M_\odot/{\rm yr}$. Similar inferences are often made from flash-ionized emission lines, and have been made for many other type II-P SNe as well (e.g., \citealt{yaron:17,moriya:18}). While the above CSM density estimates may be reasonable, we believe the extreme mass loss rates are greatly over-estimated. These mass loss rates are obtained by assuming the CSM is moving out at 10s of km/s. Instead, RSG chromospheres within several stellar radii could be approximately hydrostatic and bound, with mean outflow velocities (Figure \ref{fig:VelProf}) much slower than assumed by supernova modelers. The actual mass loss rates are correspondingly smaller. Supernova observers and modelers should stick to quoting CSM masses or densities rather than attempting to infer mass loss rates.

Another expectation of our model is that the CSM material is asymmetric and clumpy, stochastically varying in time. This is in line with observations of SN 2023ixf, where the early (and changing) continuum polarization indicates an aspherical CSM \citep{vasyleyv:23}. Spectral modeling of this event also suggests an aspherical CSM \citep{smith:23}. We suspect an aspherical CSM will also lengthen the duration of shock breakout emission \citep{goldberg:22b} and rise to V-band maximum, which may help explain the slow rise of events like SN 2023ixf relative to models (e.g., Figure \ref{fig:lightcurves} and \citealt{hosseinzadeh:23}). We note that type IIb SN 2013cu was inferred to have a smooth and spherical wind by modeling its narrow emission lines \citep{grafener:16}, but the hydrogen-poor progenitor of that event may have a different mass loss mechanism than our model for hydrogen-rich RSGs. Future hydrodynamic and radiative transfer modelling of our scenario will be needed for detailed predictions. 

An interesting and unusual feature of SN 2023ixf is that its progenitor was clearly pulsating at large amplitude \citep{kilpatrick:23,jencson:23}. Many RSGs in the Milky Way and M31 are also observed to pulsate fairly coherently, though often at somewhat smaller amplitudes \citep{soraisam:18}. Similar pulsations are thought to be crucial to the mass loss of AGB stars by driving shocks through the atmosphere \citep{hofner:18,freytag:17}. Our model could possibly account for this by replacing the convective velocity $v_{\rm con}$ with the pulsational velocity near the photosphere, which can be supersonic and hence larger than $v_{\rm con}$, driving higher mass loss rates. It is thus possible that the pulsations of SN 2023ixf's progenitor created its unusually large CSM density and mass loss rate, as has been suggested previously \citep{yooncantiello:10}.

\subsection{Caveats and uncertainties}
\label{sec:uncertainties}

The largest uncertainty of our model arises from its sensitive dependence to the convective velocity distribution. Since the chromospheric density is dominated by convective episodes with $v \gtrsim 2 v_{\rm con}$ the density and mass loss from this mechanism depend very sensitively on the high-energy tail of the convective distribution. We have assumed this tail has a Gaussian distribution, as expected for radial velocity variations in isotropic Kolmogorov turbulence. However, the vigorous near-surface convection of RSGs could behave differently. 

One difference between these regimes is the convective Mach numbers, which are small in stellar interiors but are of order unity near the surface of a RSG. Convection is neither in the low-Mach number, incompressible regime of Kolmogorov turbulence, nor in the high-Mach number and highly compressible regime of Burgers turbulence. Three-dimensional simulations (e.g., \citealt{rabatin:23,rabatin:23b} of such turbulence indicate that the convective speeds are fairly well approximated by a Maxwellian distribution, while the density is well approximated by a log-normal distribution, without strong correlations between density and velocity. However, this could be changed by effects of intermittency \citep{hopkins:13}. Additionally, RSG atmospheres are not isotropic and can reach near-Eddington luminosities, greatly changing the nature of convection (e.g., \citealt{goldberg:22}) and correlations between density and velocity. Further numerical work on non-isotropic, moderate-Mach number, convectively driven turbulence is needed to address this uncertainty.

We neglected radiative forces within the chromosphere, justified by the fact that typical opacities at these temperatures are of order $\kappa \sim 10^{-4}-10^{-2} \, {\rm cm^2}/{\rm g}$ \citep{plez:92}. The corresponding Eddington factor, i.e., the ratio between radiative acceleration and gravity, is $f_{\rm ed} = g/g_{\rm rad} = L/L_{\rm ed} = L \kappa/(4 \pi G M c) \sim 10^{-4}-10^{-2}$ at these opacities. This is approximately consistent with radiative transfer modeling of red giant atmospheres \citep{ireland:08} that find only small radiative accelerations within the chromosphere, although it has been suggested turbulent motions may decrease line shadowing and increase the effective opacity enough to have a substantial effect \citep{josselin:07}. Since our predicted mass loss rates are strongly dependent on $v_{\rm esc}$ and hence on the net effect of gravity and radiation forces, even small radiative forces could substantially increase chromospheric densities and associated mass loss rates, and should be incorporated in more sophisticated versions of our model.

We also neglected the effects of rotation and magnetic fields in our models, which is reasonable given the small rotation rates of RSGs, and the lack of strong surface fields. The cool chromospheric material is almost completely neutral, diminishing the importance of magnetic fields. Furthermore, the thermal and turbulent pressures of our time-averaged model shown in Figure \ref{fig:DensProf} are larger than the magnetic pressure of a dipole field for a surface field strength of 1 G, so fairly large magnetic fields are required in order to dominate the dynamics. \cite{thirumalai:10,thirumalai:12} investigated a hybrid magnetic Weber-Davis and dust-driven wind model for red giants and RSGs and its affect on the terminal wind speeds. However, the mass loss rate is imposed in such models (rather than being calculated from first principles), so such models do not solve the wind launching problem below the dust formation radius. 

Finally, our model for the dust-driven wind above the dust formation radius is undoubtedly too simple. More realistic dust chemistry, grain formation, and radiative driving should be developed for a better model of the wind material. Additionally, resonant drag instabilities will occur in the dust-driven ouflow \citep{hopkins:18,moseley:19,steinwandel:22}, which may affect the outflow speed and appearance. Understanding such phenomena may be needed not only for mass loss models, but also for more accurate observational mass loss rate estimates.

\section{Conclusions}

We have developed a simple model for material supported by shock waves in the chromospheres of red supergiants (RSGs). These chromospheres extend roughly from the stellar surface to the dust formation radius at $\sim$5 stellar radii. Gravity is balanced by momentum deposition from outgoing shock waves, whose effect is similar to turbulent pressure support. For our assumption of nearly constant shock speeds (justified by simulation results), the density profile is similar to that of an isothermal atmosphere with scale height $v_{\rm con}^2/g \ll R$, which would cause a steeply decreasing density for shock speeds comparable to the surface convective speeds $v_{\rm con}$. However, we also showed that averaging over a stochastically varying convective velocity spectrum greatly increases the chromospheric density, causing it to fall off less steeply with radius. 

Using the time-averaged chromospheric density at the dust formation radius, we construct a simple and analytic estimate for the mass loss rate of RSGs (equation \ref{eq:mdot}). This yields mass loss rates approximately compatible with observations of luminous RSGs (Figure \ref{fig:MassLoss}). The predicted mass loss rates scale exponentially with the ratio of $v_{\rm con}/v_{\rm esc}$, resulting in larger mass loss rates for larger radii, smaller mass, or higher convective velocities. These models can potentially explain: i) mass loss rates of $\sim 10^{-6} \, M_\odot/{\rm yr}$ for $15-20 \, M_\odot$ RSGs, ii) substantially lower mass loss rates for less massive RSGs as found by recent observations, iii) the absence of RSGs for luminosities $L \gtrsim 3 \times 10^5 \, L_{\odot}$ due to rapid boil-off of their hydrogen envelopes. Our models also predict pre-supernova mass loss rates of $\sim \! 4 \times 10^{-7}-4 \times 10^{-5} \, M_\odot$/yr, a factor 10-100 times larger than mass loss rates during core helium burning.

Our models predict chromospheric masses of $\sim \! 10^{-2} \, M_\odot$, most of which is confined within one stellar radius above the photosphere. Nonetheless, the typical chromospheric densities of our models can be a few orders of magnitude larger than predicted by a constant-velocity wind model. This chromospheric material likely helps explain i) flash-ionized emission lines observed in the first week of many type II SNe ii) early peaks in the light curves of these SNe that cannot be explained for bare RSG models iii) extended shock breakout emission (and larger shock breakout radius) of type II SNe. 

Because convective turbulence varies stochastically, we expect RSGs chromospheres to vary stochastically as well, meaning they are likely to be spatially asymmetric at any moment in time, and their total mass varies greatly in time. This may account for the time-varying molecular emission that is interferometrically observed above the surfaces of RSGs such as Betelgeuse and Antares. It may also explain the great dimming event of Betelgeuse as a moment of enhanced convective speeds, leading to enhanced mass loss, dust formation, and obscuration. Such stochasticity is also expected for type II SNe, meaning that the CSM densities (and flash ionized line strengths, etc.) will likely vary greatly from one SN to the next, even for progenitors of the same mass. 

Our chromospheric model is extremely simple and will need to be refined and improved for more robust predictions. Since the chromospheric mass scales exponentially with the convective velocity $v_{\rm con}$, the most important uncertainty of our model is the actual velocity spectrum of convective motions and their radial dependence within the chromosphere. The model will also be improved with a better treatment of dust formation and radiative driving of the wind around the dust formation radius. These uncertainties can be addressed with future simulations of RSG chromospheres, continued observations of RSGs and their mass loss rates, and early observations of type II SNe.

\section*{Acknowledgments}

We thank Luc Dessart, Jared Goldberg, Shazrene Mohamed, Konstantin Batygin, and Lars Bildsten for useful discussions, and Kostas Antoniadis for supplying mass loss measurements. This work benefited from collaborative meetings supported by the Gordon and Betty Moore Foundation (grant GBMF5076). This research was supported by the Munich Institute for Astro-, Particle and BioPhysics (MIAPbP) which is funded by the Deutsche Forschungsgemeinschaft (DFG, German Research Foundation) under Germany´s Excellence Strategy – EXC-2094 – 390783311. This research is also supported by the Sherman Fairchild Postdoctoral Fellowship at the California Institute of Technology.

\bibliographystyle{apsrev4-1}
\bibliography{MassiveWavesBib,bib}

\begin{thebibliography}{109}%
\makeatletter
\providecommand \@ifxundefined [1]{%
 \@ifx{#1\undefined}
}%
\providecommand \@ifnum [1]{%
 \ifnum #1\expandafter \@firstoftwo
 \else \expandafter \@secondoftwo
 \fi
}%
\providecommand \@ifx [1]{%
 \ifx #1\expandafter \@firstoftwo
 \else \expandafter \@secondoftwo
 \fi
}%
\providecommand \natexlab [1]{#1}%
\providecommand \enquote  [1]{``#1''}%
\providecommand \bibnamefont  [1]{#1}%
\providecommand \bibfnamefont [1]{#1}%
\providecommand \citenamefont [1]{#1}%
\providecommand \href@noop [0]{\@secondoftwo}%
\providecommand \href [0]{\begingroup \@sanitize@url \@href}%
\providecommand \@href[1]{\@@startlink{#1}\@@href}%
\providecommand \@@href[1]{\endgroup#1\@@endlink}%
\providecommand \@sanitize@url [0]{\catcode `\\12\catcode `\$12\catcode `\&12\catcode `\#12\catcode `\^12\catcode `\_12\catcode `\%12\relax}%
\providecommand \@@startlink[1]{}%
\providecommand \@@endlink[0]{}%
\providecommand \url  [0]{\begingroup\@sanitize@url \@url }%
\providecommand \@url [1]{\endgroup\@href {#1}{\urlprefix }}%
\providecommand \urlprefix  [0]{URL }%
\providecommand \Eprint [0]{\href }%
\providecommand \doibase [0]{http://dx.doi.org/}%
\providecommand \selectlanguage [0]{\@gobble}%
\providecommand \bibinfo  [0]{\@secondoftwo}%
\providecommand \bibfield  [0]{\@secondoftwo}%
\providecommand \translation [1]{[#1]}%
\providecommand \BibitemOpen [0]{}%
\providecommand \bibitemStop [0]{}%
\providecommand \bibitemNoStop [0]{.\EOS\space}%
\providecommand \EOS [0]{\spacefactor3000\relax}%
\providecommand \BibitemShut  [1]{\csname bibitem#1\endcsname}%
\let\auto@bib@innerbib\@empty
\bibitem [{\citenamefont {{Reimers}}(1975)}]{reimers:75}%
  \BibitemOpen
  \bibfield  {author} {\bibinfo {author} {\bibfnamefont {D.}~\bibnamefont {{Reimers}}},\ }\href@noop {} {\bibfield  {journal} {\bibinfo  {journal} {Memoires of the Societe Royale des Sciences de Liege}\ }\textbf {\bibinfo {volume} {8}},\ \bibinfo {pages} {369} (\bibinfo {year} {1975})}\BibitemShut {NoStop}%
\bibitem [{\citenamefont {{Antoniadis}}\ \emph {et~al.}(2024)\citenamefont {{Antoniadis}}, \citenamefont {{Bonanos}}, \citenamefont {{de Wit}}, \citenamefont {{Zapartas}}, \citenamefont {{Munoz-Sanchez}},\ and\ \citenamefont {{Maravelias}}}]{antoniadis:24}%
  \BibitemOpen
  \bibfield  {author} {\bibinfo {author} {\bibfnamefont {K.}~\bibnamefont {{Antoniadis}}}, \bibinfo {author} {\bibfnamefont {A.~Z.}\ \bibnamefont {{Bonanos}}}, \bibinfo {author} {\bibfnamefont {S.}~\bibnamefont {{de Wit}}}, \bibinfo {author} {\bibfnamefont {E.}~\bibnamefont {{Zapartas}}}, \bibinfo {author} {\bibfnamefont {G.}~\bibnamefont {{Munoz-Sanchez}}}, \ and\ \bibinfo {author} {\bibfnamefont {G.}~\bibnamefont {{Maravelias}}},\ }\href {\doibase 10.48550/arXiv.2401.15163} {\bibfield  {journal} {\bibinfo  {journal} {arXiv e-prints}\ ,\ \bibinfo {eid} {arXiv:2401.15163}} (\bibinfo {year} {2024})},\ \Eprint {http://arxiv.org/abs/2401.15163} {arXiv:2401.15163 [astro-ph.SR]} \BibitemShut {NoStop}%
\bibitem [{\citenamefont {{Levesque}}\ \emph {et~al.}(2005)\citenamefont {{Levesque}}, \citenamefont {{Massey}}, \citenamefont {{Olsen}}, \citenamefont {{Plez}}, \citenamefont {{Josselin}}, \citenamefont {{Maeder}},\ and\ \citenamefont {{Meynet}}}]{levesque:05}%
  \BibitemOpen
  \bibfield  {author} {\bibinfo {author} {\bibfnamefont {E.~M.}\ \bibnamefont {{Levesque}}}, \bibinfo {author} {\bibfnamefont {P.}~\bibnamefont {{Massey}}}, \bibinfo {author} {\bibfnamefont {K.~A.~G.}\ \bibnamefont {{Olsen}}}, \bibinfo {author} {\bibfnamefont {B.}~\bibnamefont {{Plez}}}, \bibinfo {author} {\bibfnamefont {E.}~\bibnamefont {{Josselin}}}, \bibinfo {author} {\bibfnamefont {A.}~\bibnamefont {{Maeder}}}, \ and\ \bibinfo {author} {\bibfnamefont {G.}~\bibnamefont {{Meynet}}},\ }\href {\doibase 10.1086/430901} {\bibfield  {journal} {\bibinfo  {journal} {\apj}\ }\textbf {\bibinfo {volume} {628}},\ \bibinfo {pages} {973} (\bibinfo {year} {2005})},\ \Eprint {http://arxiv.org/abs/astro-ph/0504337} {arXiv:astro-ph/0504337 [astro-ph]} \BibitemShut {NoStop}%
\bibitem [{\citenamefont {{Hoyle}}\ and\ \citenamefont {{Wickramasinghe}}(1962)}]{hoyle:62}%
  \BibitemOpen
  \bibfield  {author} {\bibinfo {author} {\bibfnamefont {F.}~\bibnamefont {{Hoyle}}}\ and\ \bibinfo {author} {\bibfnamefont {N.~C.}\ \bibnamefont {{Wickramasinghe}}},\ }\href {\doibase 10.1093/mnras/124.5.417} {\bibfield  {journal} {\bibinfo  {journal} {\mnras}\ }\textbf {\bibinfo {volume} {124}},\ \bibinfo {pages} {417} (\bibinfo {year} {1962})}\BibitemShut {NoStop}%
\bibitem [{\citenamefont {{Field}}(1974)}]{field:74}%
  \BibitemOpen
  \bibfield  {author} {\bibinfo {author} {\bibfnamefont {G.~B.}\ \bibnamefont {{Field}}},\ }\href {\doibase 10.1086/152654} {\bibfield  {journal} {\bibinfo  {journal} {\apj}\ }\textbf {\bibinfo {volume} {187}},\ \bibinfo {pages} {453} (\bibinfo {year} {1974})}\BibitemShut {NoStop}%
\bibitem [{\citenamefont {{H{\"o}fner}}\ and\ \citenamefont {{Olofsson}}(2018)}]{hofner:18}%
  \BibitemOpen
  \bibfield  {author} {\bibinfo {author} {\bibfnamefont {S.}~\bibnamefont {{H{\"o}fner}}}\ and\ \bibinfo {author} {\bibfnamefont {H.}~\bibnamefont {{Olofsson}}},\ }\href {\doibase 10.1007/s00159-017-0106-5} {\bibfield  {journal} {\bibinfo  {journal} {\aapr}\ }\textbf {\bibinfo {volume} {26}},\ \bibinfo {eid} {1} (\bibinfo {year} {2018})}\BibitemShut {NoStop}%
\bibitem [{\citenamefont {{H{\"o}fner}}\ \emph {et~al.}(2003)\citenamefont {{H{\"o}fner}}, \citenamefont {{Gautschy-Loidl}}, \citenamefont {{Aringer}},\ and\ \citenamefont {{J{\o}rgensen}}}]{hofner:03}%
  \BibitemOpen
  \bibfield  {author} {\bibinfo {author} {\bibfnamefont {S.}~\bibnamefont {{H{\"o}fner}}}, \bibinfo {author} {\bibfnamefont {R.}~\bibnamefont {{Gautschy-Loidl}}}, \bibinfo {author} {\bibfnamefont {B.}~\bibnamefont {{Aringer}}}, \ and\ \bibinfo {author} {\bibfnamefont {U.~G.}\ \bibnamefont {{J{\o}rgensen}}},\ }\href {\doibase 10.1051/0004-6361:20021757} {\bibfield  {journal} {\bibinfo  {journal} {\aap}\ }\textbf {\bibinfo {volume} {399}},\ \bibinfo {pages} {589} (\bibinfo {year} {2003})}\BibitemShut {NoStop}%
\bibitem [{\citenamefont {{Cranmer}}\ and\ \citenamefont {{Saar}}(2011)}]{cranmer:11}%
  \BibitemOpen
  \bibfield  {author} {\bibinfo {author} {\bibfnamefont {S.~R.}\ \bibnamefont {{Cranmer}}}\ and\ \bibinfo {author} {\bibfnamefont {S.~H.}\ \bibnamefont {{Saar}}},\ }\href {\doibase 10.1088/0004-637X/741/1/54} {\bibfield  {journal} {\bibinfo  {journal} {\apj}\ }\textbf {\bibinfo {volume} {741}},\ \bibinfo {eid} {54} (\bibinfo {year} {2011})},\ \Eprint {http://arxiv.org/abs/1108.4369} {arXiv:1108.4369 [astro-ph.SR]} \BibitemShut {NoStop}%
\bibitem [{\citenamefont {{Freytag}}\ and\ \citenamefont {{H{\"o}fner}}(2008)}]{freytag:08}%
  \BibitemOpen
  \bibfield  {author} {\bibinfo {author} {\bibfnamefont {B.}~\bibnamefont {{Freytag}}}\ and\ \bibinfo {author} {\bibfnamefont {S.}~\bibnamefont {{H{\"o}fner}}},\ }\href {\doibase 10.1051/0004-6361:20078096} {\bibfield  {journal} {\bibinfo  {journal} {\aap}\ }\textbf {\bibinfo {volume} {483}},\ \bibinfo {pages} {571} (\bibinfo {year} {2008})}\BibitemShut {NoStop}%
\bibitem [{\citenamefont {{Freytag}}\ \emph {et~al.}(2017)\citenamefont {{Freytag}}, \citenamefont {{Liljegren}},\ and\ \citenamefont {{H{\"o}fner}}}]{freytag:17}%
  \BibitemOpen
  \bibfield  {author} {\bibinfo {author} {\bibfnamefont {B.}~\bibnamefont {{Freytag}}}, \bibinfo {author} {\bibfnamefont {S.}~\bibnamefont {{Liljegren}}}, \ and\ \bibinfo {author} {\bibfnamefont {S.}~\bibnamefont {{H{\"o}fner}}},\ }\href {\doibase 10.1051/0004-6361/201629594} {\bibfield  {journal} {\bibinfo  {journal} {\aap}\ }\textbf {\bibinfo {volume} {600}},\ \bibinfo {eid} {A137} (\bibinfo {year} {2017})},\ \Eprint {http://arxiv.org/abs/1702.05433} {arXiv:1702.05433 [astro-ph.SR]} \BibitemShut {NoStop}%
\bibitem [{\citenamefont {{Freytag}}\ and\ \citenamefont {{H{\"o}fner}}(2023)}]{freytag:23}%
  \BibitemOpen
  \bibfield  {author} {\bibinfo {author} {\bibfnamefont {B.}~\bibnamefont {{Freytag}}}\ and\ \bibinfo {author} {\bibfnamefont {S.}~\bibnamefont {{H{\"o}fner}}},\ }\href {\doibase 10.1051/0004-6361/202244992} {\bibfield  {journal} {\bibinfo  {journal} {\aap}\ }\textbf {\bibinfo {volume} {669}},\ \bibinfo {eid} {A155} (\bibinfo {year} {2023})},\ \Eprint {http://arxiv.org/abs/2301.11836} {arXiv:2301.11836 [astro-ph.SR]} \BibitemShut {NoStop}%
\bibitem [{\citenamefont {{Yaron}}\ \emph {et~al.}(2017)\citenamefont {{Yaron}}, \citenamefont {{Perley}}, \citenamefont {{Gal-Yam}}, \citenamefont {{Groh}}, \citenamefont {{Horesh}}, \citenamefont {{Ofek}}, \citenamefont {{Kulkarni}}, \citenamefont {{Sollerman}}, \citenamefont {{Fransson}}, \citenamefont {{Rubin}}, \citenamefont {{Szabo}}, \citenamefont {{Sapir}}, \citenamefont {{Taddia}}, \citenamefont {{Cenko}}, \citenamefont {{Valenti}}, \citenamefont {{Arcavi}}, \citenamefont {{Howell}}, \citenamefont {{Kasliwal}}, \citenamefont {{Vreeswijk}}, \citenamefont {{Khazov}}, \citenamefont {{Fox}}, \citenamefont {{Cao}}, \citenamefont {{Gnat}}, \citenamefont {{Kelly}}, \citenamefont {{Nugent}}, \citenamefont {{Filippenko}}, \citenamefont {{Laher}}, \citenamefont {{Wozniak}}, \citenamefont {{Lee}}, \citenamefont {{Rebbapragada}}, \citenamefont {{Maguire}}, \citenamefont {{Sullivan}},\ and\ \citenamefont {{Soumagnac}}}]{yaron:17}%
  \BibitemOpen
  \bibfield  {author} {\bibinfo {author} {\bibfnamefont {O.}~\bibnamefont {{Yaron}}}, \bibinfo {author} {\bibfnamefont {D.~A.}\ \bibnamefont {{Perley}}}, \bibinfo {author} {\bibfnamefont {A.}~\bibnamefont {{Gal-Yam}}}, \bibinfo {author} {\bibfnamefont {J.~H.}\ \bibnamefont {{Groh}}}, \bibinfo {author} {\bibfnamefont {A.}~\bibnamefont {{Horesh}}}, \bibinfo {author} {\bibfnamefont {E.~O.}\ \bibnamefont {{Ofek}}}, \bibinfo {author} {\bibfnamefont {S.~R.}\ \bibnamefont {{Kulkarni}}}, \bibinfo {author} {\bibfnamefont {J.}~\bibnamefont {{Sollerman}}}, \bibinfo {author} {\bibfnamefont {C.}~\bibnamefont {{Fransson}}}, \bibinfo {author} {\bibfnamefont {A.}~\bibnamefont {{Rubin}}}, \bibinfo {author} {\bibfnamefont {P.}~\bibnamefont {{Szabo}}}, \bibinfo {author} {\bibfnamefont {N.}~\bibnamefont {{Sapir}}}, \bibinfo {author} {\bibfnamefont {F.}~\bibnamefont {{Taddia}}}, \bibinfo {author} {\bibfnamefont {S.~B.}\ \bibnamefont {{Cenko}}}, \bibinfo {author} {\bibfnamefont {S.}~\bibnamefont {{Valenti}}}, \bibinfo {author}
  {\bibfnamefont {I.}~\bibnamefont {{Arcavi}}}, \bibinfo {author} {\bibfnamefont {D.~A.}\ \bibnamefont {{Howell}}}, \bibinfo {author} {\bibfnamefont {M.~M.}\ \bibnamefont {{Kasliwal}}}, \bibinfo {author} {\bibfnamefont {P.~M.}\ \bibnamefont {{Vreeswijk}}}, \bibinfo {author} {\bibfnamefont {D.}~\bibnamefont {{Khazov}}}, \bibinfo {author} {\bibfnamefont {O.~D.}\ \bibnamefont {{Fox}}}, \bibinfo {author} {\bibfnamefont {Y.}~\bibnamefont {{Cao}}}, \bibinfo {author} {\bibfnamefont {O.}~\bibnamefont {{Gnat}}}, \bibinfo {author} {\bibfnamefont {P.~L.}\ \bibnamefont {{Kelly}}}, \bibinfo {author} {\bibfnamefont {P.~E.}\ \bibnamefont {{Nugent}}}, \bibinfo {author} {\bibfnamefont {A.~V.}\ \bibnamefont {{Filippenko}}}, \bibinfo {author} {\bibfnamefont {R.~R.}\ \bibnamefont {{Laher}}}, \bibinfo {author} {\bibfnamefont {P.~R.}\ \bibnamefont {{Wozniak}}}, \bibinfo {author} {\bibfnamefont {W.~H.}\ \bibnamefont {{Lee}}}, \bibinfo {author} {\bibfnamefont {U.~D.}\ \bibnamefont {{Rebbapragada}}}, \bibinfo {author} {\bibfnamefont
  {K.}~\bibnamefont {{Maguire}}}, \bibinfo {author} {\bibfnamefont {M.}~\bibnamefont {{Sullivan}}}, \ and\ \bibinfo {author} {\bibfnamefont {M.~T.}\ \bibnamefont {{Soumagnac}}},\ }\href@noop {} {\bibfield  {journal} {\bibinfo  {journal} {ArXiv e-prints}\ } (\bibinfo {year} {2017})},\ \Eprint {http://arxiv.org/abs/1701.02596} {arXiv:1701.02596 [astro-ph.HE]} \BibitemShut {NoStop}%
\bibitem [{\citenamefont {{Hiramatsu}}\ \emph {et~al.}(2021)\citenamefont {{Hiramatsu}}, \citenamefont {{Howell}}, \citenamefont {{Van Dyk}}, \citenamefont {{Goldberg}}, \citenamefont {{Maeda}}, \citenamefont {{Moriya}}, \citenamefont {{Tominaga}}, \citenamefont {{Nomoto}}, \citenamefont {{Hosseinzadeh}}, \citenamefont {{Arcavi}}, \citenamefont {{McCully}}, \citenamefont {{Burke}}, \citenamefont {{Bostroem}}, \citenamefont {{Valenti}}, \citenamefont {{Dong}}, \citenamefont {{Brown}}, \citenamefont {{Andrews}}, \citenamefont {{Bilinski}}, \citenamefont {{Williams}}, \citenamefont {{Smith}}, \citenamefont {{Smith}}, \citenamefont {{Sand}}, \citenamefont {{Anand}}, \citenamefont {{Xu}}, \citenamefont {{Filippenko}}, \citenamefont {{Bersten}}, \citenamefont {{Folatelli}}, \citenamefont {{Kelly}}, \citenamefont {{Noguchi}},\ and\ \citenamefont {{Itagaki}}}]{hiramatsu:21}%
  \BibitemOpen
  \bibfield  {author} {\bibinfo {author} {\bibfnamefont {D.}~\bibnamefont {{Hiramatsu}}}, \bibinfo {author} {\bibfnamefont {D.~A.}\ \bibnamefont {{Howell}}}, \bibinfo {author} {\bibfnamefont {S.~D.}\ \bibnamefont {{Van Dyk}}}, \bibinfo {author} {\bibfnamefont {J.~A.}\ \bibnamefont {{Goldberg}}}, \bibinfo {author} {\bibfnamefont {K.}~\bibnamefont {{Maeda}}}, \bibinfo {author} {\bibfnamefont {T.~J.}\ \bibnamefont {{Moriya}}}, \bibinfo {author} {\bibfnamefont {N.}~\bibnamefont {{Tominaga}}}, \bibinfo {author} {\bibfnamefont {K.}~\bibnamefont {{Nomoto}}}, \bibinfo {author} {\bibfnamefont {G.}~\bibnamefont {{Hosseinzadeh}}}, \bibinfo {author} {\bibfnamefont {I.}~\bibnamefont {{Arcavi}}}, \bibinfo {author} {\bibfnamefont {C.}~\bibnamefont {{McCully}}}, \bibinfo {author} {\bibfnamefont {J.}~\bibnamefont {{Burke}}}, \bibinfo {author} {\bibfnamefont {K.~A.}\ \bibnamefont {{Bostroem}}}, \bibinfo {author} {\bibfnamefont {S.}~\bibnamefont {{Valenti}}}, \bibinfo {author} {\bibfnamefont {Y.}~\bibnamefont {{Dong}}},
  \bibinfo {author} {\bibfnamefont {P.~J.}\ \bibnamefont {{Brown}}}, \bibinfo {author} {\bibfnamefont {J.~E.}\ \bibnamefont {{Andrews}}}, \bibinfo {author} {\bibfnamefont {C.}~\bibnamefont {{Bilinski}}}, \bibinfo {author} {\bibfnamefont {G.~G.}\ \bibnamefont {{Williams}}}, \bibinfo {author} {\bibfnamefont {P.~S.}\ \bibnamefont {{Smith}}}, \bibinfo {author} {\bibfnamefont {N.}~\bibnamefont {{Smith}}}, \bibinfo {author} {\bibfnamefont {D.~J.}\ \bibnamefont {{Sand}}}, \bibinfo {author} {\bibfnamefont {G.~S.}\ \bibnamefont {{Anand}}}, \bibinfo {author} {\bibfnamefont {C.}~\bibnamefont {{Xu}}}, \bibinfo {author} {\bibfnamefont {A.~V.}\ \bibnamefont {{Filippenko}}}, \bibinfo {author} {\bibfnamefont {M.~C.}\ \bibnamefont {{Bersten}}}, \bibinfo {author} {\bibfnamefont {G.}~\bibnamefont {{Folatelli}}}, \bibinfo {author} {\bibfnamefont {P.~L.}\ \bibnamefont {{Kelly}}}, \bibinfo {author} {\bibfnamefont {T.}~\bibnamefont {{Noguchi}}}, \ and\ \bibinfo {author} {\bibfnamefont {K.}~\bibnamefont {{Itagaki}}},\ }\href
  {\doibase 10.1038/s41550-021-01384-2} {\bibfield  {journal} {\bibinfo  {journal} {Nature Astronomy}\ }\textbf {\bibinfo {volume} {5}},\ \bibinfo {pages} {903} (\bibinfo {year} {2021})},\ \Eprint {http://arxiv.org/abs/2011.02176} {arXiv:2011.02176 [astro-ph.HE]} \BibitemShut {NoStop}%
\bibitem [{\citenamefont {{Bruch}}\ \emph {et~al.}(2021)\citenamefont {{Bruch}}, \citenamefont {{Gal-Yam}}, \citenamefont {{Schulze}}, \citenamefont {{Yaron}}, \citenamefont {{Yang}}, \citenamefont {{Soumagnac}}, \citenamefont {{Rigault}}, \citenamefont {{Strotjohann}}, \citenamefont {{Ofek}}, \citenamefont {{Sollerman}}, \citenamefont {{Masci}}, \citenamefont {{Barbarino}}, \citenamefont {{Ho}}, \citenamefont {{Fremling}}, \citenamefont {{Perley}}, \citenamefont {{Nordin}}, \citenamefont {{Cenko}}, \citenamefont {{Adams}}, \citenamefont {{Adreoni}}, \citenamefont {{Bellm}}, \citenamefont {{Blagorodnova}}, \citenamefont {{Bulla}}, \citenamefont {{Burdge}}, \citenamefont {{De}}, \citenamefont {{Dhawan}}, \citenamefont {{Drake}}, \citenamefont {{Duev}}, \citenamefont {{Dugas}}, \citenamefont {{Graham}}, \citenamefont {{Graham}}, \citenamefont {{Irani}}, \citenamefont {{Jencson}}, \citenamefont {{Karamehmetoglu}}, \citenamefont {{Kasliwal}}, \citenamefont {{Kim}}, \citenamefont {{Kulkarni}}, \citenamefont
  {{Kupfer}}, \citenamefont {{Liang}}, \citenamefont {{Mahabal}}, \citenamefont {{Miller}}, \citenamefont {{Prince}}, \citenamefont {{Riddle}}, \citenamefont {{Sharma}}, \citenamefont {{Smith}}, \citenamefont {{Taddia}}, \citenamefont {{Taggart}}, \citenamefont {{Walters}},\ and\ \citenamefont {{Yan}}}]{bruch:21}%
  \BibitemOpen
  \bibfield  {author} {\bibinfo {author} {\bibfnamefont {R.~J.}\ \bibnamefont {{Bruch}}}, \bibinfo {author} {\bibfnamefont {A.}~\bibnamefont {{Gal-Yam}}}, \bibinfo {author} {\bibfnamefont {S.}~\bibnamefont {{Schulze}}}, \bibinfo {author} {\bibfnamefont {O.}~\bibnamefont {{Yaron}}}, \bibinfo {author} {\bibfnamefont {Y.}~\bibnamefont {{Yang}}}, \bibinfo {author} {\bibfnamefont {M.}~\bibnamefont {{Soumagnac}}}, \bibinfo {author} {\bibfnamefont {M.}~\bibnamefont {{Rigault}}}, \bibinfo {author} {\bibfnamefont {N.~L.}\ \bibnamefont {{Strotjohann}}}, \bibinfo {author} {\bibfnamefont {E.}~\bibnamefont {{Ofek}}}, \bibinfo {author} {\bibfnamefont {J.}~\bibnamefont {{Sollerman}}}, \bibinfo {author} {\bibfnamefont {F.~J.}\ \bibnamefont {{Masci}}}, \bibinfo {author} {\bibfnamefont {C.}~\bibnamefont {{Barbarino}}}, \bibinfo {author} {\bibfnamefont {A.~Y.~Q.}\ \bibnamefont {{Ho}}}, \bibinfo {author} {\bibfnamefont {C.}~\bibnamefont {{Fremling}}}, \bibinfo {author} {\bibfnamefont {D.}~\bibnamefont {{Perley}}}, \bibinfo
  {author} {\bibfnamefont {J.}~\bibnamefont {{Nordin}}}, \bibinfo {author} {\bibfnamefont {S.~B.}\ \bibnamefont {{Cenko}}}, \bibinfo {author} {\bibfnamefont {S.}~\bibnamefont {{Adams}}}, \bibinfo {author} {\bibfnamefont {I.}~\bibnamefont {{Adreoni}}}, \bibinfo {author} {\bibfnamefont {E.~C.}\ \bibnamefont {{Bellm}}}, \bibinfo {author} {\bibfnamefont {N.}~\bibnamefont {{Blagorodnova}}}, \bibinfo {author} {\bibfnamefont {M.}~\bibnamefont {{Bulla}}}, \bibinfo {author} {\bibfnamefont {K.}~\bibnamefont {{Burdge}}}, \bibinfo {author} {\bibfnamefont {K.}~\bibnamefont {{De}}}, \bibinfo {author} {\bibfnamefont {S.}~\bibnamefont {{Dhawan}}}, \bibinfo {author} {\bibfnamefont {A.~J.}\ \bibnamefont {{Drake}}}, \bibinfo {author} {\bibfnamefont {D.~A.}\ \bibnamefont {{Duev}}}, \bibinfo {author} {\bibfnamefont {A.}~\bibnamefont {{Dugas}}}, \bibinfo {author} {\bibfnamefont {M.}~\bibnamefont {{Graham}}}, \bibinfo {author} {\bibfnamefont {M.~L.}\ \bibnamefont {{Graham}}}, \bibinfo {author} {\bibfnamefont {I.}~\bibnamefont
  {{Irani}}}, \bibinfo {author} {\bibfnamefont {J.}~\bibnamefont {{Jencson}}}, \bibinfo {author} {\bibfnamefont {E.}~\bibnamefont {{Karamehmetoglu}}}, \bibinfo {author} {\bibfnamefont {M.}~\bibnamefont {{Kasliwal}}}, \bibinfo {author} {\bibfnamefont {Y.-L.}\ \bibnamefont {{Kim}}}, \bibinfo {author} {\bibfnamefont {S.}~\bibnamefont {{Kulkarni}}}, \bibinfo {author} {\bibfnamefont {T.}~\bibnamefont {{Kupfer}}}, \bibinfo {author} {\bibfnamefont {J.}~\bibnamefont {{Liang}}}, \bibinfo {author} {\bibfnamefont {A.}~\bibnamefont {{Mahabal}}}, \bibinfo {author} {\bibfnamefont {A.~A.}\ \bibnamefont {{Miller}}}, \bibinfo {author} {\bibfnamefont {T.~A.}\ \bibnamefont {{Prince}}}, \bibinfo {author} {\bibfnamefont {R.}~\bibnamefont {{Riddle}}}, \bibinfo {author} {\bibfnamefont {Y.}~\bibnamefont {{Sharma}}}, \bibinfo {author} {\bibfnamefont {R.}~\bibnamefont {{Smith}}}, \bibinfo {author} {\bibfnamefont {F.}~\bibnamefont {{Taddia}}}, \bibinfo {author} {\bibfnamefont {K.}~\bibnamefont {{Taggart}}}, \bibinfo {author}
  {\bibfnamefont {R.}~\bibnamefont {{Walters}}}, \ and\ \bibinfo {author} {\bibfnamefont {L.}~\bibnamefont {{Yan}}},\ }\href {\doibase 10.3847/1538-4357/abef05} {\bibfield  {journal} {\bibinfo  {journal} {\apj}\ }\textbf {\bibinfo {volume} {912}},\ \bibinfo {eid} {46} (\bibinfo {year} {2021})},\ \Eprint {http://arxiv.org/abs/2008.09986} {arXiv:2008.09986 [astro-ph.HE]} \BibitemShut {NoStop}%
\bibitem [{\citenamefont {{Li}}\ \emph {et~al.}(2023)\citenamefont {{Li}}, \citenamefont {{Hu}}, \citenamefont {{Li}}, \citenamefont {{Yang}}, \citenamefont {{Wang}}, \citenamefont {{Yan}}, \citenamefont {{Hu}}, \citenamefont {{Zhang}}, \citenamefont {{Mao}}, \citenamefont {{Riise}}, \citenamefont {{Gao}}, \citenamefont {{Sun}}, \citenamefont {{Liu}}, \citenamefont {{Xiong}}, \citenamefont {{Wang}}, \citenamefont {{Mo}}, \citenamefont {{Iskandar}}, \citenamefont {{Xi}}, \citenamefont {{Xiang}}, \citenamefont {{Wang}}, \citenamefont {{Sun}}, \citenamefont {{Zhang}}, \citenamefont {{Chen}}, \citenamefont {{Lin}}, \citenamefont {{Guo}}, \citenamefont {{Liu}}, \citenamefont {{Cai}}, \citenamefont {{Zhou}}, \citenamefont {{Zhao}}, \citenamefont {{Chen}}, \citenamefont {{Zheng}}, \citenamefont {{Li}}, \citenamefont {{Zhang}}, \citenamefont {{Xu}}, \citenamefont {{Lyu}}, \citenamefont {{Castro-Tirado}}, \citenamefont {{Chufarin}}, \citenamefont {{Potapov}}, \citenamefont {{Ionov}}, \citenamefont {{Korotkiy}},
  \citenamefont {{Nazarov}}, \citenamefont {{Sokolovsky}}, \citenamefont {{Hamann}},\ and\ \citenamefont {{Herman}}}]{li:23}%
  \BibitemOpen
  \bibfield  {author} {\bibinfo {author} {\bibfnamefont {G.}~\bibnamefont {{Li}}}, \bibinfo {author} {\bibfnamefont {M.}~\bibnamefont {{Hu}}}, \bibinfo {author} {\bibfnamefont {W.}~\bibnamefont {{Li}}}, \bibinfo {author} {\bibfnamefont {Y.}~\bibnamefont {{Yang}}}, \bibinfo {author} {\bibfnamefont {X.}~\bibnamefont {{Wang}}}, \bibinfo {author} {\bibfnamefont {S.}~\bibnamefont {{Yan}}}, \bibinfo {author} {\bibfnamefont {L.}~\bibnamefont {{Hu}}}, \bibinfo {author} {\bibfnamefont {J.}~\bibnamefont {{Zhang}}}, \bibinfo {author} {\bibfnamefont {Y.}~\bibnamefont {{Mao}}}, \bibinfo {author} {\bibfnamefont {H.}~\bibnamefont {{Riise}}}, \bibinfo {author} {\bibfnamefont {X.}~\bibnamefont {{Gao}}}, \bibinfo {author} {\bibfnamefont {T.}~\bibnamefont {{Sun}}}, \bibinfo {author} {\bibfnamefont {J.}~\bibnamefont {{Liu}}}, \bibinfo {author} {\bibfnamefont {D.}~\bibnamefont {{Xiong}}}, \bibinfo {author} {\bibfnamefont {L.}~\bibnamefont {{Wang}}}, \bibinfo {author} {\bibfnamefont {J.}~\bibnamefont {{Mo}}}, \bibinfo {author}
  {\bibfnamefont {A.}~\bibnamefont {{Iskandar}}}, \bibinfo {author} {\bibfnamefont {G.}~\bibnamefont {{Xi}}}, \bibinfo {author} {\bibfnamefont {D.}~\bibnamefont {{Xiang}}}, \bibinfo {author} {\bibfnamefont {L.}~\bibnamefont {{Wang}}}, \bibinfo {author} {\bibfnamefont {G.}~\bibnamefont {{Sun}}}, \bibinfo {author} {\bibfnamefont {K.}~\bibnamefont {{Zhang}}}, \bibinfo {author} {\bibfnamefont {J.}~\bibnamefont {{Chen}}}, \bibinfo {author} {\bibfnamefont {W.}~\bibnamefont {{Lin}}}, \bibinfo {author} {\bibfnamefont {F.}~\bibnamefont {{Guo}}}, \bibinfo {author} {\bibfnamefont {Q.}~\bibnamefont {{Liu}}}, \bibinfo {author} {\bibfnamefont {G.}~\bibnamefont {{Cai}}}, \bibinfo {author} {\bibfnamefont {W.}~\bibnamefont {{Zhou}}}, \bibinfo {author} {\bibfnamefont {J.}~\bibnamefont {{Zhao}}}, \bibinfo {author} {\bibfnamefont {J.}~\bibnamefont {{Chen}}}, \bibinfo {author} {\bibfnamefont {X.}~\bibnamefont {{Zheng}}}, \bibinfo {author} {\bibfnamefont {K.}~\bibnamefont {{Li}}}, \bibinfo {author} {\bibfnamefont {M.}~\bibnamefont
  {{Zhang}}}, \bibinfo {author} {\bibfnamefont {S.}~\bibnamefont {{Xu}}}, \bibinfo {author} {\bibfnamefont {X.}~\bibnamefont {{Lyu}}}, \bibinfo {author} {\bibfnamefont {A.~J.}\ \bibnamefont {{Castro-Tirado}}}, \bibinfo {author} {\bibfnamefont {V.}~\bibnamefont {{Chufarin}}}, \bibinfo {author} {\bibfnamefont {N.}~\bibnamefont {{Potapov}}}, \bibinfo {author} {\bibfnamefont {I.}~\bibnamefont {{Ionov}}}, \bibinfo {author} {\bibfnamefont {S.}~\bibnamefont {{Korotkiy}}}, \bibinfo {author} {\bibfnamefont {S.}~\bibnamefont {{Nazarov}}}, \bibinfo {author} {\bibfnamefont {K.}~\bibnamefont {{Sokolovsky}}}, \bibinfo {author} {\bibfnamefont {N.}~\bibnamefont {{Hamann}}}, \ and\ \bibinfo {author} {\bibfnamefont {E.}~\bibnamefont {{Herman}}},\ }\href {\doibase 10.48550/arXiv.2311.14409} {\bibfield  {journal} {\bibinfo  {journal} {arXiv e-prints}\ ,\ \bibinfo {eid} {arXiv:2311.14409}} (\bibinfo {year} {2023})},\ \Eprint {http://arxiv.org/abs/2311.14409} {arXiv:2311.14409 [astro-ph.HE]} \BibitemShut {NoStop}%
\bibitem [{\citenamefont {{Meza Retamal}}\ \emph {et~al.}(2024)\citenamefont {{Meza Retamal}}, \citenamefont {{Dong}}, \citenamefont {{Bostroem}}, \citenamefont {{Valenti}}, \citenamefont {{Galbany}}, \citenamefont {{Pearson}}, \citenamefont {{Hosseinzadeh}}, \citenamefont {{Andrews}}, \citenamefont {{Sand}}, \citenamefont {{Jencson}}, \citenamefont {{Janzen}}, \citenamefont {{Lundquist}}, \citenamefont {{Hoang}}, \citenamefont {{Wyatt}}, \citenamefont {{Brown}}, \citenamefont {{Howell}}, \citenamefont {{Newsome}}, \citenamefont {{Padilla Gonzalez}}, \citenamefont {{Pellegrino}}, \citenamefont {{Terreran}}, \citenamefont {{Kouprianov}}, \citenamefont {{Hiramatsu}}, \citenamefont {{Jha}}, \citenamefont {{Smith}}, \citenamefont {{Haislip}}, \citenamefont {{Reichart}}, \citenamefont {{Shrestha}},\ and\ \citenamefont {{Fabi{\'a}n Rosales-Ortega}}}]{mezaretamal:24}%
  \BibitemOpen
  \bibfield  {author} {\bibinfo {author} {\bibfnamefont {N.}~\bibnamefont {{Meza Retamal}}}, \bibinfo {author} {\bibfnamefont {Y.}~\bibnamefont {{Dong}}}, \bibinfo {author} {\bibfnamefont {K.~A.}\ \bibnamefont {{Bostroem}}}, \bibinfo {author} {\bibfnamefont {S.}~\bibnamefont {{Valenti}}}, \bibinfo {author} {\bibfnamefont {L.}~\bibnamefont {{Galbany}}}, \bibinfo {author} {\bibfnamefont {J.}~\bibnamefont {{Pearson}}}, \bibinfo {author} {\bibfnamefont {G.}~\bibnamefont {{Hosseinzadeh}}}, \bibinfo {author} {\bibfnamefont {J.~E.}\ \bibnamefont {{Andrews}}}, \bibinfo {author} {\bibfnamefont {D.~J.}\ \bibnamefont {{Sand}}}, \bibinfo {author} {\bibfnamefont {J.~E.}\ \bibnamefont {{Jencson}}}, \bibinfo {author} {\bibfnamefont {D.}~\bibnamefont {{Janzen}}}, \bibinfo {author} {\bibfnamefont {M.~J.}\ \bibnamefont {{Lundquist}}}, \bibinfo {author} {\bibfnamefont {E.~T.}\ \bibnamefont {{Hoang}}}, \bibinfo {author} {\bibfnamefont {S.}~\bibnamefont {{Wyatt}}}, \bibinfo {author} {\bibfnamefont {P.~J.}\ \bibnamefont
  {{Brown}}}, \bibinfo {author} {\bibfnamefont {D.~A.}\ \bibnamefont {{Howell}}}, \bibinfo {author} {\bibfnamefont {M.}~\bibnamefont {{Newsome}}}, \bibinfo {author} {\bibfnamefont {E.}~\bibnamefont {{Padilla Gonzalez}}}, \bibinfo {author} {\bibfnamefont {C.}~\bibnamefont {{Pellegrino}}}, \bibinfo {author} {\bibfnamefont {G.}~\bibnamefont {{Terreran}}}, \bibinfo {author} {\bibfnamefont {V.}~\bibnamefont {{Kouprianov}}}, \bibinfo {author} {\bibfnamefont {D.}~\bibnamefont {{Hiramatsu}}}, \bibinfo {author} {\bibfnamefont {S.~W.}\ \bibnamefont {{Jha}}}, \bibinfo {author} {\bibfnamefont {N.}~\bibnamefont {{Smith}}}, \bibinfo {author} {\bibfnamefont {J.}~\bibnamefont {{Haislip}}}, \bibinfo {author} {\bibfnamefont {D.~E.}\ \bibnamefont {{Reichart}}}, \bibinfo {author} {\bibfnamefont {M.}~\bibnamefont {{Shrestha}}}, \ and\ \bibinfo {author} {\bibfnamefont {F.}~\bibnamefont {{Fabi{\'a}n Rosales-Ortega}}},\ }\href {\doibase 10.48550/arXiv.2401.04027} {\bibfield  {journal} {\bibinfo  {journal} {arXiv e-prints}\ ,\
  \bibinfo {eid} {arXiv:2401.04027}} (\bibinfo {year} {2024})},\ \Eprint {http://arxiv.org/abs/2401.04027} {arXiv:2401.04027 [astro-ph.HE]} \BibitemShut {NoStop}%
\bibitem [{\citenamefont {{Khazov}}\ \emph {et~al.}(2016)\citenamefont {{Khazov}}, \citenamefont {{Yaron}}, \citenamefont {{Gal-Yam}}, \citenamefont {{Manulis}}, \citenamefont {{Rubin}}, \citenamefont {{Kulkarni}}, \citenamefont {{Arcavi}}, \citenamefont {{Kasliwal}}, \citenamefont {{Ofek}}, \citenamefont {{Cao}}, \citenamefont {{Perley}}, \citenamefont {{Sollerman}}, \citenamefont {{Horesh}}, \citenamefont {{Sullivan}}, \citenamefont {{Filippenko}}, \citenamefont {{Nugent}}, \citenamefont {{Howell}}, \citenamefont {{Cenko}}, \citenamefont {{Silverman}}, \citenamefont {{Ebeling}}, \citenamefont {{Taddia}}, \citenamefont {{Johansson}}, \citenamefont {{Laher}}, \citenamefont {{Surace}}, \citenamefont {{Rebbapragada}}, \citenamefont {{Wozniak}},\ and\ \citenamefont {{Matheson}}}]{khazov:16}%
  \BibitemOpen
  \bibfield  {author} {\bibinfo {author} {\bibfnamefont {D.}~\bibnamefont {{Khazov}}}, \bibinfo {author} {\bibfnamefont {O.}~\bibnamefont {{Yaron}}}, \bibinfo {author} {\bibfnamefont {A.}~\bibnamefont {{Gal-Yam}}}, \bibinfo {author} {\bibfnamefont {I.}~\bibnamefont {{Manulis}}}, \bibinfo {author} {\bibfnamefont {A.}~\bibnamefont {{Rubin}}}, \bibinfo {author} {\bibfnamefont {S.~R.}\ \bibnamefont {{Kulkarni}}}, \bibinfo {author} {\bibfnamefont {I.}~\bibnamefont {{Arcavi}}}, \bibinfo {author} {\bibfnamefont {M.~M.}\ \bibnamefont {{Kasliwal}}}, \bibinfo {author} {\bibfnamefont {E.~O.}\ \bibnamefont {{Ofek}}}, \bibinfo {author} {\bibfnamefont {Y.}~\bibnamefont {{Cao}}}, \bibinfo {author} {\bibfnamefont {D.}~\bibnamefont {{Perley}}}, \bibinfo {author} {\bibfnamefont {J.}~\bibnamefont {{Sollerman}}}, \bibinfo {author} {\bibfnamefont {A.}~\bibnamefont {{Horesh}}}, \bibinfo {author} {\bibfnamefont {M.}~\bibnamefont {{Sullivan}}}, \bibinfo {author} {\bibfnamefont {A.~V.}\ \bibnamefont {{Filippenko}}}, \bibinfo {author}
  {\bibfnamefont {P.~E.}\ \bibnamefont {{Nugent}}}, \bibinfo {author} {\bibfnamefont {D.~A.}\ \bibnamefont {{Howell}}}, \bibinfo {author} {\bibfnamefont {S.~B.}\ \bibnamefont {{Cenko}}}, \bibinfo {author} {\bibfnamefont {J.~M.}\ \bibnamefont {{Silverman}}}, \bibinfo {author} {\bibfnamefont {H.}~\bibnamefont {{Ebeling}}}, \bibinfo {author} {\bibfnamefont {F.}~\bibnamefont {{Taddia}}}, \bibinfo {author} {\bibfnamefont {J.}~\bibnamefont {{Johansson}}}, \bibinfo {author} {\bibfnamefont {R.~R.}\ \bibnamefont {{Laher}}}, \bibinfo {author} {\bibfnamefont {J.}~\bibnamefont {{Surace}}}, \bibinfo {author} {\bibfnamefont {U.~D.}\ \bibnamefont {{Rebbapragada}}}, \bibinfo {author} {\bibfnamefont {P.~R.}\ \bibnamefont {{Wozniak}}}, \ and\ \bibinfo {author} {\bibfnamefont {T.}~\bibnamefont {{Matheson}}},\ }\href {\doibase 10.3847/0004-637X/818/1/3} {\bibfield  {journal} {\bibinfo  {journal} {\apj}\ }\textbf {\bibinfo {volume} {818}},\ \bibinfo {eid} {3} (\bibinfo {year} {2016})},\ \Eprint {http://arxiv.org/abs/1512.00846}
  {arXiv:1512.00846 [astro-ph.HE]} \BibitemShut {NoStop}%
\bibitem [{\citenamefont {{Bruch}}\ \emph {et~al.}(2023)\citenamefont {{Bruch}}, \citenamefont {{Gal-Yam}}, \citenamefont {{Yaron}}, \citenamefont {{Chen}}, \citenamefont {{Strotjohann}}, \citenamefont {{Irani}}, \citenamefont {{Zimmerman}}, \citenamefont {{Schulze}}, \citenamefont {{Yang}}, \citenamefont {{Kim}}, \citenamefont {{Bulla}}, \citenamefont {{Sollerman}}, \citenamefont {{Rigault}}, \citenamefont {{Ofek}}, \citenamefont {{Soumagnac}}, \citenamefont {{Masci}}, \citenamefont {{Fremling}}, \citenamefont {{Perley}}, \citenamefont {{Nordin}}, \citenamefont {{Cenko}}, \citenamefont {{Ho}}, \citenamefont {{Adams}}, \citenamefont {{Adreoni}}, \citenamefont {{Bellm}}, \citenamefont {{Blagorodnova}}, \citenamefont {{Burdge}}, \citenamefont {{De}}, \citenamefont {{Dekany}}, \citenamefont {{Dhawan}}, \citenamefont {{Drake}}, \citenamefont {{Duev}}, \citenamefont {{Graham}}, \citenamefont {{Graham}}, \citenamefont {{Jencson}}, \citenamefont {{Karamehmetoglu}}, \citenamefont {{Kasliwal}}, \citenamefont
  {{Kulkarni}}, \citenamefont {{Miller}}, \citenamefont {{Neill}}, \citenamefont {{Prince}}, \citenamefont {{Riddle}}, \citenamefont {{Rusholme}}, \citenamefont {{Sharma}}, \citenamefont {{Smith}}, \citenamefont {{Sravan}}, \citenamefont {{Taggart}}, \citenamefont {{Walters}},\ and\ \citenamefont {{Yan}}}]{bruch:23}%
  \BibitemOpen
  \bibfield  {author} {\bibinfo {author} {\bibfnamefont {R.~J.}\ \bibnamefont {{Bruch}}}, \bibinfo {author} {\bibfnamefont {A.}~\bibnamefont {{Gal-Yam}}}, \bibinfo {author} {\bibfnamefont {O.}~\bibnamefont {{Yaron}}}, \bibinfo {author} {\bibfnamefont {P.}~\bibnamefont {{Chen}}}, \bibinfo {author} {\bibfnamefont {N.~L.}\ \bibnamefont {{Strotjohann}}}, \bibinfo {author} {\bibfnamefont {I.}~\bibnamefont {{Irani}}}, \bibinfo {author} {\bibfnamefont {E.}~\bibnamefont {{Zimmerman}}}, \bibinfo {author} {\bibfnamefont {S.}~\bibnamefont {{Schulze}}}, \bibinfo {author} {\bibfnamefont {Y.}~\bibnamefont {{Yang}}}, \bibinfo {author} {\bibfnamefont {Y.-L.}\ \bibnamefont {{Kim}}}, \bibinfo {author} {\bibfnamefont {M.}~\bibnamefont {{Bulla}}}, \bibinfo {author} {\bibfnamefont {J.}~\bibnamefont {{Sollerman}}}, \bibinfo {author} {\bibfnamefont {M.}~\bibnamefont {{Rigault}}}, \bibinfo {author} {\bibfnamefont {E.}~\bibnamefont {{Ofek}}}, \bibinfo {author} {\bibfnamefont {M.}~\bibnamefont {{Soumagnac}}}, \bibinfo {author}
  {\bibfnamefont {F.~J.}\ \bibnamefont {{Masci}}}, \bibinfo {author} {\bibfnamefont {C.}~\bibnamefont {{Fremling}}}, \bibinfo {author} {\bibfnamefont {D.}~\bibnamefont {{Perley}}}, \bibinfo {author} {\bibfnamefont {J.}~\bibnamefont {{Nordin}}}, \bibinfo {author} {\bibfnamefont {S.~B.}\ \bibnamefont {{Cenko}}}, \bibinfo {author} {\bibfnamefont {A.~Y.~Q.}\ \bibnamefont {{Ho}}}, \bibinfo {author} {\bibfnamefont {S.}~\bibnamefont {{Adams}}}, \bibinfo {author} {\bibfnamefont {I.}~\bibnamefont {{Adreoni}}}, \bibinfo {author} {\bibfnamefont {E.~C.}\ \bibnamefont {{Bellm}}}, \bibinfo {author} {\bibfnamefont {N.}~\bibnamefont {{Blagorodnova}}}, \bibinfo {author} {\bibfnamefont {K.}~\bibnamefont {{Burdge}}}, \bibinfo {author} {\bibfnamefont {K.}~\bibnamefont {{De}}}, \bibinfo {author} {\bibfnamefont {R.~G.}\ \bibnamefont {{Dekany}}}, \bibinfo {author} {\bibfnamefont {S.}~\bibnamefont {{Dhawan}}}, \bibinfo {author} {\bibfnamefont {A.~J.}\ \bibnamefont {{Drake}}}, \bibinfo {author} {\bibfnamefont {D.~A.}\ \bibnamefont
  {{Duev}}}, \bibinfo {author} {\bibfnamefont {M.}~\bibnamefont {{Graham}}}, \bibinfo {author} {\bibfnamefont {M.~L.}\ \bibnamefont {{Graham}}}, \bibinfo {author} {\bibfnamefont {J.}~\bibnamefont {{Jencson}}}, \bibinfo {author} {\bibfnamefont {E.}~\bibnamefont {{Karamehmetoglu}}}, \bibinfo {author} {\bibfnamefont {M.~M.}\ \bibnamefont {{Kasliwal}}}, \bibinfo {author} {\bibfnamefont {S.}~\bibnamefont {{Kulkarni}}}, \bibinfo {author} {\bibfnamefont {A.~A.}\ \bibnamefont {{Miller}}}, \bibinfo {author} {\bibfnamefont {J.~D.}\ \bibnamefont {{Neill}}}, \bibinfo {author} {\bibfnamefont {T.~A.}\ \bibnamefont {{Prince}}}, \bibinfo {author} {\bibfnamefont {R.}~\bibnamefont {{Riddle}}}, \bibinfo {author} {\bibfnamefont {B.}~\bibnamefont {{Rusholme}}}, \bibinfo {author} {\bibfnamefont {Y.}~\bibnamefont {{Sharma}}}, \bibinfo {author} {\bibfnamefont {R.}~\bibnamefont {{Smith}}}, \bibinfo {author} {\bibfnamefont {N.}~\bibnamefont {{Sravan}}}, \bibinfo {author} {\bibfnamefont {K.}~\bibnamefont {{Taggart}}}, \bibinfo {author}
  {\bibfnamefont {R.}~\bibnamefont {{Walters}}}, \ and\ \bibinfo {author} {\bibfnamefont {L.}~\bibnamefont {{Yan}}},\ }\href {\doibase 10.3847/1538-4357/acd8be} {\bibfield  {journal} {\bibinfo  {journal} {\apj}\ }\textbf {\bibinfo {volume} {952}},\ \bibinfo {eid} {119} (\bibinfo {year} {2023})},\ \Eprint {http://arxiv.org/abs/2212.03313} {arXiv:2212.03313 [astro-ph.HE]} \BibitemShut {NoStop}%
\bibitem [{\citenamefont {{Dessart}}\ \emph {et~al.}(2017)\citenamefont {{Dessart}}, \citenamefont {{Hillier}},\ and\ \citenamefont {{Audit}}}]{dessart:17}%
  \BibitemOpen
  \bibfield  {author} {\bibinfo {author} {\bibfnamefont {L.}~\bibnamefont {{Dessart}}}, \bibinfo {author} {\bibfnamefont {D.~J.}\ \bibnamefont {{Hillier}}}, \ and\ \bibinfo {author} {\bibfnamefont {E.}~\bibnamefont {{Audit}}},\ }\href@noop {} {\bibfield  {journal} {\bibinfo  {journal} {ArXiv e-prints}\ } (\bibinfo {year} {2017})},\ \Eprint {http://arxiv.org/abs/1704.01697} {arXiv:1704.01697 [astro-ph.SR]} \BibitemShut {NoStop}%
\bibitem [{\citenamefont {{Moriya}}\ \emph {et~al.}(2017)\citenamefont {{Moriya}}, \citenamefont {{Yoon}}, \citenamefont {{Gr{\"a}fener}},\ and\ \citenamefont {{Blinnikov}}}]{moriya:17}%
  \BibitemOpen
  \bibfield  {author} {\bibinfo {author} {\bibfnamefont {T.~J.}\ \bibnamefont {{Moriya}}}, \bibinfo {author} {\bibfnamefont {S.-C.}\ \bibnamefont {{Yoon}}}, \bibinfo {author} {\bibfnamefont {G.}~\bibnamefont {{Gr{\"a}fener}}}, \ and\ \bibinfo {author} {\bibfnamefont {S.~I.}\ \bibnamefont {{Blinnikov}}},\ }\href@noop {} {\bibfield  {journal} {\bibinfo  {journal} {ArXiv e-prints}\ } (\bibinfo {year} {2017})},\ \Eprint {http://arxiv.org/abs/1703.03084} {arXiv:1703.03084 [astro-ph.HE]} \BibitemShut {NoStop}%
\bibitem [{\citenamefont {{Morozova}}\ \emph {et~al.}(2017)\citenamefont {{Morozova}}, \citenamefont {{Piro}},\ and\ \citenamefont {{Valenti}}}]{morozova:17}%
  \BibitemOpen
  \bibfield  {author} {\bibinfo {author} {\bibfnamefont {V.}~\bibnamefont {{Morozova}}}, \bibinfo {author} {\bibfnamefont {A.~L.}\ \bibnamefont {{Piro}}}, \ and\ \bibinfo {author} {\bibfnamefont {S.}~\bibnamefont {{Valenti}}},\ }\href {\doibase 10.3847/1538-4357/aa6251} {\bibfield  {journal} {\bibinfo  {journal} {\apj}\ }\textbf {\bibinfo {volume} {838}},\ \bibinfo {eid} {28} (\bibinfo {year} {2017})},\ \Eprint {http://arxiv.org/abs/1610.08054} {arXiv:1610.08054 [astro-ph.HE]} \BibitemShut {NoStop}%
\bibitem [{\citenamefont {{Moriya}}\ \emph {et~al.}(2018)\citenamefont {{Moriya}}, \citenamefont {{F{\"o}rster}}, \citenamefont {{Yoon}}, \citenamefont {{Gr{\"a}fener}},\ and\ \citenamefont {{Blinnikov}}}]{moriya:18}%
  \BibitemOpen
  \bibfield  {author} {\bibinfo {author} {\bibfnamefont {T.~J.}\ \bibnamefont {{Moriya}}}, \bibinfo {author} {\bibfnamefont {F.}~\bibnamefont {{F{\"o}rster}}}, \bibinfo {author} {\bibfnamefont {S.-C.}\ \bibnamefont {{Yoon}}}, \bibinfo {author} {\bibfnamefont {G.}~\bibnamefont {{Gr{\"a}fener}}}, \ and\ \bibinfo {author} {\bibfnamefont {S.~I.}\ \bibnamefont {{Blinnikov}}},\ }\href {\doibase 10.1093/mnras/sty475} {\bibfield  {journal} {\bibinfo  {journal} {\mnras}\ }\textbf {\bibinfo {volume} {476}},\ \bibinfo {pages} {2840} (\bibinfo {year} {2018})},\ \Eprint {http://arxiv.org/abs/1802.07752} {arXiv:1802.07752 [astro-ph.HE]} \BibitemShut {NoStop}%
\bibitem [{\citenamefont {{Morozova}}\ \emph {et~al.}(2018)\citenamefont {{Morozova}}, \citenamefont {{Piro}},\ and\ \citenamefont {{Valenti}}}]{morozova:18}%
  \BibitemOpen
  \bibfield  {author} {\bibinfo {author} {\bibfnamefont {V.}~\bibnamefont {{Morozova}}}, \bibinfo {author} {\bibfnamefont {A.~L.}\ \bibnamefont {{Piro}}}, \ and\ \bibinfo {author} {\bibfnamefont {S.}~\bibnamefont {{Valenti}}},\ }\href {\doibase 10.3847/1538-4357/aab9a6} {\bibfield  {journal} {\bibinfo  {journal} {\apj}\ }\textbf {\bibinfo {volume} {858}},\ \bibinfo {eid} {15} (\bibinfo {year} {2018})},\ \Eprint {http://arxiv.org/abs/1709.04928} {arXiv:1709.04928 [astro-ph.HE]} \BibitemShut {NoStop}%
\bibitem [{\citenamefont {{Woosley}}\ and\ \citenamefont {{Heger}}(2015)}]{woosleyheger:15}%
  \BibitemOpen
  \bibfield  {author} {\bibinfo {author} {\bibfnamefont {S.~E.}\ \bibnamefont {{Woosley}}}\ and\ \bibinfo {author} {\bibfnamefont {A.}~\bibnamefont {{Heger}}},\ }\href {\doibase 10.1088/0004-637X/810/1/34} {\bibfield  {journal} {\bibinfo  {journal} {\apj}\ }\textbf {\bibinfo {volume} {810}},\ \bibinfo {eid} {34} (\bibinfo {year} {2015})},\ \Eprint {http://arxiv.org/abs/1505.06712} {arXiv:1505.06712 [astro-ph.SR]} \BibitemShut {NoStop}%
\bibitem [{\citenamefont {{Fuller}}(2017)}]{fuller:17}%
  \BibitemOpen
  \bibfield  {author} {\bibinfo {author} {\bibfnamefont {J.}~\bibnamefont {{Fuller}}},\ }\href {\doibase 10.1093/mnras/stx1314} {\bibfield  {journal} {\bibinfo  {journal} {\mnras}\ }\textbf {\bibinfo {volume} {470}},\ \bibinfo {pages} {1642} (\bibinfo {year} {2017})},\ \Eprint {http://arxiv.org/abs/1704.08696} {arXiv:1704.08696 [astro-ph.SR]} \BibitemShut {NoStop}%
\bibitem [{\citenamefont {{Goldberg}}\ \emph {et~al.}(2022{\natexlab{a}})\citenamefont {{Goldberg}}, \citenamefont {{Jiang}},\ and\ \citenamefont {{Bildsten}}}]{goldberg:22}%
  \BibitemOpen
  \bibfield  {author} {\bibinfo {author} {\bibfnamefont {J.~A.}\ \bibnamefont {{Goldberg}}}, \bibinfo {author} {\bibfnamefont {Y.-F.}\ \bibnamefont {{Jiang}}}, \ and\ \bibinfo {author} {\bibfnamefont {L.}~\bibnamefont {{Bildsten}}},\ }\href {\doibase 10.3847/1538-4357/ac5ab3} {\bibfield  {journal} {\bibinfo  {journal} {\apj}\ }\textbf {\bibinfo {volume} {929}},\ \bibinfo {eid} {156} (\bibinfo {year} {2022}{\natexlab{a}})},\ \Eprint {http://arxiv.org/abs/2110.03261} {arXiv:2110.03261 [astro-ph.SR]} \BibitemShut {NoStop}%
\bibitem [{\citenamefont {{Bertschinger}}\ and\ \citenamefont {{Chevalier}}(1985)}]{bertschinger:85}%
  \BibitemOpen
  \bibfield  {author} {\bibinfo {author} {\bibfnamefont {E.}~\bibnamefont {{Bertschinger}}}\ and\ \bibinfo {author} {\bibfnamefont {R.~A.}\ \bibnamefont {{Chevalier}}},\ }\href {\doibase 10.1086/163690} {\bibfield  {journal} {\bibinfo  {journal} {\apj}\ }\textbf {\bibinfo {volume} {299}},\ \bibinfo {pages} {167} (\bibinfo {year} {1985})}\BibitemShut {NoStop}%
\bibitem [{\citenamefont {{Schirrmacher}}\ \emph {et~al.}(2003)\citenamefont {{Schirrmacher}}, \citenamefont {{Woitke}},\ and\ \citenamefont {{Sedlmayr}}}]{schirrmacher:03}%
  \BibitemOpen
  \bibfield  {author} {\bibinfo {author} {\bibfnamefont {V.}~\bibnamefont {{Schirrmacher}}}, \bibinfo {author} {\bibfnamefont {P.}~\bibnamefont {{Woitke}}}, \ and\ \bibinfo {author} {\bibfnamefont {E.}~\bibnamefont {{Sedlmayr}}},\ }\href {\doibase 10.1051/0004-6361:20030444} {\bibfield  {journal} {\bibinfo  {journal} {\aap}\ }\textbf {\bibinfo {volume} {404}},\ \bibinfo {pages} {267} (\bibinfo {year} {2003})}\BibitemShut {NoStop}%
\bibitem [{\citenamefont {{Shu}}(1992)}]{shu:92}%
  \BibitemOpen
  \bibfield  {author} {\bibinfo {author} {\bibfnamefont {F.~H.}\ \bibnamefont {{Shu}}},\ }\href@noop {} {\emph {\bibinfo {title} {{The physics of astrophysics. Volume II: Gas dynamics.}}}}\ (\bibinfo {year} {1992})\BibitemShut {NoStop}%
\bibitem [{\citenamefont {{Coughlin}}\ \emph {et~al.}(2018)\citenamefont {{Coughlin}}, \citenamefont {{Quataert}},\ and\ \citenamefont {{Ro}}}]{coughlin:18}%
  \BibitemOpen
  \bibfield  {author} {\bibinfo {author} {\bibfnamefont {E.~R.}\ \bibnamefont {{Coughlin}}}, \bibinfo {author} {\bibfnamefont {E.}~\bibnamefont {{Quataert}}}, \ and\ \bibinfo {author} {\bibfnamefont {S.}~\bibnamefont {{Ro}}},\ }\href {\doibase 10.3847/1538-4357/aad198} {\bibfield  {journal} {\bibinfo  {journal} {\apj}\ }\textbf {\bibinfo {volume} {863}},\ \bibinfo {eid} {158} (\bibinfo {year} {2018})},\ \Eprint {http://arxiv.org/abs/1805.06456} {arXiv:1805.06456 [astro-ph.HE]} \BibitemShut {NoStop}%
\bibitem [{\citenamefont {{Matzner}}\ and\ \citenamefont {{McKee}}(1999)}]{matzner:99}%
  \BibitemOpen
  \bibfield  {author} {\bibinfo {author} {\bibfnamefont {C.~D.}\ \bibnamefont {{Matzner}}}\ and\ \bibinfo {author} {\bibfnamefont {C.~F.}\ \bibnamefont {{McKee}}},\ }\href {\doibase 10.1086/306571} {\bibfield  {journal} {\bibinfo  {journal} {\apj}\ }\textbf {\bibinfo {volume} {510}},\ \bibinfo {pages} {379} (\bibinfo {year} {1999})},\ \Eprint {http://arxiv.org/abs/astro-ph/9807046} {arXiv:astro-ph/9807046 [astro-ph]} \BibitemShut {NoStop}%
\bibitem [{\citenamefont {Sakurai}(1960)}]{sakurai:60}%
  \BibitemOpen
  \bibfield  {author} {\bibinfo {author} {\bibfnamefont {A.}~\bibnamefont {Sakurai}},\ }\href {\doibase 10.1002/cpa.3160130303} {\bibfield  {journal} {\bibinfo  {journal} {Communs. Pure and Appl. Math.}\ }\textbf {\bibinfo {volume} {13}} (\bibinfo {year} {1960}),\ 10.1002/cpa.3160130303}\BibitemShut {NoStop}%
\bibitem [{\citenamefont {Whitham}(2011)}]{whitham:74}%
  \BibitemOpen
  \bibfield  {author} {\bibinfo {author} {\bibfnamefont {G.}~\bibnamefont {Whitham}},\ }\href {https://books.google.com/books?id=84Pulkf-Oa8C} {\emph {\bibinfo {title} {Linear and Nonlinear Waves}}},\ Pure and Applied Mathematics: A Wiley Series of Texts, Monographs and Tracts\ (\bibinfo  {publisher} {Wiley},\ \bibinfo {year} {2011})\BibitemShut {NoStop}%
\bibitem [{\citenamefont {{Paxton}}\ \emph {et~al.}(2011)\citenamefont {{Paxton}}, \citenamefont {{Bildsten}}, \citenamefont {{Dotter}}, \citenamefont {{Herwig}}, \citenamefont {{Lesaffre}},\ and\ \citenamefont {{Timmes}}}]{paxton:11}%
  \BibitemOpen
  \bibfield  {author} {\bibinfo {author} {\bibfnamefont {B.}~\bibnamefont {{Paxton}}}, \bibinfo {author} {\bibfnamefont {L.}~\bibnamefont {{Bildsten}}}, \bibinfo {author} {\bibfnamefont {A.}~\bibnamefont {{Dotter}}}, \bibinfo {author} {\bibfnamefont {F.}~\bibnamefont {{Herwig}}}, \bibinfo {author} {\bibfnamefont {P.}~\bibnamefont {{Lesaffre}}}, \ and\ \bibinfo {author} {\bibfnamefont {F.}~\bibnamefont {{Timmes}}},\ }\href {\doibase 10.1088/0067-0049/192/1/3} {\bibfield  {journal} {\bibinfo  {journal} {\apjs}\ }\textbf {\bibinfo {volume} {192}},\ \bibinfo {eid} {3} (\bibinfo {year} {2011})},\ \Eprint {http://arxiv.org/abs/1009.1622} {arXiv:1009.1622 [astro-ph.SR]} \BibitemShut {NoStop}%
\bibitem [{\citenamefont {{Paxton}}\ \emph {et~al.}(2013)\citenamefont {{Paxton}}, \citenamefont {{Cantiello}}, \citenamefont {{Arras}}, \citenamefont {{Bildsten}}, \citenamefont {{Brown}}, \citenamefont {{Dotter}}, \citenamefont {{Mankovich}}, \citenamefont {{Montgomery}}, \citenamefont {{Stello}}, \citenamefont {{Timmes}},\ and\ \citenamefont {{Townsend}}}]{paxton:13}%
  \BibitemOpen
  \bibfield  {author} {\bibinfo {author} {\bibfnamefont {B.}~\bibnamefont {{Paxton}}}, \bibinfo {author} {\bibfnamefont {M.}~\bibnamefont {{Cantiello}}}, \bibinfo {author} {\bibfnamefont {P.}~\bibnamefont {{Arras}}}, \bibinfo {author} {\bibfnamefont {L.}~\bibnamefont {{Bildsten}}}, \bibinfo {author} {\bibfnamefont {E.~F.}\ \bibnamefont {{Brown}}}, \bibinfo {author} {\bibfnamefont {A.}~\bibnamefont {{Dotter}}}, \bibinfo {author} {\bibfnamefont {C.}~\bibnamefont {{Mankovich}}}, \bibinfo {author} {\bibfnamefont {M.~H.}\ \bibnamefont {{Montgomery}}}, \bibinfo {author} {\bibfnamefont {D.}~\bibnamefont {{Stello}}}, \bibinfo {author} {\bibfnamefont {F.~X.}\ \bibnamefont {{Timmes}}}, \ and\ \bibinfo {author} {\bibfnamefont {R.}~\bibnamefont {{Townsend}}},\ }\href {\doibase 10.1088/0067-0049/208/1/4} {\bibfield  {journal} {\bibinfo  {journal} {\apjs}\ }\textbf {\bibinfo {volume} {208}},\ \bibinfo {eid} {4} (\bibinfo {year} {2013})},\ \Eprint {http://arxiv.org/abs/1301.0319} {arXiv:1301.0319 [astro-ph.SR]} \BibitemShut
  {NoStop}%
\bibitem [{\citenamefont {{Paxton}}\ \emph {et~al.}(2015)\citenamefont {{Paxton}}, \citenamefont {{Marchant}}, \citenamefont {{Schwab}}, \citenamefont {{Bauer}}, \citenamefont {{Bildsten}}, \citenamefont {{Cantiello}}, \citenamefont {{Dessart}}, \citenamefont {{Farmer}}, \citenamefont {{Hu}}, \citenamefont {{Langer}}, \citenamefont {{Townsend}}, \citenamefont {{Townsley}},\ and\ \citenamefont {{Timmes}}}]{paxton:15}%
  \BibitemOpen
  \bibfield  {author} {\bibinfo {author} {\bibfnamefont {B.}~\bibnamefont {{Paxton}}}, \bibinfo {author} {\bibfnamefont {P.}~\bibnamefont {{Marchant}}}, \bibinfo {author} {\bibfnamefont {J.}~\bibnamefont {{Schwab}}}, \bibinfo {author} {\bibfnamefont {E.~B.}\ \bibnamefont {{Bauer}}}, \bibinfo {author} {\bibfnamefont {L.}~\bibnamefont {{Bildsten}}}, \bibinfo {author} {\bibfnamefont {M.}~\bibnamefont {{Cantiello}}}, \bibinfo {author} {\bibfnamefont {L.}~\bibnamefont {{Dessart}}}, \bibinfo {author} {\bibfnamefont {R.}~\bibnamefont {{Farmer}}}, \bibinfo {author} {\bibfnamefont {H.}~\bibnamefont {{Hu}}}, \bibinfo {author} {\bibfnamefont {N.}~\bibnamefont {{Langer}}}, \bibinfo {author} {\bibfnamefont {R.~H.~D.}\ \bibnamefont {{Townsend}}}, \bibinfo {author} {\bibfnamefont {D.~M.}\ \bibnamefont {{Townsley}}}, \ and\ \bibinfo {author} {\bibfnamefont {F.~X.}\ \bibnamefont {{Timmes}}},\ }\href {\doibase 10.1088/0067-0049/220/1/15} {\bibfield  {journal} {\bibinfo  {journal} {\apjs}\ }\textbf {\bibinfo {volume} {220}},\
  \bibinfo {eid} {15} (\bibinfo {year} {2015})},\ \Eprint {http://arxiv.org/abs/1506.03146} {arXiv:1506.03146 [astro-ph.SR]} \BibitemShut {NoStop}%
\bibitem [{\citenamefont {{Chandrasekhar}}(1934)}]{chandrasekhar:34}%
  \BibitemOpen
  \bibfield  {author} {\bibinfo {author} {\bibfnamefont {S.}~\bibnamefont {{Chandrasekhar}}},\ }\href {\doibase 10.1093/mnras/94.5.444} {\bibfield  {journal} {\bibinfo  {journal} {\mnras}\ }\textbf {\bibinfo {volume} {94}},\ \bibinfo {pages} {444} (\bibinfo {year} {1934})}\BibitemShut {NoStop}%
\bibitem [{\citenamefont {{Bowen}}(1988)}]{bowen:88}%
  \BibitemOpen
  \bibfield  {author} {\bibinfo {author} {\bibfnamefont {G.~H.}\ \bibnamefont {{Bowen}}},\ }\href {\doibase 10.1086/166378} {\bibfield  {journal} {\bibinfo  {journal} {\apj}\ }\textbf {\bibinfo {volume} {329}},\ \bibinfo {pages} {299} (\bibinfo {year} {1988})}\BibitemShut {NoStop}%
\bibitem [{\citenamefont {{Winters}}\ \emph {et~al.}(1997)\citenamefont {{Winters}}, \citenamefont {{Fleischer}}, \citenamefont {{Le Bertre}},\ and\ \citenamefont {{Sedlmayr}}}]{winters:97}%
  \BibitemOpen
  \bibfield  {author} {\bibinfo {author} {\bibfnamefont {J.~M.}\ \bibnamefont {{Winters}}}, \bibinfo {author} {\bibfnamefont {A.~J.}\ \bibnamefont {{Fleischer}}}, \bibinfo {author} {\bibfnamefont {T.}~\bibnamefont {{Le Bertre}}}, \ and\ \bibinfo {author} {\bibfnamefont {E.}~\bibnamefont {{Sedlmayr}}},\ }\href@noop {} {\bibfield  {journal} {\bibinfo  {journal} {\aap}\ }\textbf {\bibinfo {volume} {326}},\ \bibinfo {pages} {305} (\bibinfo {year} {1997})}\BibitemShut {NoStop}%
\bibitem [{\citenamefont {{Sjouwerman}}\ \emph {et~al.}(1998)\citenamefont {{Sjouwerman}}, \citenamefont {{van Langevelde}}, \citenamefont {{Winnberg}},\ and\ \citenamefont {{Habing}}}]{sjouwerman:98}%
  \BibitemOpen
  \bibfield  {author} {\bibinfo {author} {\bibfnamefont {L.~O.}\ \bibnamefont {{Sjouwerman}}}, \bibinfo {author} {\bibfnamefont {H.~J.}\ \bibnamefont {{van Langevelde}}}, \bibinfo {author} {\bibfnamefont {A.}~\bibnamefont {{Winnberg}}}, \ and\ \bibinfo {author} {\bibfnamefont {H.~J.}\ \bibnamefont {{Habing}}},\ }\href {\doibase 10.1051/aas:1998127} {\bibfield  {journal} {\bibinfo  {journal} {\aaps}\ }\textbf {\bibinfo {volume} {128}},\ \bibinfo {pages} {35} (\bibinfo {year} {1998})}\BibitemShut {NoStop}%
\bibitem [{\citenamefont {{van Loon}}\ \emph {et~al.}(2001)\citenamefont {{van Loon}}, \citenamefont {{Zijlstra}}, \citenamefont {{Bujarrabal}},\ and\ \citenamefont {{Nyman}}}]{vanloon:01}%
  \BibitemOpen
  \bibfield  {author} {\bibinfo {author} {\bibfnamefont {J.~T.}\ \bibnamefont {{van Loon}}}, \bibinfo {author} {\bibfnamefont {A.~A.}\ \bibnamefont {{Zijlstra}}}, \bibinfo {author} {\bibfnamefont {V.}~\bibnamefont {{Bujarrabal}}}, \ and\ \bibinfo {author} {\bibfnamefont {L.~{\r{A}}.}\ \bibnamefont {{Nyman}}},\ }\href {\doibase 10.1051/0004-6361:20010052} {\bibfield  {journal} {\bibinfo  {journal} {\aap}\ }\textbf {\bibinfo {volume} {368}},\ \bibinfo {pages} {950} (\bibinfo {year} {2001})},\ \Eprint {http://arxiv.org/abs/astro-ph/0101125} {arXiv:astro-ph/0101125 [astro-ph]} \BibitemShut {NoStop}%
\bibitem [{\citenamefont {{Marshall}}\ \emph {et~al.}(2004)\citenamefont {{Marshall}}, \citenamefont {{van Loon}}, \citenamefont {{Matsuura}}, \citenamefont {{Wood}}, \citenamefont {{Zijlstra}},\ and\ \citenamefont {{Whitelock}}}]{marshall:04}%
  \BibitemOpen
  \bibfield  {author} {\bibinfo {author} {\bibfnamefont {J.~R.}\ \bibnamefont {{Marshall}}}, \bibinfo {author} {\bibfnamefont {J.~T.}\ \bibnamefont {{van Loon}}}, \bibinfo {author} {\bibfnamefont {M.}~\bibnamefont {{Matsuura}}}, \bibinfo {author} {\bibfnamefont {P.~R.}\ \bibnamefont {{Wood}}}, \bibinfo {author} {\bibfnamefont {A.~A.}\ \bibnamefont {{Zijlstra}}}, \ and\ \bibinfo {author} {\bibfnamefont {P.~A.}\ \bibnamefont {{Whitelock}}},\ }\href {\doibase 10.1111/j.1365-2966.2004.08417.x} {\bibfield  {journal} {\bibinfo  {journal} {\mnras}\ }\textbf {\bibinfo {volume} {355}},\ \bibinfo {pages} {1348} (\bibinfo {year} {2004})},\ \Eprint {http://arxiv.org/abs/astro-ph/0410120} {arXiv:astro-ph/0410120 [astro-ph]} \BibitemShut {NoStop}%
\bibitem [{\citenamefont {{Mauron}}\ and\ \citenamefont {{Josselin}}(2011)}]{mauron:11}%
  \BibitemOpen
  \bibfield  {author} {\bibinfo {author} {\bibfnamefont {N.}~\bibnamefont {{Mauron}}}\ and\ \bibinfo {author} {\bibfnamefont {E.}~\bibnamefont {{Josselin}}},\ }\href {\doibase 10.1051/0004-6361/201013993} {\bibfield  {journal} {\bibinfo  {journal} {\aap}\ }\textbf {\bibinfo {volume} {526}},\ \bibinfo {eid} {A156} (\bibinfo {year} {2011})},\ \Eprint {http://arxiv.org/abs/1010.5369} {arXiv:1010.5369 [astro-ph.SR]} \BibitemShut {NoStop}%
\bibitem [{\citenamefont {{de Jager}}\ \emph {et~al.}(1988)\citenamefont {{de Jager}}, \citenamefont {{Nieuwenhuijzen}},\ and\ \citenamefont {{van der Hucht}}}]{dejager:98}%
  \BibitemOpen
  \bibfield  {author} {\bibinfo {author} {\bibfnamefont {C.}~\bibnamefont {{de Jager}}}, \bibinfo {author} {\bibfnamefont {H.}~\bibnamefont {{Nieuwenhuijzen}}}, \ and\ \bibinfo {author} {\bibfnamefont {K.~A.}\ \bibnamefont {{van der Hucht}}},\ }\href@noop {} {\bibfield  {journal} {\bibinfo  {journal} {\aaps}\ }\textbf {\bibinfo {volume} {72}},\ \bibinfo {pages} {259} (\bibinfo {year} {1988})}\BibitemShut {NoStop}%
\bibitem [{\citenamefont {{Beasor}}\ \emph {et~al.}(2020)\citenamefont {{Beasor}}, \citenamefont {{Davies}}, \citenamefont {{Smith}}, \citenamefont {{van Loon}}, \citenamefont {{Gehrz}},\ and\ \citenamefont {{Figer}}}]{beasor:20}%
  \BibitemOpen
  \bibfield  {author} {\bibinfo {author} {\bibfnamefont {E.~R.}\ \bibnamefont {{Beasor}}}, \bibinfo {author} {\bibfnamefont {B.}~\bibnamefont {{Davies}}}, \bibinfo {author} {\bibfnamefont {N.}~\bibnamefont {{Smith}}}, \bibinfo {author} {\bibfnamefont {J.~T.}\ \bibnamefont {{van Loon}}}, \bibinfo {author} {\bibfnamefont {R.~D.}\ \bibnamefont {{Gehrz}}}, \ and\ \bibinfo {author} {\bibfnamefont {D.~F.}\ \bibnamefont {{Figer}}},\ }\href {\doibase 10.1093/mnras/staa255} {\bibfield  {journal} {\bibinfo  {journal} {\mnras}\ }\textbf {\bibinfo {volume} {492}},\ \bibinfo {pages} {5994} (\bibinfo {year} {2020})},\ \Eprint {http://arxiv.org/abs/2001.07222} {arXiv:2001.07222 [astro-ph.SR]} \BibitemShut {NoStop}%
\bibitem [{\citenamefont {{Wen}}\ \emph {et~al.}(2024)\citenamefont {{Wen}}, \citenamefont {{Gao}}, \citenamefont {{Yang}}, \citenamefont {{Chen}}, \citenamefont {{Ren}}, \citenamefont {{Wang}},\ and\ \citenamefont {{Jiang}}}]{wen:24}%
  \BibitemOpen
  \bibfield  {author} {\bibinfo {author} {\bibfnamefont {J.}~\bibnamefont {{Wen}}}, \bibinfo {author} {\bibfnamefont {J.}~\bibnamefont {{Gao}}}, \bibinfo {author} {\bibfnamefont {M.}~\bibnamefont {{Yang}}}, \bibinfo {author} {\bibfnamefont {B.}~\bibnamefont {{Chen}}}, \bibinfo {author} {\bibfnamefont {Y.}~\bibnamefont {{Ren}}}, \bibinfo {author} {\bibfnamefont {T.}~\bibnamefont {{Wang}}}, \ and\ \bibinfo {author} {\bibfnamefont {B.}~\bibnamefont {{Jiang}}},\ }\href {\doibase 10.3847/1538-3881/ad12bf} {\bibfield  {journal} {\bibinfo  {journal} {\aj}\ }\textbf {\bibinfo {volume} {167}},\ \bibinfo {eid} {51} (\bibinfo {year} {2024})},\ \Eprint {http://arxiv.org/abs/2401.03778} {arXiv:2401.03778 [astro-ph.GA]} \BibitemShut {NoStop}%
\bibitem [{\citenamefont {{Yang}}\ \emph {et~al.}(2023)\citenamefont {{Yang}}, \citenamefont {{Bonanos}}, \citenamefont {{Jiang}}, \citenamefont {{Zapartas}}, \citenamefont {{Gao}}, \citenamefont {{Ren}}, \citenamefont {{Lam}}, \citenamefont {{Wang}}, \citenamefont {{Maravelias}}, \citenamefont {{Gavras}}, \citenamefont {{Wang}}, \citenamefont {{Chen}}, \citenamefont {{Tramper}}, \citenamefont {{de Wit}}, \citenamefont {{Chen}}, \citenamefont {{Wen}}, \citenamefont {{Liu}}, \citenamefont {{Tian}}, \citenamefont {{Antoniadis}},\ and\ \citenamefont {{Luo}}}]{yang:23}%
  \BibitemOpen
  \bibfield  {author} {\bibinfo {author} {\bibfnamefont {M.}~\bibnamefont {{Yang}}}, \bibinfo {author} {\bibfnamefont {A.~Z.}\ \bibnamefont {{Bonanos}}}, \bibinfo {author} {\bibfnamefont {B.}~\bibnamefont {{Jiang}}}, \bibinfo {author} {\bibfnamefont {E.}~\bibnamefont {{Zapartas}}}, \bibinfo {author} {\bibfnamefont {J.}~\bibnamefont {{Gao}}}, \bibinfo {author} {\bibfnamefont {Y.}~\bibnamefont {{Ren}}}, \bibinfo {author} {\bibfnamefont {M.~I.}\ \bibnamefont {{Lam}}}, \bibinfo {author} {\bibfnamefont {T.}~\bibnamefont {{Wang}}}, \bibinfo {author} {\bibfnamefont {G.}~\bibnamefont {{Maravelias}}}, \bibinfo {author} {\bibfnamefont {P.}~\bibnamefont {{Gavras}}}, \bibinfo {author} {\bibfnamefont {S.}~\bibnamefont {{Wang}}}, \bibinfo {author} {\bibfnamefont {X.}~\bibnamefont {{Chen}}}, \bibinfo {author} {\bibfnamefont {F.}~\bibnamefont {{Tramper}}}, \bibinfo {author} {\bibfnamefont {S.}~\bibnamefont {{de Wit}}}, \bibinfo {author} {\bibfnamefont {B.}~\bibnamefont {{Chen}}}, \bibinfo {author} {\bibfnamefont
  {J.}~\bibnamefont {{Wen}}}, \bibinfo {author} {\bibfnamefont {J.}~\bibnamefont {{Liu}}}, \bibinfo {author} {\bibfnamefont {H.}~\bibnamefont {{Tian}}}, \bibinfo {author} {\bibfnamefont {K.}~\bibnamefont {{Antoniadis}}}, \ and\ \bibinfo {author} {\bibfnamefont {C.}~\bibnamefont {{Luo}}},\ }\href {\doibase 10.1051/0004-6361/202244770} {\bibfield  {journal} {\bibinfo  {journal} {\aap}\ }\textbf {\bibinfo {volume} {676}},\ \bibinfo {eid} {A84} (\bibinfo {year} {2023})},\ \Eprint {http://arxiv.org/abs/2304.01835} {arXiv:2304.01835 [astro-ph.SR]} \BibitemShut {NoStop}%
\bibitem [{\citenamefont {{Schr{\"o}der}}\ and\ \citenamefont {{Cuntz}}(2005)}]{schroder:05}%
  \BibitemOpen
  \bibfield  {author} {\bibinfo {author} {\bibfnamefont {K.~P.}\ \bibnamefont {{Schr{\"o}der}}}\ and\ \bibinfo {author} {\bibfnamefont {M.}~\bibnamefont {{Cuntz}}},\ }\href {\doibase 10.1086/491579} {\bibfield  {journal} {\bibinfo  {journal} {\apjl}\ }\textbf {\bibinfo {volume} {630}},\ \bibinfo {pages} {L73} (\bibinfo {year} {2005})},\ \Eprint {http://arxiv.org/abs/astro-ph/0507598} {arXiv:astro-ph/0507598 [astro-ph]} \BibitemShut {NoStop}%
\bibitem [{\citenamefont {{Beasor}}\ and\ \citenamefont {{Smith}}(2022)}]{beasor:22}%
  \BibitemOpen
  \bibfield  {author} {\bibinfo {author} {\bibfnamefont {E.~R.}\ \bibnamefont {{Beasor}}}\ and\ \bibinfo {author} {\bibfnamefont {N.}~\bibnamefont {{Smith}}},\ }\href {\doibase 10.3847/1538-4357/ac6dcf} {\bibfield  {journal} {\bibinfo  {journal} {\apj}\ }\textbf {\bibinfo {volume} {933}},\ \bibinfo {eid} {41} (\bibinfo {year} {2022})},\ \Eprint {http://arxiv.org/abs/2205.02207} {arXiv:2205.02207 [astro-ph.SR]} \BibitemShut {NoStop}%
\bibitem [{\citenamefont {{Davies}}\ \emph {et~al.}(2018)\citenamefont {{Davies}}, \citenamefont {{Crowther}},\ and\ \citenamefont {{Beasor}}}]{davies:18}%
  \BibitemOpen
  \bibfield  {author} {\bibinfo {author} {\bibfnamefont {B.}~\bibnamefont {{Davies}}}, \bibinfo {author} {\bibfnamefont {P.~A.}\ \bibnamefont {{Crowther}}}, \ and\ \bibinfo {author} {\bibfnamefont {E.~R.}\ \bibnamefont {{Beasor}}},\ }\href {\doibase 10.1093/mnras/sty1302} {\bibfield  {journal} {\bibinfo  {journal} {\mnras}\ }\textbf {\bibinfo {volume} {478}},\ \bibinfo {pages} {3138} (\bibinfo {year} {2018})},\ \Eprint {http://arxiv.org/abs/1804.06417} {arXiv:1804.06417 [astro-ph.SR]} \BibitemShut {NoStop}%
\bibitem [{\citenamefont {{Vink}}\ and\ \citenamefont {{Sabhahit}}(2023)}]{vink:23}%
  \BibitemOpen
  \bibfield  {author} {\bibinfo {author} {\bibfnamefont {J.~S.}\ \bibnamefont {{Vink}}}\ and\ \bibinfo {author} {\bibfnamefont {G.~N.}\ \bibnamefont {{Sabhahit}}},\ }\href {\doibase 10.1051/0004-6361/202347801} {\bibfield  {journal} {\bibinfo  {journal} {\aap}\ }\textbf {\bibinfo {volume} {678}},\ \bibinfo {eid} {L3} (\bibinfo {year} {2023})},\ \Eprint {http://arxiv.org/abs/2309.08657} {arXiv:2309.08657 [astro-ph.SR]} \BibitemShut {NoStop}%
\bibitem [{\citenamefont {{Dessart}}\ and\ \citenamefont {{Jacobson-Gal{\'a}n}}(2023)}]{dessart:23}%
  \BibitemOpen
  \bibfield  {author} {\bibinfo {author} {\bibfnamefont {L.}~\bibnamefont {{Dessart}}}\ and\ \bibinfo {author} {\bibfnamefont {W.~V.}\ \bibnamefont {{Jacobson-Gal{\'a}n}}},\ }\href {\doibase 10.1051/0004-6361/202346754} {\bibfield  {journal} {\bibinfo  {journal} {\aap}\ }\textbf {\bibinfo {volume} {677}},\ \bibinfo {eid} {A105} (\bibinfo {year} {2023})},\ \Eprint {http://arxiv.org/abs/2307.08584} {arXiv:2307.08584 [astro-ph.SR]} \BibitemShut {NoStop}%
\bibitem [{\citenamefont {{Zimmerman}}\ \emph {et~al.}(2023)\citenamefont {{Zimmerman}}, \citenamefont {{Irani}}, \citenamefont {{Chen}}, \citenamefont {{Gal-Yam}}, \citenamefont {{Schulze}}, \citenamefont {{Perley}}, \citenamefont {{Sollerman}}, \citenamefont {{Filippenko}}, \citenamefont {{Shenar}}, \citenamefont {{Yaron}}, \citenamefont {{Shahaf}}, \citenamefont {{Bruch}}, \citenamefont {{Ofek}}, \citenamefont {{De Cia}}, \citenamefont {{Brink}}, \citenamefont {{Yang}}, \citenamefont {{Vasylyev}}, \citenamefont {{Ben Ami}}, \citenamefont {{Aubert}}, \citenamefont {{Badash}}, \citenamefont {{Bloom}}, \citenamefont {{Brown}}, \citenamefont {{De}}, \citenamefont {{Dimitriadis}}, \citenamefont {{Fransson}}, \citenamefont {{Fremling}}, \citenamefont {{Hinds}}, \citenamefont {{Horesh}}, \citenamefont {{Johansson}}, \citenamefont {{Kasliwal}}, \citenamefont {{Kulkarni}}, \citenamefont {{Kushnir}}, \citenamefont {{Martin}}, \citenamefont {{Matuzewski}}, \citenamefont {{McGurk}}, \citenamefont {{Miller}},
  \citenamefont {{Morag}}, \citenamefont {{Neil}}, \citenamefont {{Nugent}}, \citenamefont {{Post}}, \citenamefont {{Prusinski}}, \citenamefont {{Qin}}, \citenamefont {{Raichoor}}, \citenamefont {{Riddle}}, \citenamefont {{Rowe}}, \citenamefont {{Rusholme}}, \citenamefont {{Sfaradi}}, \citenamefont {{Sjoberg}}, \citenamefont {{Soumagnac}}, \citenamefont {{Stein}}, \citenamefont {{Strotjohann}}, \citenamefont {{Terwel}}, \citenamefont {{Wasserman}}, \citenamefont {{Wise}}, \citenamefont {{Wold}}, \citenamefont {{Yan}},\ and\ \citenamefont {{Zhang}}}]{zimmerman:23}%
  \BibitemOpen
  \bibfield  {author} {\bibinfo {author} {\bibfnamefont {E.~A.}\ \bibnamefont {{Zimmerman}}}, \bibinfo {author} {\bibfnamefont {I.}~\bibnamefont {{Irani}}}, \bibinfo {author} {\bibfnamefont {P.}~\bibnamefont {{Chen}}}, \bibinfo {author} {\bibfnamefont {A.}~\bibnamefont {{Gal-Yam}}}, \bibinfo {author} {\bibfnamefont {S.}~\bibnamefont {{Schulze}}}, \bibinfo {author} {\bibfnamefont {D.~A.}\ \bibnamefont {{Perley}}}, \bibinfo {author} {\bibfnamefont {J.}~\bibnamefont {{Sollerman}}}, \bibinfo {author} {\bibfnamefont {A.~V.}\ \bibnamefont {{Filippenko}}}, \bibinfo {author} {\bibfnamefont {T.}~\bibnamefont {{Shenar}}}, \bibinfo {author} {\bibfnamefont {O.}~\bibnamefont {{Yaron}}}, \bibinfo {author} {\bibfnamefont {S.}~\bibnamefont {{Shahaf}}}, \bibinfo {author} {\bibfnamefont {R.~J.}\ \bibnamefont {{Bruch}}}, \bibinfo {author} {\bibfnamefont {E.~O.}\ \bibnamefont {{Ofek}}}, \bibinfo {author} {\bibfnamefont {A.}~\bibnamefont {{De Cia}}}, \bibinfo {author} {\bibfnamefont {T.~G.}\ \bibnamefont {{Brink}}}, \bibinfo
  {author} {\bibfnamefont {Y.}~\bibnamefont {{Yang}}}, \bibinfo {author} {\bibfnamefont {S.~S.}\ \bibnamefont {{Vasylyev}}}, \bibinfo {author} {\bibfnamefont {S.}~\bibnamefont {{Ben Ami}}}, \bibinfo {author} {\bibfnamefont {M.}~\bibnamefont {{Aubert}}}, \bibinfo {author} {\bibfnamefont {A.}~\bibnamefont {{Badash}}}, \bibinfo {author} {\bibfnamefont {J.~S.}\ \bibnamefont {{Bloom}}}, \bibinfo {author} {\bibfnamefont {P.~J.}\ \bibnamefont {{Brown}}}, \bibinfo {author} {\bibfnamefont {K.}~\bibnamefont {{De}}}, \bibinfo {author} {\bibfnamefont {G.}~\bibnamefont {{Dimitriadis}}}, \bibinfo {author} {\bibfnamefont {C.}~\bibnamefont {{Fransson}}}, \bibinfo {author} {\bibfnamefont {C.}~\bibnamefont {{Fremling}}}, \bibinfo {author} {\bibfnamefont {K.}~\bibnamefont {{Hinds}}}, \bibinfo {author} {\bibfnamefont {A.}~\bibnamefont {{Horesh}}}, \bibinfo {author} {\bibfnamefont {J.~P.}\ \bibnamefont {{Johansson}}}, \bibinfo {author} {\bibfnamefont {M.~M.}\ \bibnamefont {{Kasliwal}}}, \bibinfo {author} {\bibfnamefont {S.~R.}\
  \bibnamefont {{Kulkarni}}}, \bibinfo {author} {\bibfnamefont {D.}~\bibnamefont {{Kushnir}}}, \bibinfo {author} {\bibfnamefont {C.}~\bibnamefont {{Martin}}}, \bibinfo {author} {\bibfnamefont {M.}~\bibnamefont {{Matuzewski}}}, \bibinfo {author} {\bibfnamefont {R.~C.}\ \bibnamefont {{McGurk}}}, \bibinfo {author} {\bibfnamefont {A.~A.}\ \bibnamefont {{Miller}}}, \bibinfo {author} {\bibfnamefont {J.}~\bibnamefont {{Morag}}}, \bibinfo {author} {\bibfnamefont {J.~D.}\ \bibnamefont {{Neil}}}, \bibinfo {author} {\bibfnamefont {P.~E.}\ \bibnamefont {{Nugent}}}, \bibinfo {author} {\bibfnamefont {R.~S.}\ \bibnamefont {{Post}}}, \bibinfo {author} {\bibfnamefont {N.~Z.}\ \bibnamefont {{Prusinski}}}, \bibinfo {author} {\bibfnamefont {Y.}~\bibnamefont {{Qin}}}, \bibinfo {author} {\bibfnamefont {A.}~\bibnamefont {{Raichoor}}}, \bibinfo {author} {\bibfnamefont {R.}~\bibnamefont {{Riddle}}}, \bibinfo {author} {\bibfnamefont {M.}~\bibnamefont {{Rowe}}}, \bibinfo {author} {\bibfnamefont {B.}~\bibnamefont {{Rusholme}}}, \bibinfo
  {author} {\bibfnamefont {I.}~\bibnamefont {{Sfaradi}}}, \bibinfo {author} {\bibfnamefont {K.~M.}\ \bibnamefont {{Sjoberg}}}, \bibinfo {author} {\bibfnamefont {M.}~\bibnamefont {{Soumagnac}}}, \bibinfo {author} {\bibfnamefont {R.~D.}\ \bibnamefont {{Stein}}}, \bibinfo {author} {\bibfnamefont {N.~L.}\ \bibnamefont {{Strotjohann}}}, \bibinfo {author} {\bibfnamefont {J.~H.}\ \bibnamefont {{Terwel}}}, \bibinfo {author} {\bibfnamefont {T.}~\bibnamefont {{Wasserman}}}, \bibinfo {author} {\bibfnamefont {J.}~\bibnamefont {{Wise}}}, \bibinfo {author} {\bibfnamefont {A.}~\bibnamefont {{Wold}}}, \bibinfo {author} {\bibfnamefont {L.}~\bibnamefont {{Yan}}}, \ and\ \bibinfo {author} {\bibfnamefont {K.}~\bibnamefont {{Zhang}}},\ }\href {\doibase 10.48550/arXiv.2310.10727} {\bibfield  {journal} {\bibinfo  {journal} {arXiv e-prints}\ ,\ \bibinfo {eid} {arXiv:2310.10727}} (\bibinfo {year} {2023})},\ \Eprint {http://arxiv.org/abs/2310.10727} {arXiv:2310.10727 [astro-ph.HE]} \BibitemShut {NoStop}%
\bibitem [{\citenamefont {{Irani}}\ \emph {et~al.}(2023)\citenamefont {{Irani}}, \citenamefont {{Morag}}, \citenamefont {{Gal-Yam}}, \citenamefont {{Waxman}}, \citenamefont {{Schulze}}, \citenamefont {{Sollerman}}, \citenamefont {{Hinds}}, \citenamefont {{Perley}}, \citenamefont {{Chen}}, \citenamefont {{Strotjohann}}, \citenamefont {{Yaron}}, \citenamefont {{Zimmerman}}, \citenamefont {{Bruch}}, \citenamefont {{Ofek}}, \citenamefont {{Soumagnac}}, \citenamefont {{Yang}}, \citenamefont {{Groom}}, \citenamefont {{Masci}}, \citenamefont {{Riddle}}, \citenamefont {{Bellm}},\ and\ \citenamefont {{Hale}}}]{irani:23}%
  \BibitemOpen
  \bibfield  {author} {\bibinfo {author} {\bibfnamefont {I.}~\bibnamefont {{Irani}}}, \bibinfo {author} {\bibfnamefont {J.}~\bibnamefont {{Morag}}}, \bibinfo {author} {\bibfnamefont {A.}~\bibnamefont {{Gal-Yam}}}, \bibinfo {author} {\bibfnamefont {E.}~\bibnamefont {{Waxman}}}, \bibinfo {author} {\bibfnamefont {S.}~\bibnamefont {{Schulze}}}, \bibinfo {author} {\bibfnamefont {J.}~\bibnamefont {{Sollerman}}}, \bibinfo {author} {\bibfnamefont {K.-R.}\ \bibnamefont {{Hinds}}}, \bibinfo {author} {\bibfnamefont {D.~A.}\ \bibnamefont {{Perley}}}, \bibinfo {author} {\bibfnamefont {P.}~\bibnamefont {{Chen}}}, \bibinfo {author} {\bibfnamefont {N.~L.}\ \bibnamefont {{Strotjohann}}}, \bibinfo {author} {\bibfnamefont {O.}~\bibnamefont {{Yaron}}}, \bibinfo {author} {\bibfnamefont {E.~A.}\ \bibnamefont {{Zimmerman}}}, \bibinfo {author} {\bibfnamefont {R.}~\bibnamefont {{Bruch}}}, \bibinfo {author} {\bibfnamefont {E.~O.}\ \bibnamefont {{Ofek}}}, \bibinfo {author} {\bibfnamefont {M.~T.}\ \bibnamefont {{Soumagnac}}}, \bibinfo
  {author} {\bibfnamefont {Y.}~\bibnamefont {{Yang}}}, \bibinfo {author} {\bibfnamefont {S.~L.}\ \bibnamefont {{Groom}}}, \bibinfo {author} {\bibfnamefont {F.~J.}\ \bibnamefont {{Masci}}}, \bibinfo {author} {\bibfnamefont {R.}~\bibnamefont {{Riddle}}}, \bibinfo {author} {\bibfnamefont {E.~C.}\ \bibnamefont {{Bellm}}}, \ and\ \bibinfo {author} {\bibfnamefont {D.}~\bibnamefont {{Hale}}},\ }\href {\doibase 10.48550/arXiv.2310.16885} {\bibfield  {journal} {\bibinfo  {journal} {arXiv e-prints}\ ,\ \bibinfo {eid} {arXiv:2310.16885}} (\bibinfo {year} {2023})},\ \Eprint {http://arxiv.org/abs/2310.16885} {arXiv:2310.16885 [astro-ph.HE]} \BibitemShut {NoStop}%
\bibitem [{\citenamefont {{Gr{\"a}fener}}\ and\ \citenamefont {{Vink}}(2016)}]{grafener:16}%
  \BibitemOpen
  \bibfield  {author} {\bibinfo {author} {\bibfnamefont {G.}~\bibnamefont {{Gr{\"a}fener}}}\ and\ \bibinfo {author} {\bibfnamefont {J.~S.}\ \bibnamefont {{Vink}}},\ }\href {\doibase 10.1093/mnras/stv2283} {\bibfield  {journal} {\bibinfo  {journal} {\mnras}\ }\textbf {\bibinfo {volume} {455}},\ \bibinfo {pages} {112} (\bibinfo {year} {2016})},\ \Eprint {http://arxiv.org/abs/1510.00013} {arXiv:1510.00013 [astro-ph.SR]} \BibitemShut {NoStop}%
\bibitem [{\citenamefont {{Tsuna}}\ \emph {et~al.}(2023)\citenamefont {{Tsuna}}, \citenamefont {{Murase}},\ and\ \citenamefont {{Moriya}}}]{tsuna:23}%
  \BibitemOpen
  \bibfield  {author} {\bibinfo {author} {\bibfnamefont {D.}~\bibnamefont {{Tsuna}}}, \bibinfo {author} {\bibfnamefont {K.}~\bibnamefont {{Murase}}}, \ and\ \bibinfo {author} {\bibfnamefont {T.~J.}\ \bibnamefont {{Moriya}}},\ }\href {\doibase 10.3847/1538-4357/acdb71} {\bibfield  {journal} {\bibinfo  {journal} {\apj}\ }\textbf {\bibinfo {volume} {952}},\ \bibinfo {eid} {115} (\bibinfo {year} {2023})},\ \Eprint {http://arxiv.org/abs/2301.10667} {arXiv:2301.10667 [astro-ph.HE]} \BibitemShut {NoStop}%
\bibitem [{\citenamefont {{Morozova}}\ \emph {et~al.}(2015)\citenamefont {{Morozova}}, \citenamefont {{Piro}}, \citenamefont {{Renzo}}, \citenamefont {{Ott}}, \citenamefont {{Clausen}}, \citenamefont {{Couch}}, \citenamefont {{Ellis}},\ and\ \citenamefont {{Roberts}}}]{morozova:15}%
  \BibitemOpen
  \bibfield  {author} {\bibinfo {author} {\bibfnamefont {V.}~\bibnamefont {{Morozova}}}, \bibinfo {author} {\bibfnamefont {A.~L.}\ \bibnamefont {{Piro}}}, \bibinfo {author} {\bibfnamefont {M.}~\bibnamefont {{Renzo}}}, \bibinfo {author} {\bibfnamefont {C.~D.}\ \bibnamefont {{Ott}}}, \bibinfo {author} {\bibfnamefont {D.}~\bibnamefont {{Clausen}}}, \bibinfo {author} {\bibfnamefont {S.~M.}\ \bibnamefont {{Couch}}}, \bibinfo {author} {\bibfnamefont {J.}~\bibnamefont {{Ellis}}}, \ and\ \bibinfo {author} {\bibfnamefont {L.~F.}\ \bibnamefont {{Roberts}}},\ }\href {\doibase 10.1088/0004-637X/814/1/63} {\bibfield  {journal} {\bibinfo  {journal} {\apj}\ }\textbf {\bibinfo {volume} {814}},\ \bibinfo {eid} {63} (\bibinfo {year} {2015})},\ \Eprint {http://arxiv.org/abs/1505.06746} {arXiv:1505.06746 [astro-ph.HE]} \BibitemShut {NoStop}%
\bibitem [{\citenamefont {{Goldberg}}\ \emph {et~al.}(2022{\natexlab{b}})\citenamefont {{Goldberg}}, \citenamefont {{Jiang}},\ and\ \citenamefont {{Bildsten}}}]{goldberg:22b}%
  \BibitemOpen
  \bibfield  {author} {\bibinfo {author} {\bibfnamefont {J.~A.}\ \bibnamefont {{Goldberg}}}, \bibinfo {author} {\bibfnamefont {Y.-F.}\ \bibnamefont {{Jiang}}}, \ and\ \bibinfo {author} {\bibfnamefont {L.}~\bibnamefont {{Bildsten}}},\ }\href {\doibase 10.3847/1538-4357/ac75e3} {\bibfield  {journal} {\bibinfo  {journal} {\apj}\ }\textbf {\bibinfo {volume} {933}},\ \bibinfo {eid} {164} (\bibinfo {year} {2022}{\natexlab{b}})},\ \Eprint {http://arxiv.org/abs/2206.04134} {arXiv:2206.04134 [astro-ph.SR]} \BibitemShut {NoStop}%
\bibitem [{\citenamefont {{Nakar}}\ and\ \citenamefont {{Sari}}(2010)}]{nakar:2010}%
  \BibitemOpen
  \bibfield  {author} {\bibinfo {author} {\bibfnamefont {E.}~\bibnamefont {{Nakar}}}\ and\ \bibinfo {author} {\bibfnamefont {R.}~\bibnamefont {{Sari}}},\ }\href {\doibase 10.1088/0004-637X/725/1/904} {\bibfield  {journal} {\bibinfo  {journal} {\apj}\ }\textbf {\bibinfo {volume} {725}},\ \bibinfo {pages} {904} (\bibinfo {year} {2010})},\ \Eprint {http://arxiv.org/abs/1004.2496} {arXiv:1004.2496 [astro-ph.HE]} \BibitemShut {NoStop}%
\bibitem [{\citenamefont {{Svirski}}\ \emph {et~al.}(2012)\citenamefont {{Svirski}}, \citenamefont {{Nakar}},\ and\ \citenamefont {{Sari}}}]{svirski:2012}%
  \BibitemOpen
  \bibfield  {author} {\bibinfo {author} {\bibfnamefont {G.}~\bibnamefont {{Svirski}}}, \bibinfo {author} {\bibfnamefont {E.}~\bibnamefont {{Nakar}}}, \ and\ \bibinfo {author} {\bibfnamefont {R.}~\bibnamefont {{Sari}}},\ }\href {\doibase 10.1088/0004-637X/759/2/108} {\bibfield  {journal} {\bibinfo  {journal} {\apj}\ }\textbf {\bibinfo {volume} {759}},\ \bibinfo {eid} {108} (\bibinfo {year} {2012})},\ \Eprint {http://arxiv.org/abs/1202.3437} {arXiv:1202.3437 [astro-ph.HE]} \BibitemShut {NoStop}%
\bibitem [{\citenamefont {{Haynie}}\ and\ \citenamefont {{Piro}}(2021)}]{haynie:2021}%
  \BibitemOpen
  \bibfield  {author} {\bibinfo {author} {\bibfnamefont {A.}~\bibnamefont {{Haynie}}}\ and\ \bibinfo {author} {\bibfnamefont {A.~L.}\ \bibnamefont {{Piro}}},\ }\href {\doibase 10.3847/1538-4357/abe938} {\bibfield  {journal} {\bibinfo  {journal} {\apj}\ }\textbf {\bibinfo {volume} {910}},\ \bibinfo {eid} {128} (\bibinfo {year} {2021})},\ \Eprint {http://arxiv.org/abs/2011.01937} {arXiv:2011.01937 [astro-ph.HE]} \BibitemShut {NoStop}%
\bibitem [{\citenamefont {{Tsuna}}\ \emph {et~al.}(2021)\citenamefont {{Tsuna}}, \citenamefont {{Kashiyama}},\ and\ \citenamefont {{Shigeyama}}}]{tsuna:21}%
  \BibitemOpen
  \bibfield  {author} {\bibinfo {author} {\bibfnamefont {D.}~\bibnamefont {{Tsuna}}}, \bibinfo {author} {\bibfnamefont {K.}~\bibnamefont {{Kashiyama}}}, \ and\ \bibinfo {author} {\bibfnamefont {T.}~\bibnamefont {{Shigeyama}}},\ }\href {\doibase 10.3847/1538-4357/abfaf8} {\bibfield  {journal} {\bibinfo  {journal} {\apj}\ }\textbf {\bibinfo {volume} {914}},\ \bibinfo {eid} {64} (\bibinfo {year} {2021})},\ \Eprint {http://arxiv.org/abs/2103.08338} {arXiv:2103.08338 [astro-ph.HE]} \BibitemShut {NoStop}%
\bibitem [{\citenamefont {{Ohnaka}}\ \emph {et~al.}(2017)\citenamefont {{Ohnaka}}, \citenamefont {{Weigelt}},\ and\ \citenamefont {{Hofmann}}}]{ohnaka:17}%
  \BibitemOpen
  \bibfield  {author} {\bibinfo {author} {\bibfnamefont {K.}~\bibnamefont {{Ohnaka}}}, \bibinfo {author} {\bibfnamefont {G.}~\bibnamefont {{Weigelt}}}, \ and\ \bibinfo {author} {\bibfnamefont {K.~H.}\ \bibnamefont {{Hofmann}}},\ }\href {\doibase 10.1038/nature23445} {\bibfield  {journal} {\bibinfo  {journal} {\nat}\ }\textbf {\bibinfo {volume} {548}},\ \bibinfo {pages} {310} (\bibinfo {year} {2017})},\ \Eprint {http://arxiv.org/abs/1708.06372} {arXiv:1708.06372 [astro-ph.SR]} \BibitemShut {NoStop}%
\bibitem [{\citenamefont {{L{\'o}pez Ariste}}\ \emph {et~al.}(2022)\citenamefont {{L{\'o}pez Ariste}}, \citenamefont {{Georgiev}}, \citenamefont {{Mathias}}, \citenamefont {{L{\`e}bre}}, \citenamefont {{Wavasseur}}, \citenamefont {{Josselin}}, \citenamefont {{Konstantinova-Antova}},\ and\ \citenamefont {{Roudier}}}]{lopez:22}%
  \BibitemOpen
  \bibfield  {author} {\bibinfo {author} {\bibfnamefont {A.}~\bibnamefont {{L{\'o}pez Ariste}}}, \bibinfo {author} {\bibfnamefont {S.}~\bibnamefont {{Georgiev}}}, \bibinfo {author} {\bibfnamefont {P.}~\bibnamefont {{Mathias}}}, \bibinfo {author} {\bibfnamefont {A.}~\bibnamefont {{L{\`e}bre}}}, \bibinfo {author} {\bibfnamefont {M.}~\bibnamefont {{Wavasseur}}}, \bibinfo {author} {\bibfnamefont {E.}~\bibnamefont {{Josselin}}}, \bibinfo {author} {\bibfnamefont {R.}~\bibnamefont {{Konstantinova-Antova}}}, \ and\ \bibinfo {author} {\bibfnamefont {T.}~\bibnamefont {{Roudier}}},\ }\href {\doibase 10.1051/0004-6361/202142271} {\bibfield  {journal} {\bibinfo  {journal} {\aap}\ }\textbf {\bibinfo {volume} {661}},\ \bibinfo {eid} {A91} (\bibinfo {year} {2022})},\ \Eprint {http://arxiv.org/abs/2202.12011} {arXiv:2202.12011 [astro-ph.SR]} \BibitemShut {NoStop}%
\bibitem [{\citenamefont {{Josselin}}\ and\ \citenamefont {{Plez}}(2007)}]{josselin:07}%
  \BibitemOpen
  \bibfield  {author} {\bibinfo {author} {\bibfnamefont {E.}~\bibnamefont {{Josselin}}}\ and\ \bibinfo {author} {\bibfnamefont {B.}~\bibnamefont {{Plez}}},\ }\href {\doibase 10.1051/0004-6361:20066353} {\bibfield  {journal} {\bibinfo  {journal} {\aap}\ }\textbf {\bibinfo {volume} {469}},\ \bibinfo {pages} {671} (\bibinfo {year} {2007})},\ \Eprint {http://arxiv.org/abs/0705.0266} {arXiv:0705.0266 [astro-ph]} \BibitemShut {NoStop}%
\bibitem [{\citenamefont {{Arroyo-Torres}}\ \emph {et~al.}(2015)\citenamefont {{Arroyo-Torres}}, \citenamefont {{Wittkowski}}, \citenamefont {{Chiavassa}}, \citenamefont {{Scholz}}, \citenamefont {{Freytag}}, \citenamefont {{Marcaide}}, \citenamefont {{Hauschildt}}, \citenamefont {{Wood}},\ and\ \citenamefont {{Abellan}}}]{arroyo-torres:15}%
  \BibitemOpen
  \bibfield  {author} {\bibinfo {author} {\bibfnamefont {B.}~\bibnamefont {{Arroyo-Torres}}}, \bibinfo {author} {\bibfnamefont {M.}~\bibnamefont {{Wittkowski}}}, \bibinfo {author} {\bibfnamefont {A.}~\bibnamefont {{Chiavassa}}}, \bibinfo {author} {\bibfnamefont {M.}~\bibnamefont {{Scholz}}}, \bibinfo {author} {\bibfnamefont {B.}~\bibnamefont {{Freytag}}}, \bibinfo {author} {\bibfnamefont {J.~M.}\ \bibnamefont {{Marcaide}}}, \bibinfo {author} {\bibfnamefont {P.~H.}\ \bibnamefont {{Hauschildt}}}, \bibinfo {author} {\bibfnamefont {P.~R.}\ \bibnamefont {{Wood}}}, \ and\ \bibinfo {author} {\bibfnamefont {F.~J.}\ \bibnamefont {{Abellan}}},\ }\href {\doibase 10.1051/0004-6361/201425212} {\bibfield  {journal} {\bibinfo  {journal} {\aap}\ }\textbf {\bibinfo {volume} {575}},\ \bibinfo {eid} {A50} (\bibinfo {year} {2015})},\ \Eprint {http://arxiv.org/abs/1501.01560} {arXiv:1501.01560 [astro-ph.SR]} \BibitemShut {NoStop}%
\bibitem [{\citenamefont {{Khouri}}\ \emph {et~al.}(2024)\citenamefont {{Khouri}}, \citenamefont {{Olofsson}}, \citenamefont {{Vlemmings}}, \citenamefont {{Schirmer}}, \citenamefont {{Tafoya}}, \citenamefont {{Maercker}}, \citenamefont {{De Beck}}, \citenamefont {{Nyman}},\ and\ \citenamefont {{Saberi}}}]{khouri:24}%
  \BibitemOpen
  \bibfield  {author} {\bibinfo {author} {\bibfnamefont {T.}~\bibnamefont {{Khouri}}}, \bibinfo {author} {\bibfnamefont {H.}~\bibnamefont {{Olofsson}}}, \bibinfo {author} {\bibfnamefont {W.~H.~T.}\ \bibnamefont {{Vlemmings}}}, \bibinfo {author} {\bibfnamefont {T.}~\bibnamefont {{Schirmer}}}, \bibinfo {author} {\bibfnamefont {D.}~\bibnamefont {{Tafoya}}}, \bibinfo {author} {\bibfnamefont {M.}~\bibnamefont {{Maercker}}}, \bibinfo {author} {\bibfnamefont {E.}~\bibnamefont {{De Beck}}}, \bibinfo {author} {\bibfnamefont {L.~{\r{A}}.}\ \bibnamefont {{Nyman}}}, \ and\ \bibinfo {author} {\bibfnamefont {M.}~\bibnamefont {{Saberi}}},\ }\href {\doibase 10.48550/arXiv.2402.13676} {\bibfield  {journal} {\bibinfo  {journal} {arXiv e-prints}\ ,\ \bibinfo {eid} {arXiv:2402.13676}} (\bibinfo {year} {2024})},\ \Eprint {http://arxiv.org/abs/2402.13676} {arXiv:2402.13676 [astro-ph.SR]} \BibitemShut {NoStop}%
\bibitem [{\citenamefont {{Montarg{\`e}s}}\ \emph {et~al.}(2021)\citenamefont {{Montarg{\`e}s}}, \citenamefont {{Cannon}}, \citenamefont {{Lagadec}}, \citenamefont {{de Koter}}, \citenamefont {{Kervella}}, \citenamefont {{Sanchez-Bermudez}}, \citenamefont {{Paladini}}, \citenamefont {{Cantalloube}}, \citenamefont {{Decin}}, \citenamefont {{Scicluna}}, \citenamefont {{Kravchenko}}, \citenamefont {{Dupree}}, \citenamefont {{Ridgway}}, \citenamefont {{Wittkowski}}, \citenamefont {{Anugu}}, \citenamefont {{Norris}}, \citenamefont {{Rau}}, \citenamefont {{Perrin}}, \citenamefont {{Chiavassa}}, \citenamefont {{Kraus}}, \citenamefont {{Monnier}}, \citenamefont {{Millour}}, \citenamefont {{Le Bouquin}}, \citenamefont {{Haubois}}, \citenamefont {{Lopez}}, \citenamefont {{Stee}},\ and\ \citenamefont {{Danchi}}}]{montarges:21}%
  \BibitemOpen
  \bibfield  {author} {\bibinfo {author} {\bibfnamefont {M.}~\bibnamefont {{Montarg{\`e}s}}}, \bibinfo {author} {\bibfnamefont {E.}~\bibnamefont {{Cannon}}}, \bibinfo {author} {\bibfnamefont {E.}~\bibnamefont {{Lagadec}}}, \bibinfo {author} {\bibfnamefont {A.}~\bibnamefont {{de Koter}}}, \bibinfo {author} {\bibfnamefont {P.}~\bibnamefont {{Kervella}}}, \bibinfo {author} {\bibfnamefont {J.}~\bibnamefont {{Sanchez-Bermudez}}}, \bibinfo {author} {\bibfnamefont {C.}~\bibnamefont {{Paladini}}}, \bibinfo {author} {\bibfnamefont {F.}~\bibnamefont {{Cantalloube}}}, \bibinfo {author} {\bibfnamefont {L.}~\bibnamefont {{Decin}}}, \bibinfo {author} {\bibfnamefont {P.}~\bibnamefont {{Scicluna}}}, \bibinfo {author} {\bibfnamefont {K.}~\bibnamefont {{Kravchenko}}}, \bibinfo {author} {\bibfnamefont {A.~K.}\ \bibnamefont {{Dupree}}}, \bibinfo {author} {\bibfnamefont {S.}~\bibnamefont {{Ridgway}}}, \bibinfo {author} {\bibfnamefont {M.}~\bibnamefont {{Wittkowski}}}, \bibinfo {author} {\bibfnamefont {N.}~\bibnamefont {{Anugu}}},
  \bibinfo {author} {\bibfnamefont {R.}~\bibnamefont {{Norris}}}, \bibinfo {author} {\bibfnamefont {G.}~\bibnamefont {{Rau}}}, \bibinfo {author} {\bibfnamefont {G.}~\bibnamefont {{Perrin}}}, \bibinfo {author} {\bibfnamefont {A.}~\bibnamefont {{Chiavassa}}}, \bibinfo {author} {\bibfnamefont {S.}~\bibnamefont {{Kraus}}}, \bibinfo {author} {\bibfnamefont {J.~D.}\ \bibnamefont {{Monnier}}}, \bibinfo {author} {\bibfnamefont {F.}~\bibnamefont {{Millour}}}, \bibinfo {author} {\bibfnamefont {J.~B.}\ \bibnamefont {{Le Bouquin}}}, \bibinfo {author} {\bibfnamefont {X.}~\bibnamefont {{Haubois}}}, \bibinfo {author} {\bibfnamefont {B.}~\bibnamefont {{Lopez}}}, \bibinfo {author} {\bibfnamefont {P.}~\bibnamefont {{Stee}}}, \ and\ \bibinfo {author} {\bibfnamefont {W.}~\bibnamefont {{Danchi}}},\ }\href {\doibase 10.1038/s41586-021-03546-8} {\bibfield  {journal} {\bibinfo  {journal} {\nat}\ }\textbf {\bibinfo {volume} {594}},\ \bibinfo {pages} {365} (\bibinfo {year} {2021})},\ \Eprint {http://arxiv.org/abs/2201.10551}
  {arXiv:2201.10551 [astro-ph.SR]} \BibitemShut {NoStop}%
\bibitem [{\citenamefont {{Taniguchi}}\ \emph {et~al.}(2022)\citenamefont {{Taniguchi}}, \citenamefont {{Yamazaki}},\ and\ \citenamefont {{Uno}}}]{taniguchi:22}%
  \BibitemOpen
  \bibfield  {author} {\bibinfo {author} {\bibfnamefont {D.}~\bibnamefont {{Taniguchi}}}, \bibinfo {author} {\bibfnamefont {K.}~\bibnamefont {{Yamazaki}}}, \ and\ \bibinfo {author} {\bibfnamefont {S.}~\bibnamefont {{Uno}}},\ }\href {\doibase 10.1038/s41550-022-01680-5} {\bibfield  {journal} {\bibinfo  {journal} {Nature Astronomy}\ }\textbf {\bibinfo {volume} {6}},\ \bibinfo {pages} {930} (\bibinfo {year} {2022})},\ \Eprint {http://arxiv.org/abs/2205.14165} {arXiv:2205.14165 [astro-ph.SR]} \BibitemShut {NoStop}%
\bibitem [{\citenamefont {{Dupree}}\ \emph {et~al.}(2022)\citenamefont {{Dupree}}, \citenamefont {{Strassmeier}}, \citenamefont {{Calderwood}}, \citenamefont {{Granzer}}, \citenamefont {{Weber}}, \citenamefont {{Kravchenko}}, \citenamefont {{Matthews}}, \citenamefont {{Montarg{\`e}s}}, \citenamefont {{Tappin}},\ and\ \citenamefont {{Thompson}}}]{dupree:22}%
  \BibitemOpen
  \bibfield  {author} {\bibinfo {author} {\bibfnamefont {A.~K.}\ \bibnamefont {{Dupree}}}, \bibinfo {author} {\bibfnamefont {K.~G.}\ \bibnamefont {{Strassmeier}}}, \bibinfo {author} {\bibfnamefont {T.}~\bibnamefont {{Calderwood}}}, \bibinfo {author} {\bibfnamefont {T.}~\bibnamefont {{Granzer}}}, \bibinfo {author} {\bibfnamefont {M.}~\bibnamefont {{Weber}}}, \bibinfo {author} {\bibfnamefont {K.}~\bibnamefont {{Kravchenko}}}, \bibinfo {author} {\bibfnamefont {L.~D.}\ \bibnamefont {{Matthews}}}, \bibinfo {author} {\bibfnamefont {M.}~\bibnamefont {{Montarg{\`e}s}}}, \bibinfo {author} {\bibfnamefont {J.}~\bibnamefont {{Tappin}}}, \ and\ \bibinfo {author} {\bibfnamefont {W.~T.}\ \bibnamefont {{Thompson}}},\ }\href {\doibase 10.3847/1538-4357/ac7853} {\bibfield  {journal} {\bibinfo  {journal} {\apj}\ }\textbf {\bibinfo {volume} {936}},\ \bibinfo {eid} {18} (\bibinfo {year} {2022})},\ \Eprint {http://arxiv.org/abs/2208.01676} {arXiv:2208.01676 [astro-ph.SR]} \BibitemShut {NoStop}%
\bibitem [{\citenamefont {{Jadlovsk{\'y}}}\ \emph {et~al.}(2023)\citenamefont {{Jadlovsk{\'y}}}, \citenamefont {{Krti{\v{c}}ka}}, \citenamefont {{Paunzen}},\ and\ \citenamefont {{{\v{S}}tefl}}}]{jadlovsky:23}%
  \BibitemOpen
  \bibfield  {author} {\bibinfo {author} {\bibfnamefont {D.}~\bibnamefont {{Jadlovsk{\'y}}}}, \bibinfo {author} {\bibfnamefont {J.}~\bibnamefont {{Krti{\v{c}}ka}}}, \bibinfo {author} {\bibfnamefont {E.}~\bibnamefont {{Paunzen}}}, \ and\ \bibinfo {author} {\bibfnamefont {V.}~\bibnamefont {{{\v{S}}tefl}}},\ }\href {\doibase 10.1016/j.newast.2022.101962} {\bibfield  {journal} {\bibinfo  {journal} {\na}\ }\textbf {\bibinfo {volume} {99}},\ \bibinfo {eid} {101962} (\bibinfo {year} {2023})},\ \Eprint {http://arxiv.org/abs/2211.04380} {arXiv:2211.04380 [astro-ph.SR]} \BibitemShut {NoStop}%
\bibitem [{\citenamefont {{Kravchenko}}\ \emph {et~al.}(2021)\citenamefont {{Kravchenko}}, \citenamefont {{Jorissen}}, \citenamefont {{Van Eck}}, \citenamefont {{Merle}}, \citenamefont {{Chiavassa}}, \citenamefont {{Paladini}}, \citenamefont {{Freytag}}, \citenamefont {{Plez}}, \citenamefont {{Montarg{\`e}s}},\ and\ \citenamefont {{Van Winckel}}}]{kravchenko:21}%
  \BibitemOpen
  \bibfield  {author} {\bibinfo {author} {\bibfnamefont {K.}~\bibnamefont {{Kravchenko}}}, \bibinfo {author} {\bibfnamefont {A.}~\bibnamefont {{Jorissen}}}, \bibinfo {author} {\bibfnamefont {S.}~\bibnamefont {{Van Eck}}}, \bibinfo {author} {\bibfnamefont {T.}~\bibnamefont {{Merle}}}, \bibinfo {author} {\bibfnamefont {A.}~\bibnamefont {{Chiavassa}}}, \bibinfo {author} {\bibfnamefont {C.}~\bibnamefont {{Paladini}}}, \bibinfo {author} {\bibfnamefont {B.}~\bibnamefont {{Freytag}}}, \bibinfo {author} {\bibfnamefont {B.}~\bibnamefont {{Plez}}}, \bibinfo {author} {\bibfnamefont {M.}~\bibnamefont {{Montarg{\`e}s}}}, \ and\ \bibinfo {author} {\bibfnamefont {H.}~\bibnamefont {{Van Winckel}}},\ }\href {\doibase 10.1051/0004-6361/202039801} {\bibfield  {journal} {\bibinfo  {journal} {\aap}\ }\textbf {\bibinfo {volume} {650}},\ \bibinfo {eid} {L17} (\bibinfo {year} {2021})},\ \Eprint {http://arxiv.org/abs/2104.08105} {arXiv:2104.08105 [astro-ph.SR]} \BibitemShut {NoStop}%
\bibitem [{\citenamefont {{Jadlovsk{\'y}}}\ \emph {et~al.}(2024)\citenamefont {{Jadlovsk{\'y}}}, \citenamefont {{Granzer}}, \citenamefont {{Weber}}, \citenamefont {{Kravchenko}}, \citenamefont {{Krti{\v{c}}ka}}, \citenamefont {{Dupree}}, \citenamefont {{Chiavassa}}, \citenamefont {{Strassmeier}},\ and\ \citenamefont {{Poppenh{\"a}ger}}}]{jadlovsky:24}%
  \BibitemOpen
  \bibfield  {author} {\bibinfo {author} {\bibfnamefont {D.}~\bibnamefont {{Jadlovsk{\'y}}}}, \bibinfo {author} {\bibfnamefont {T.}~\bibnamefont {{Granzer}}}, \bibinfo {author} {\bibfnamefont {M.}~\bibnamefont {{Weber}}}, \bibinfo {author} {\bibfnamefont {K.}~\bibnamefont {{Kravchenko}}}, \bibinfo {author} {\bibfnamefont {J.}~\bibnamefont {{Krti{\v{c}}ka}}}, \bibinfo {author} {\bibfnamefont {A.~K.}\ \bibnamefont {{Dupree}}}, \bibinfo {author} {\bibfnamefont {A.}~\bibnamefont {{Chiavassa}}}, \bibinfo {author} {\bibfnamefont {K.~G.}\ \bibnamefont {{Strassmeier}}}, \ and\ \bibinfo {author} {\bibfnamefont {K.}~\bibnamefont {{Poppenh{\"a}ger}}},\ }\href {\doibase 10.1051/0004-6361/202348846} {\bibfield  {journal} {\bibinfo  {journal} {\aap}\ }\textbf {\bibinfo {volume} {685}},\ \bibinfo {eid} {A124} (\bibinfo {year} {2024})},\ \Eprint {http://arxiv.org/abs/2312.02816} {arXiv:2312.02816 [astro-ph.SR]} \BibitemShut {NoStop}%
\bibitem [{\citenamefont {{Soker}}(2021)}]{soker:21}%
  \BibitemOpen
  \bibfield  {author} {\bibinfo {author} {\bibfnamefont {N.}~\bibnamefont {{Soker}}},\ }\href {\doibase 10.3847/1538-4357/abca8f} {\bibfield  {journal} {\bibinfo  {journal} {\apj}\ }\textbf {\bibinfo {volume} {906}},\ \bibinfo {eid} {1} (\bibinfo {year} {2021})},\ \Eprint {http://arxiv.org/abs/2010.01531} {arXiv:2010.01531 [astro-ph.HE]} \BibitemShut {NoStop}%
\bibitem [{\citenamefont {{Kee}}\ \emph {et~al.}(2021)\citenamefont {{Kee}}, \citenamefont {{Sundqvist}}, \citenamefont {{Decin}}, \citenamefont {{de Koter}},\ and\ \citenamefont {{Sana}}}]{kee:21}%
  \BibitemOpen
  \bibfield  {author} {\bibinfo {author} {\bibfnamefont {N.~D.}\ \bibnamefont {{Kee}}}, \bibinfo {author} {\bibfnamefont {J.~O.}\ \bibnamefont {{Sundqvist}}}, \bibinfo {author} {\bibfnamefont {L.}~\bibnamefont {{Decin}}}, \bibinfo {author} {\bibfnamefont {A.}~\bibnamefont {{de Koter}}}, \ and\ \bibinfo {author} {\bibfnamefont {H.}~\bibnamefont {{Sana}}},\ }\href {\doibase 10.1051/0004-6361/202039224} {\bibfield  {journal} {\bibinfo  {journal} {\aap}\ }\textbf {\bibinfo {volume} {646}},\ \bibinfo {eid} {A180} (\bibinfo {year} {2021})},\ \Eprint {http://arxiv.org/abs/2101.03070} {arXiv:2101.03070 [astro-ph.SR]} \BibitemShut {NoStop}%
\bibitem [{\citenamefont {{Hill}}\ and\ \citenamefont {{Willson}}(1979)}]{hill:1979}%
  \BibitemOpen
  \bibfield  {author} {\bibinfo {author} {\bibfnamefont {S.~J.}\ \bibnamefont {{Hill}}}\ and\ \bibinfo {author} {\bibfnamefont {L.~A.}\ \bibnamefont {{Willson}}},\ }\href {\doibase 10.1086/157038} {\bibfield  {journal} {\bibinfo  {journal} {\apj}\ }\textbf {\bibinfo {volume} {229}},\ \bibinfo {pages} {1029} (\bibinfo {year} {1979})}\BibitemShut {NoStop}%
\bibitem [{\citenamefont {{Chiavassa}}\ \emph {et~al.}(2011{\natexlab{a}})\citenamefont {{Chiavassa}}, \citenamefont {{Freytag}}, \citenamefont {{Masseron}},\ and\ \citenamefont {{Plez}}}]{chinavassa:11}%
  \BibitemOpen
  \bibfield  {author} {\bibinfo {author} {\bibfnamefont {A.}~\bibnamefont {{Chiavassa}}}, \bibinfo {author} {\bibfnamefont {B.}~\bibnamefont {{Freytag}}}, \bibinfo {author} {\bibfnamefont {T.}~\bibnamefont {{Masseron}}}, \ and\ \bibinfo {author} {\bibfnamefont {B.}~\bibnamefont {{Plez}}},\ }\href {\doibase 10.1051/0004-6361/201117463} {\bibfield  {journal} {\bibinfo  {journal} {\aap}\ }\textbf {\bibinfo {volume} {535}},\ \bibinfo {eid} {A22} (\bibinfo {year} {2011}{\natexlab{a}})},\ \Eprint {http://arxiv.org/abs/1109.3619} {arXiv:1109.3619 [astro-ph.SR]} \BibitemShut {NoStop}%
\bibitem [{\citenamefont {{Chiavassa}}\ \emph {et~al.}(2011{\natexlab{b}})\citenamefont {{Chiavassa}}, \citenamefont {{Pasquato}}, \citenamefont {{Jorissen}}, \citenamefont {{Sacuto}}, \citenamefont {{Babusiaux}}, \citenamefont {{Freytag}}, \citenamefont {{Ludwig}}, \citenamefont {{Cruzal{\`e}bes}}, \citenamefont {{Rabbia}}, \citenamefont {{Spang}},\ and\ \citenamefont {{Chesneau}}}]{chinavassa:11b}%
  \BibitemOpen
  \bibfield  {author} {\bibinfo {author} {\bibfnamefont {A.}~\bibnamefont {{Chiavassa}}}, \bibinfo {author} {\bibfnamefont {E.}~\bibnamefont {{Pasquato}}}, \bibinfo {author} {\bibfnamefont {A.}~\bibnamefont {{Jorissen}}}, \bibinfo {author} {\bibfnamefont {S.}~\bibnamefont {{Sacuto}}}, \bibinfo {author} {\bibfnamefont {C.}~\bibnamefont {{Babusiaux}}}, \bibinfo {author} {\bibfnamefont {B.}~\bibnamefont {{Freytag}}}, \bibinfo {author} {\bibfnamefont {H.~G.}\ \bibnamefont {{Ludwig}}}, \bibinfo {author} {\bibfnamefont {P.}~\bibnamefont {{Cruzal{\`e}bes}}}, \bibinfo {author} {\bibfnamefont {Y.}~\bibnamefont {{Rabbia}}}, \bibinfo {author} {\bibfnamefont {A.}~\bibnamefont {{Spang}}}, \ and\ \bibinfo {author} {\bibfnamefont {O.}~\bibnamefont {{Chesneau}}},\ }\href {\doibase 10.1051/0004-6361/201015768} {\bibfield  {journal} {\bibinfo  {journal} {\aap}\ }\textbf {\bibinfo {volume} {528}},\ \bibinfo {eid} {A120} (\bibinfo {year} {2011}{\natexlab{b}})},\ \Eprint {http://arxiv.org/abs/1012.5234} {arXiv:1012.5234
  [astro-ph.SR]} \BibitemShut {NoStop}%
\bibitem [{\citenamefont {{Ma}}\ \emph {et~al.}(2023)\citenamefont {{Ma}}, \citenamefont {{Chiavassa}}, \citenamefont {{de Mink}}, \citenamefont {{Valli}}, \citenamefont {{Justham}},\ and\ \citenamefont {{Freytag}}}]{ma:23}%
  \BibitemOpen
  \bibfield  {author} {\bibinfo {author} {\bibfnamefont {J.-Z.}\ \bibnamefont {{Ma}}}, \bibinfo {author} {\bibfnamefont {A.}~\bibnamefont {{Chiavassa}}}, \bibinfo {author} {\bibfnamefont {S.~E.}\ \bibnamefont {{de Mink}}}, \bibinfo {author} {\bibfnamefont {R.}~\bibnamefont {{Valli}}}, \bibinfo {author} {\bibfnamefont {S.}~\bibnamefont {{Justham}}}, \ and\ \bibinfo {author} {\bibfnamefont {B.}~\bibnamefont {{Freytag}}},\ }\href {\doibase 10.48550/arXiv.2311.16885} {\bibfield  {journal} {\bibinfo  {journal} {arXiv e-prints}\ ,\ \bibinfo {eid} {arXiv:2311.16885}} (\bibinfo {year} {2023})},\ \Eprint {http://arxiv.org/abs/2311.16885} {arXiv:2311.16885 [astro-ph.SR]} \BibitemShut {NoStop}%
\bibitem [{\citenamefont {{Chiavassa}}\ \emph {et~al.}(2024)\citenamefont {{Chiavassa}}, \citenamefont {{Kravchenko}},\ and\ \citenamefont {{Goldberg}}}]{chiavassa:24}%
  \BibitemOpen
  \bibfield  {author} {\bibinfo {author} {\bibfnamefont {A.}~\bibnamefont {{Chiavassa}}}, \bibinfo {author} {\bibfnamefont {K.}~\bibnamefont {{Kravchenko}}}, \ and\ \bibinfo {author} {\bibfnamefont {J.~A.}\ \bibnamefont {{Goldberg}}},\ }\href@noop {} {\bibfield  {journal} {\bibinfo  {journal} {arXiv e-prints}\ ,\ \bibinfo {eid} {arXiv:2402.00187}} (\bibinfo {year} {2024})},\ \Eprint {http://arxiv.org/abs/2402.00187} {arXiv:2402.00187 [astro-ph.SR]} \BibitemShut {NoStop}%
\bibitem [{\citenamefont {{Wedemeyer}}\ \emph {et~al.}(2017)\citenamefont {{Wedemeyer}}, \citenamefont {{Ku{\v{c}}inskas}}, \citenamefont {{Klevas}},\ and\ \citenamefont {{Ludwig}}}]{wedemeyer:17}%
  \BibitemOpen
  \bibfield  {author} {\bibinfo {author} {\bibfnamefont {S.}~\bibnamefont {{Wedemeyer}}}, \bibinfo {author} {\bibfnamefont {A.}~\bibnamefont {{Ku{\v{c}}inskas}}}, \bibinfo {author} {\bibfnamefont {J.}~\bibnamefont {{Klevas}}}, \ and\ \bibinfo {author} {\bibfnamefont {H.-G.}\ \bibnamefont {{Ludwig}}},\ }\href {\doibase 10.1051/0004-6361/201730405} {\bibfield  {journal} {\bibinfo  {journal} {\aap}\ }\textbf {\bibinfo {volume} {606}},\ \bibinfo {eid} {A26} (\bibinfo {year} {2017})},\ \Eprint {http://arxiv.org/abs/1705.09641} {arXiv:1705.09641 [astro-ph.SR]} \BibitemShut {NoStop}%
\bibitem [{\citenamefont {{van Loon}}\ \emph {et~al.}(2005)\citenamefont {{van Loon}}, \citenamefont {{Cioni}}, \citenamefont {{Zijlstra}},\ and\ \citenamefont {{Loup}}}]{vanloon:05}%
  \BibitemOpen
  \bibfield  {author} {\bibinfo {author} {\bibfnamefont {J.~T.}\ \bibnamefont {{van Loon}}}, \bibinfo {author} {\bibfnamefont {M.~R.~L.}\ \bibnamefont {{Cioni}}}, \bibinfo {author} {\bibfnamefont {A.~A.}\ \bibnamefont {{Zijlstra}}}, \ and\ \bibinfo {author} {\bibfnamefont {C.}~\bibnamefont {{Loup}}},\ }\href {\doibase 10.1051/0004-6361:20042555} {\bibfield  {journal} {\bibinfo  {journal} {\aap}\ }\textbf {\bibinfo {volume} {438}},\ \bibinfo {pages} {273} (\bibinfo {year} {2005})},\ \Eprint {http://arxiv.org/abs/astro-ph/0504379} {arXiv:astro-ph/0504379 [astro-ph]} \BibitemShut {NoStop}%
\bibitem [{\citenamefont {{Morozova}}\ \emph {et~al.}(2020)\citenamefont {{Morozova}}, \citenamefont {{Piro}}, \citenamefont {{Fuller}},\ and\ \citenamefont {{Van Dyk}}}]{morozova:20}%
  \BibitemOpen
  \bibfield  {author} {\bibinfo {author} {\bibfnamefont {V.}~\bibnamefont {{Morozova}}}, \bibinfo {author} {\bibfnamefont {A.~L.}\ \bibnamefont {{Piro}}}, \bibinfo {author} {\bibfnamefont {J.}~\bibnamefont {{Fuller}}}, \ and\ \bibinfo {author} {\bibfnamefont {S.~D.}\ \bibnamefont {{Van Dyk}}},\ }\href {\doibase 10.3847/2041-8213/ab77c8} {\bibfield  {journal} {\bibinfo  {journal} {\apjl}\ }\textbf {\bibinfo {volume} {891}},\ \bibinfo {eid} {L32} (\bibinfo {year} {2020})},\ \Eprint {http://arxiv.org/abs/1912.10050} {arXiv:1912.10050 [astro-ph.HE]} \BibitemShut {NoStop}%
\bibitem [{\citenamefont {{Jacobson-Gal{\'a}n}}\ \emph {et~al.}(2022)\citenamefont {{Jacobson-Gal{\'a}n}}, \citenamefont {{Dessart}}, \citenamefont {{Jones}}, \citenamefont {{Margutti}}, \citenamefont {{Coppejans}}, \citenamefont {{Dimitriadis}}, \citenamefont {{Foley}}, \citenamefont {{Kilpatrick}}, \citenamefont {{Matthews}}, \citenamefont {{Rest}}, \citenamefont {{Terreran}}, \citenamefont {{Aleo}}, \citenamefont {{Auchettl}}, \citenamefont {{Blanchard}}, \citenamefont {{Coulter}}, \citenamefont {{Davis}}, \citenamefont {{de Boer}}, \citenamefont {{DeMarchi}}, \citenamefont {{Drout}}, \citenamefont {{Earl}}, \citenamefont {{Gagliano}}, \citenamefont {{Gall}}, \citenamefont {{Hjorth}}, \citenamefont {{Huber}}, \citenamefont {{Ibik}}, \citenamefont {{Milisavljevic}}, \citenamefont {{Pan}}, \citenamefont {{Rest}}, \citenamefont {{Ridden-Harper}}, \citenamefont {{Rojas-Bravo}}, \citenamefont {{Siebert}}, \citenamefont {{Smith}}, \citenamefont {{Taggart}}, \citenamefont {{Tinyanont}}, \citenamefont {{Wang}},\
  and\ \citenamefont {{Zenati}}}]{jacobson-galan:22}%
  \BibitemOpen
  \bibfield  {author} {\bibinfo {author} {\bibfnamefont {W.~V.}\ \bibnamefont {{Jacobson-Gal{\'a}n}}}, \bibinfo {author} {\bibfnamefont {L.}~\bibnamefont {{Dessart}}}, \bibinfo {author} {\bibfnamefont {D.~O.}\ \bibnamefont {{Jones}}}, \bibinfo {author} {\bibfnamefont {R.}~\bibnamefont {{Margutti}}}, \bibinfo {author} {\bibfnamefont {D.~L.}\ \bibnamefont {{Coppejans}}}, \bibinfo {author} {\bibfnamefont {G.}~\bibnamefont {{Dimitriadis}}}, \bibinfo {author} {\bibfnamefont {R.~J.}\ \bibnamefont {{Foley}}}, \bibinfo {author} {\bibfnamefont {C.~D.}\ \bibnamefont {{Kilpatrick}}}, \bibinfo {author} {\bibfnamefont {D.~J.}\ \bibnamefont {{Matthews}}}, \bibinfo {author} {\bibfnamefont {S.}~\bibnamefont {{Rest}}}, \bibinfo {author} {\bibfnamefont {G.}~\bibnamefont {{Terreran}}}, \bibinfo {author} {\bibfnamefont {P.~D.}\ \bibnamefont {{Aleo}}}, \bibinfo {author} {\bibfnamefont {K.}~\bibnamefont {{Auchettl}}}, \bibinfo {author} {\bibfnamefont {P.~K.}\ \bibnamefont {{Blanchard}}}, \bibinfo {author} {\bibfnamefont {D.~A.}\
  \bibnamefont {{Coulter}}}, \bibinfo {author} {\bibfnamefont {K.~W.}\ \bibnamefont {{Davis}}}, \bibinfo {author} {\bibfnamefont {T.~J.~L.}\ \bibnamefont {{de Boer}}}, \bibinfo {author} {\bibfnamefont {L.}~\bibnamefont {{DeMarchi}}}, \bibinfo {author} {\bibfnamefont {M.~R.}\ \bibnamefont {{Drout}}}, \bibinfo {author} {\bibfnamefont {N.}~\bibnamefont {{Earl}}}, \bibinfo {author} {\bibfnamefont {A.}~\bibnamefont {{Gagliano}}}, \bibinfo {author} {\bibfnamefont {C.}~\bibnamefont {{Gall}}}, \bibinfo {author} {\bibfnamefont {J.}~\bibnamefont {{Hjorth}}}, \bibinfo {author} {\bibfnamefont {M.~E.}\ \bibnamefont {{Huber}}}, \bibinfo {author} {\bibfnamefont {A.~L.}\ \bibnamefont {{Ibik}}}, \bibinfo {author} {\bibfnamefont {D.}~\bibnamefont {{Milisavljevic}}}, \bibinfo {author} {\bibfnamefont {Y.~C.}\ \bibnamefont {{Pan}}}, \bibinfo {author} {\bibfnamefont {A.}~\bibnamefont {{Rest}}}, \bibinfo {author} {\bibfnamefont {R.}~\bibnamefont {{Ridden-Harper}}}, \bibinfo {author} {\bibfnamefont {C.}~\bibnamefont
  {{Rojas-Bravo}}}, \bibinfo {author} {\bibfnamefont {M.~R.}\ \bibnamefont {{Siebert}}}, \bibinfo {author} {\bibfnamefont {K.~W.}\ \bibnamefont {{Smith}}}, \bibinfo {author} {\bibfnamefont {K.}~\bibnamefont {{Taggart}}}, \bibinfo {author} {\bibfnamefont {S.}~\bibnamefont {{Tinyanont}}}, \bibinfo {author} {\bibfnamefont {Q.}~\bibnamefont {{Wang}}}, \ and\ \bibinfo {author} {\bibfnamefont {Y.}~\bibnamefont {{Zenati}}},\ }\href {\doibase 10.3847/1538-4357/ac3f3a} {\bibfield  {journal} {\bibinfo  {journal} {\apj}\ }\textbf {\bibinfo {volume} {924}},\ \bibinfo {eid} {15} (\bibinfo {year} {2022})},\ \Eprint {http://arxiv.org/abs/2109.12136} {arXiv:2109.12136 [astro-ph.HE]} \BibitemShut {NoStop}%
\bibitem [{\citenamefont {{Ofek}}\ \emph {et~al.}(2014)\citenamefont {{Ofek}}, \citenamefont {{Sullivan}}, \citenamefont {{Shaviv}}, \citenamefont {{Steinbok}}, \citenamefont {{Arcavi}}, \citenamefont {{Gal-Yam}}, \citenamefont {{Tal}}, \citenamefont {{Kulkarni}}, \citenamefont {{Nugent}}, \citenamefont {{Ben-Ami}}, \citenamefont {{Kasliwal}}, \citenamefont {{Cenko}}, \citenamefont {{Laher}}, \citenamefont {{Surace}}, \citenamefont {{Bloom}}, \citenamefont {{Filippenko}}, \citenamefont {{Silverman}},\ and\ \citenamefont {{Yaron}}}]{ofek:14}%
  \BibitemOpen
  \bibfield  {author} {\bibinfo {author} {\bibfnamefont {E.~O.}\ \bibnamefont {{Ofek}}}, \bibinfo {author} {\bibfnamefont {M.}~\bibnamefont {{Sullivan}}}, \bibinfo {author} {\bibfnamefont {N.~J.}\ \bibnamefont {{Shaviv}}}, \bibinfo {author} {\bibfnamefont {A.}~\bibnamefont {{Steinbok}}}, \bibinfo {author} {\bibfnamefont {I.}~\bibnamefont {{Arcavi}}}, \bibinfo {author} {\bibfnamefont {A.}~\bibnamefont {{Gal-Yam}}}, \bibinfo {author} {\bibfnamefont {D.}~\bibnamefont {{Tal}}}, \bibinfo {author} {\bibfnamefont {S.~R.}\ \bibnamefont {{Kulkarni}}}, \bibinfo {author} {\bibfnamefont {P.~E.}\ \bibnamefont {{Nugent}}}, \bibinfo {author} {\bibfnamefont {S.}~\bibnamefont {{Ben-Ami}}}, \bibinfo {author} {\bibfnamefont {M.~M.}\ \bibnamefont {{Kasliwal}}}, \bibinfo {author} {\bibfnamefont {S.~B.}\ \bibnamefont {{Cenko}}}, \bibinfo {author} {\bibfnamefont {R.}~\bibnamefont {{Laher}}}, \bibinfo {author} {\bibfnamefont {J.}~\bibnamefont {{Surace}}}, \bibinfo {author} {\bibfnamefont {J.~S.}\ \bibnamefont {{Bloom}}}, \bibinfo
  {author} {\bibfnamefont {A.~V.}\ \bibnamefont {{Filippenko}}}, \bibinfo {author} {\bibfnamefont {J.~M.}\ \bibnamefont {{Silverman}}}, \ and\ \bibinfo {author} {\bibfnamefont {O.}~\bibnamefont {{Yaron}}},\ }\href {\doibase 10.1088/0004-637X/789/2/104} {\bibfield  {journal} {\bibinfo  {journal} {\apj}\ }\textbf {\bibinfo {volume} {789}},\ \bibinfo {eid} {104} (\bibinfo {year} {2014})},\ \Eprint {http://arxiv.org/abs/1401.5468} {arXiv:1401.5468 [astro-ph.HE]} \BibitemShut {NoStop}%
\bibitem [{\citenamefont {{Strotjohann}}\ \emph {et~al.}(2021)\citenamefont {{Strotjohann}}, \citenamefont {{Ofek}}, \citenamefont {{Gal-Yam}}, \citenamefont {{Bruch}}, \citenamefont {{Schulze}}, \citenamefont {{Shaviv}}, \citenamefont {{Sollerman}}, \citenamefont {{Filippenko}}, \citenamefont {{Yaron}}, \citenamefont {{Fremling}}, \citenamefont {{Nordin}}, \citenamefont {{Kool}}, \citenamefont {{Perley}}, \citenamefont {{Ho}}, \citenamefont {{Yang}}, \citenamefont {{Yao}}, \citenamefont {{Soumagnac}}, \citenamefont {{Graham}}, \citenamefont {{Barbarino}}, \citenamefont {{Tartaglia}}, \citenamefont {{De}}, \citenamefont {{Goldstein}}, \citenamefont {{Cook}}, \citenamefont {{Brink}}, \citenamefont {{Taggart}}, \citenamefont {{Yan}}, \citenamefont {{Lunnan}}, \citenamefont {{Kasliwal}}, \citenamefont {{Kulkarni}}, \citenamefont {{Nugent}}, \citenamefont {{Masci}}, \citenamefont {{Rosnet}}, \citenamefont {{Adams}}, \citenamefont {{Andreoni}}, \citenamefont {{Bagdasaryan}}, \citenamefont {{Bellm}}, \citenamefont
  {{Burdge}}, \citenamefont {{Duev}}, \citenamefont {{Dugas}}, \citenamefont {{Frederick}}, \citenamefont {{Goldwasser}}, \citenamefont {{Hankins}}, \citenamefont {{Irani}}, \citenamefont {{Karambelkar}}, \citenamefont {{Kupfer}}, \citenamefont {{Liang}}, \citenamefont {{Neill}}, \citenamefont {{Porter}}, \citenamefont {{Riddle}}, \citenamefont {{Sharma}}, \citenamefont {{Short}}, \citenamefont {{Taddia}}, \citenamefont {{Tzanidakis}}, \citenamefont {{van Roestel}}, \citenamefont {{Walters}},\ and\ \citenamefont {{Zhuang}}}]{strotjohann:21}%
  \BibitemOpen
  \bibfield  {author} {\bibinfo {author} {\bibfnamefont {N.~L.}\ \bibnamefont {{Strotjohann}}}, \bibinfo {author} {\bibfnamefont {E.~O.}\ \bibnamefont {{Ofek}}}, \bibinfo {author} {\bibfnamefont {A.}~\bibnamefont {{Gal-Yam}}}, \bibinfo {author} {\bibfnamefont {R.}~\bibnamefont {{Bruch}}}, \bibinfo {author} {\bibfnamefont {S.}~\bibnamefont {{Schulze}}}, \bibinfo {author} {\bibfnamefont {N.}~\bibnamefont {{Shaviv}}}, \bibinfo {author} {\bibfnamefont {J.}~\bibnamefont {{Sollerman}}}, \bibinfo {author} {\bibfnamefont {A.~V.}\ \bibnamefont {{Filippenko}}}, \bibinfo {author} {\bibfnamefont {O.}~\bibnamefont {{Yaron}}}, \bibinfo {author} {\bibfnamefont {C.}~\bibnamefont {{Fremling}}}, \bibinfo {author} {\bibfnamefont {J.}~\bibnamefont {{Nordin}}}, \bibinfo {author} {\bibfnamefont {E.~C.}\ \bibnamefont {{Kool}}}, \bibinfo {author} {\bibfnamefont {D.~A.}\ \bibnamefont {{Perley}}}, \bibinfo {author} {\bibfnamefont {A.~Y.~Q.}\ \bibnamefont {{Ho}}}, \bibinfo {author} {\bibfnamefont {Y.}~\bibnamefont {{Yang}}}, \bibinfo
  {author} {\bibfnamefont {Y.}~\bibnamefont {{Yao}}}, \bibinfo {author} {\bibfnamefont {M.~T.}\ \bibnamefont {{Soumagnac}}}, \bibinfo {author} {\bibfnamefont {M.~L.}\ \bibnamefont {{Graham}}}, \bibinfo {author} {\bibfnamefont {C.}~\bibnamefont {{Barbarino}}}, \bibinfo {author} {\bibfnamefont {L.}~\bibnamefont {{Tartaglia}}}, \bibinfo {author} {\bibfnamefont {K.}~\bibnamefont {{De}}}, \bibinfo {author} {\bibfnamefont {D.~A.}\ \bibnamefont {{Goldstein}}}, \bibinfo {author} {\bibfnamefont {D.~O.}\ \bibnamefont {{Cook}}}, \bibinfo {author} {\bibfnamefont {T.~G.}\ \bibnamefont {{Brink}}}, \bibinfo {author} {\bibfnamefont {K.}~\bibnamefont {{Taggart}}}, \bibinfo {author} {\bibfnamefont {L.}~\bibnamefont {{Yan}}}, \bibinfo {author} {\bibfnamefont {R.}~\bibnamefont {{Lunnan}}}, \bibinfo {author} {\bibfnamefont {M.}~\bibnamefont {{Kasliwal}}}, \bibinfo {author} {\bibfnamefont {S.~R.}\ \bibnamefont {{Kulkarni}}}, \bibinfo {author} {\bibfnamefont {P.~E.}\ \bibnamefont {{Nugent}}}, \bibinfo {author} {\bibfnamefont
  {F.~J.}\ \bibnamefont {{Masci}}}, \bibinfo {author} {\bibfnamefont {P.}~\bibnamefont {{Rosnet}}}, \bibinfo {author} {\bibfnamefont {S.~M.}\ \bibnamefont {{Adams}}}, \bibinfo {author} {\bibfnamefont {I.}~\bibnamefont {{Andreoni}}}, \bibinfo {author} {\bibfnamefont {A.}~\bibnamefont {{Bagdasaryan}}}, \bibinfo {author} {\bibfnamefont {E.~C.}\ \bibnamefont {{Bellm}}}, \bibinfo {author} {\bibfnamefont {K.}~\bibnamefont {{Burdge}}}, \bibinfo {author} {\bibfnamefont {D.~A.}\ \bibnamefont {{Duev}}}, \bibinfo {author} {\bibfnamefont {A.}~\bibnamefont {{Dugas}}}, \bibinfo {author} {\bibfnamefont {S.}~\bibnamefont {{Frederick}}}, \bibinfo {author} {\bibfnamefont {S.}~\bibnamefont {{Goldwasser}}}, \bibinfo {author} {\bibfnamefont {M.}~\bibnamefont {{Hankins}}}, \bibinfo {author} {\bibfnamefont {I.}~\bibnamefont {{Irani}}}, \bibinfo {author} {\bibfnamefont {V.}~\bibnamefont {{Karambelkar}}}, \bibinfo {author} {\bibfnamefont {T.}~\bibnamefont {{Kupfer}}}, \bibinfo {author} {\bibfnamefont {J.}~\bibnamefont {{Liang}}},
  \bibinfo {author} {\bibfnamefont {J.~D.}\ \bibnamefont {{Neill}}}, \bibinfo {author} {\bibfnamefont {M.}~\bibnamefont {{Porter}}}, \bibinfo {author} {\bibfnamefont {R.~L.}\ \bibnamefont {{Riddle}}}, \bibinfo {author} {\bibfnamefont {Y.}~\bibnamefont {{Sharma}}}, \bibinfo {author} {\bibfnamefont {P.}~\bibnamefont {{Short}}}, \bibinfo {author} {\bibfnamefont {F.}~\bibnamefont {{Taddia}}}, \bibinfo {author} {\bibfnamefont {A.}~\bibnamefont {{Tzanidakis}}}, \bibinfo {author} {\bibfnamefont {J.}~\bibnamefont {{van Roestel}}}, \bibinfo {author} {\bibfnamefont {R.}~\bibnamefont {{Walters}}}, \ and\ \bibinfo {author} {\bibfnamefont {Z.}~\bibnamefont {{Zhuang}}},\ }\href {\doibase 10.3847/1538-4357/abd032} {\bibfield  {journal} {\bibinfo  {journal} {\apj}\ }\textbf {\bibinfo {volume} {907}},\ \bibinfo {eid} {99} (\bibinfo {year} {2021})},\ \Eprint {http://arxiv.org/abs/2010.11196} {arXiv:2010.11196 [astro-ph.HE]} \BibitemShut {NoStop}%
\bibitem [{\citenamefont {{Davies}}\ \emph {et~al.}(2022)\citenamefont {{Davies}}, \citenamefont {{Plez}},\ and\ \citenamefont {{Petrault}}}]{davies:22}%
  \BibitemOpen
  \bibfield  {author} {\bibinfo {author} {\bibfnamefont {B.}~\bibnamefont {{Davies}}}, \bibinfo {author} {\bibfnamefont {B.}~\bibnamefont {{Plez}}}, \ and\ \bibinfo {author} {\bibfnamefont {M.}~\bibnamefont {{Petrault}}},\ }\href {\doibase 10.1093/mnras/stac2427} {\bibfield  {journal} {\bibinfo  {journal} {\mnras}\ }\textbf {\bibinfo {volume} {517}},\ \bibinfo {pages} {1483} (\bibinfo {year} {2022})},\ \Eprint {http://arxiv.org/abs/2208.10883} {arXiv:2208.10883 [astro-ph.SR]} \BibitemShut {NoStop}%
\bibitem [{\citenamefont {{Zhang}}\ \emph {et~al.}(2023)\citenamefont {{Zhang}}, \citenamefont {{Lin}}, \citenamefont {{Wang}}, \citenamefont {{Zhao}}, \citenamefont {{Li}}, \citenamefont {{Liu}}, \citenamefont {{Yan}}, \citenamefont {{Xiang}}, \citenamefont {{Wang}},\ and\ \citenamefont {{Bai}}}]{zhang:23}%
  \BibitemOpen
  \bibfield  {author} {\bibinfo {author} {\bibfnamefont {J.}~\bibnamefont {{Zhang}}}, \bibinfo {author} {\bibfnamefont {H.}~\bibnamefont {{Lin}}}, \bibinfo {author} {\bibfnamefont {X.}~\bibnamefont {{Wang}}}, \bibinfo {author} {\bibfnamefont {Z.}~\bibnamefont {{Zhao}}}, \bibinfo {author} {\bibfnamefont {L.}~\bibnamefont {{Li}}}, \bibinfo {author} {\bibfnamefont {J.}~\bibnamefont {{Liu}}}, \bibinfo {author} {\bibfnamefont {S.}~\bibnamefont {{Yan}}}, \bibinfo {author} {\bibfnamefont {D.}~\bibnamefont {{Xiang}}}, \bibinfo {author} {\bibfnamefont {H.}~\bibnamefont {{Wang}}}, \ and\ \bibinfo {author} {\bibfnamefont {J.}~\bibnamefont {{Bai}}},\ }\href {\doibase 10.1016/j.scib.2023.09.015} {\bibfield  {journal} {\bibinfo  {journal} {Science Bulletin}\ }\textbf {\bibinfo {volume} {68}},\ \bibinfo {pages} {2548} (\bibinfo {year} {2023})},\ \Eprint {http://arxiv.org/abs/2309.01998} {arXiv:2309.01998 [astro-ph.HE]} \BibitemShut {NoStop}%
\bibitem [{\citenamefont {{Bostroem}}\ \emph {et~al.}(2023)\citenamefont {{Bostroem}}, \citenamefont {{Pearson}}, \citenamefont {{Shrestha}}, \citenamefont {{Sand}}, \citenamefont {{Valenti}}, \citenamefont {{Jha}}, \citenamefont {{Andrews}}, \citenamefont {{Smith}}, \citenamefont {{Terreran}}, \citenamefont {{Green}}, \citenamefont {{Dong}}, \citenamefont {{Lundquist}}, \citenamefont {{Haislip}}, \citenamefont {{Hoang}}, \citenamefont {{Hosseinzadeh}}, \citenamefont {{Janzen}}, \citenamefont {{Jencson}}, \citenamefont {{Kouprianov}}, \citenamefont {{Paraskeva}}, \citenamefont {{Meza Retamal}}, \citenamefont {{Reichart}}, \citenamefont {{Arcavi}}, \citenamefont {{Bonanos}}, \citenamefont {{Coughlin}}, \citenamefont {{Dobson}}, \citenamefont {{Farah}}, \citenamefont {{Galbany}}, \citenamefont {{Guti{\'e}rrez}}, \citenamefont {{Hawley}}, \citenamefont {{Hebb}}, \citenamefont {{Hiramatsu}}, \citenamefont {{Howell}}, \citenamefont {{Iijima}}, \citenamefont {{Ilyin}}, \citenamefont {{Jhass}}, \citenamefont
  {{McCully}}, \citenamefont {{Moran}}, \citenamefont {{Morris}}, \citenamefont {{Mura}}, \citenamefont {{M{\"u}ller-Bravo}}, \citenamefont {{Munday}}, \citenamefont {{Newsome}}, \citenamefont {{Pabst}}, \citenamefont {{Ochner}}, \citenamefont {{Gonzalez}}, \citenamefont {{Pastorello}}, \citenamefont {{Pellegrino}}, \citenamefont {{Piscarreta}}, \citenamefont {{Ravi}}, \citenamefont {{Reguitti}}, \citenamefont {{Salo}}, \citenamefont {{Vink{\'o}}}, \citenamefont {{de Vos}}, \citenamefont {{Wheeler}}, \citenamefont {{Williams}},\ and\ \citenamefont {{Wyatt}}}]{bostroem:23}%
  \BibitemOpen
  \bibfield  {author} {\bibinfo {author} {\bibfnamefont {K.~A.}\ \bibnamefont {{Bostroem}}}, \bibinfo {author} {\bibfnamefont {J.}~\bibnamefont {{Pearson}}}, \bibinfo {author} {\bibfnamefont {M.}~\bibnamefont {{Shrestha}}}, \bibinfo {author} {\bibfnamefont {D.~J.}\ \bibnamefont {{Sand}}}, \bibinfo {author} {\bibfnamefont {S.}~\bibnamefont {{Valenti}}}, \bibinfo {author} {\bibfnamefont {S.~W.}\ \bibnamefont {{Jha}}}, \bibinfo {author} {\bibfnamefont {J.~E.}\ \bibnamefont {{Andrews}}}, \bibinfo {author} {\bibfnamefont {N.}~\bibnamefont {{Smith}}}, \bibinfo {author} {\bibfnamefont {G.}~\bibnamefont {{Terreran}}}, \bibinfo {author} {\bibfnamefont {E.}~\bibnamefont {{Green}}}, \bibinfo {author} {\bibfnamefont {Y.}~\bibnamefont {{Dong}}}, \bibinfo {author} {\bibfnamefont {M.}~\bibnamefont {{Lundquist}}}, \bibinfo {author} {\bibfnamefont {J.}~\bibnamefont {{Haislip}}}, \bibinfo {author} {\bibfnamefont {E.~T.}\ \bibnamefont {{Hoang}}}, \bibinfo {author} {\bibfnamefont {G.}~\bibnamefont {{Hosseinzadeh}}}, \bibinfo
  {author} {\bibfnamefont {D.}~\bibnamefont {{Janzen}}}, \bibinfo {author} {\bibfnamefont {J.~E.}\ \bibnamefont {{Jencson}}}, \bibinfo {author} {\bibfnamefont {V.}~\bibnamefont {{Kouprianov}}}, \bibinfo {author} {\bibfnamefont {E.}~\bibnamefont {{Paraskeva}}}, \bibinfo {author} {\bibfnamefont {N.~E.}\ \bibnamefont {{Meza Retamal}}}, \bibinfo {author} {\bibfnamefont {D.~E.}\ \bibnamefont {{Reichart}}}, \bibinfo {author} {\bibfnamefont {I.}~\bibnamefont {{Arcavi}}}, \bibinfo {author} {\bibfnamefont {A.~Z.}\ \bibnamefont {{Bonanos}}}, \bibinfo {author} {\bibfnamefont {M.~W.}\ \bibnamefont {{Coughlin}}}, \bibinfo {author} {\bibfnamefont {R.}~\bibnamefont {{Dobson}}}, \bibinfo {author} {\bibfnamefont {J.}~\bibnamefont {{Farah}}}, \bibinfo {author} {\bibfnamefont {L.}~\bibnamefont {{Galbany}}}, \bibinfo {author} {\bibfnamefont {C.}~\bibnamefont {{Guti{\'e}rrez}}}, \bibinfo {author} {\bibfnamefont {S.}~\bibnamefont {{Hawley}}}, \bibinfo {author} {\bibfnamefont {L.}~\bibnamefont {{Hebb}}}, \bibinfo {author}
  {\bibfnamefont {D.}~\bibnamefont {{Hiramatsu}}}, \bibinfo {author} {\bibfnamefont {D.~A.}\ \bibnamefont {{Howell}}}, \bibinfo {author} {\bibfnamefont {T.}~\bibnamefont {{Iijima}}}, \bibinfo {author} {\bibfnamefont {I.}~\bibnamefont {{Ilyin}}}, \bibinfo {author} {\bibfnamefont {K.}~\bibnamefont {{Jhass}}}, \bibinfo {author} {\bibfnamefont {C.}~\bibnamefont {{McCully}}}, \bibinfo {author} {\bibfnamefont {S.}~\bibnamefont {{Moran}}}, \bibinfo {author} {\bibfnamefont {B.~M.}\ \bibnamefont {{Morris}}}, \bibinfo {author} {\bibfnamefont {A.~C.}\ \bibnamefont {{Mura}}}, \bibinfo {author} {\bibfnamefont {T.~E.}\ \bibnamefont {{M{\"u}ller-Bravo}}}, \bibinfo {author} {\bibfnamefont {J.}~\bibnamefont {{Munday}}}, \bibinfo {author} {\bibfnamefont {M.}~\bibnamefont {{Newsome}}}, \bibinfo {author} {\bibfnamefont {M.~T.}\ \bibnamefont {{Pabst}}}, \bibinfo {author} {\bibfnamefont {P.}~\bibnamefont {{Ochner}}}, \bibinfo {author} {\bibfnamefont {E.~P.}\ \bibnamefont {{Gonzalez}}}, \bibinfo {author} {\bibfnamefont
  {A.}~\bibnamefont {{Pastorello}}}, \bibinfo {author} {\bibfnamefont {C.}~\bibnamefont {{Pellegrino}}}, \bibinfo {author} {\bibfnamefont {L.}~\bibnamefont {{Piscarreta}}}, \bibinfo {author} {\bibfnamefont {A.~P.}\ \bibnamefont {{Ravi}}}, \bibinfo {author} {\bibfnamefont {A.}~\bibnamefont {{Reguitti}}}, \bibinfo {author} {\bibfnamefont {L.}~\bibnamefont {{Salo}}}, \bibinfo {author} {\bibfnamefont {J.}~\bibnamefont {{Vink{\'o}}}}, \bibinfo {author} {\bibfnamefont {K.}~\bibnamefont {{de Vos}}}, \bibinfo {author} {\bibfnamefont {J.~C.}\ \bibnamefont {{Wheeler}}}, \bibinfo {author} {\bibfnamefont {G.~G.}\ \bibnamefont {{Williams}}}, \ and\ \bibinfo {author} {\bibfnamefont {S.}~\bibnamefont {{Wyatt}}},\ }\href {\doibase 10.3847/2041-8213/acf9a4} {\bibfield  {journal} {\bibinfo  {journal} {\apjl}\ }\textbf {\bibinfo {volume} {956}},\ \bibinfo {eid} {L5} (\bibinfo {year} {2023})},\ \Eprint {http://arxiv.org/abs/2306.10119} {arXiv:2306.10119 [astro-ph.HE]} \BibitemShut {NoStop}%
\bibitem [{\citenamefont {Grefenstette}\ \emph {et~al.}(2023)\citenamefont {Grefenstette}, \citenamefont {Brightman}, \citenamefont {Earnshaw}, \citenamefont {Harrison},\ and\ \citenamefont {Margutti}}]{grefenstette:23}%
  \BibitemOpen
  \bibfield  {author} {\bibinfo {author} {\bibfnamefont {B.~W.}\ \bibnamefont {Grefenstette}}, \bibinfo {author} {\bibfnamefont {M.}~\bibnamefont {Brightman}}, \bibinfo {author} {\bibfnamefont {H.~P.}\ \bibnamefont {Earnshaw}}, \bibinfo {author} {\bibfnamefont {F.~A.}\ \bibnamefont {Harrison}}, \ and\ \bibinfo {author} {\bibfnamefont {R.}~\bibnamefont {Margutti}},\ }\href {\doibase 10.3847/2041-8213/acdf4e} {\bibfield  {journal} {\bibinfo  {journal} {The Astrophysical Journal Letters}\ }\textbf {\bibinfo {volume} {952}},\ \bibinfo {pages} {L3} (\bibinfo {year} {2023})}\BibitemShut {NoStop}%
\bibitem [{\citenamefont {{Jencson}}\ \emph {et~al.}(2023)\citenamefont {{Jencson}}, \citenamefont {{Pearson}}, \citenamefont {{Beasor}}, \citenamefont {{Lau}}, \citenamefont {{Andrews}}, \citenamefont {{Bostroem}}, \citenamefont {{Dong}}, \citenamefont {{Engesser}}, \citenamefont {{Gomez}}, \citenamefont {{Guolo}}, \citenamefont {{Hoang}}, \citenamefont {{Hosseinzadeh}}, \citenamefont {{Jha}}, \citenamefont {{Karambelkar}}, \citenamefont {{Kasliwal}}, \citenamefont {{Lundquist}}, \citenamefont {{Meza Retamal}}, \citenamefont {{Rest}}, \citenamefont {{Sand}}, \citenamefont {{Shahbandeh}}, \citenamefont {{Shrestha}}, \citenamefont {{Smith}}, \citenamefont {{Strader}}, \citenamefont {{Valenti}}, \citenamefont {{Wang}},\ and\ \citenamefont {{Zenati}}}]{jencson:23}%
  \BibitemOpen
  \bibfield  {author} {\bibinfo {author} {\bibfnamefont {J.~E.}\ \bibnamefont {{Jencson}}}, \bibinfo {author} {\bibfnamefont {J.}~\bibnamefont {{Pearson}}}, \bibinfo {author} {\bibfnamefont {E.~R.}\ \bibnamefont {{Beasor}}}, \bibinfo {author} {\bibfnamefont {R.~M.}\ \bibnamefont {{Lau}}}, \bibinfo {author} {\bibfnamefont {J.~E.}\ \bibnamefont {{Andrews}}}, \bibinfo {author} {\bibfnamefont {K.~A.}\ \bibnamefont {{Bostroem}}}, \bibinfo {author} {\bibfnamefont {Y.}~\bibnamefont {{Dong}}}, \bibinfo {author} {\bibfnamefont {M.}~\bibnamefont {{Engesser}}}, \bibinfo {author} {\bibfnamefont {S.}~\bibnamefont {{Gomez}}}, \bibinfo {author} {\bibfnamefont {M.}~\bibnamefont {{Guolo}}}, \bibinfo {author} {\bibfnamefont {E.}~\bibnamefont {{Hoang}}}, \bibinfo {author} {\bibfnamefont {G.}~\bibnamefont {{Hosseinzadeh}}}, \bibinfo {author} {\bibfnamefont {S.~W.}\ \bibnamefont {{Jha}}}, \bibinfo {author} {\bibfnamefont {V.}~\bibnamefont {{Karambelkar}}}, \bibinfo {author} {\bibfnamefont {M.~M.}\ \bibnamefont {{Kasliwal}}},
  \bibinfo {author} {\bibfnamefont {M.}~\bibnamefont {{Lundquist}}}, \bibinfo {author} {\bibfnamefont {N.~E.}\ \bibnamefont {{Meza Retamal}}}, \bibinfo {author} {\bibfnamefont {A.}~\bibnamefont {{Rest}}}, \bibinfo {author} {\bibfnamefont {D.~J.}\ \bibnamefont {{Sand}}}, \bibinfo {author} {\bibfnamefont {M.}~\bibnamefont {{Shahbandeh}}}, \bibinfo {author} {\bibfnamefont {M.}~\bibnamefont {{Shrestha}}}, \bibinfo {author} {\bibfnamefont {N.}~\bibnamefont {{Smith}}}, \bibinfo {author} {\bibfnamefont {J.}~\bibnamefont {{Strader}}}, \bibinfo {author} {\bibfnamefont {S.}~\bibnamefont {{Valenti}}}, \bibinfo {author} {\bibfnamefont {Q.}~\bibnamefont {{Wang}}}, \ and\ \bibinfo {author} {\bibfnamefont {Y.}~\bibnamefont {{Zenati}}},\ }\href {\doibase 10.3847/2041-8213/ace618} {\bibfield  {journal} {\bibinfo  {journal} {\apjl}\ }\textbf {\bibinfo {volume} {952}},\ \bibinfo {eid} {L30} (\bibinfo {year} {2023})},\ \Eprint {http://arxiv.org/abs/2306.08678} {arXiv:2306.08678 [astro-ph.SR]} \BibitemShut {NoStop}%
\bibitem [{\citenamefont {{Jacobson-Gal{\'a}n}}\ \emph {et~al.}(2023)\citenamefont {{Jacobson-Gal{\'a}n}}, \citenamefont {{Dessart}}, \citenamefont {{Margutti}}, \citenamefont {{Chornock}}, \citenamefont {{Foley}}, \citenamefont {{Kilpatrick}}, \citenamefont {{Jones}}, \citenamefont {{Taggart}}, \citenamefont {{Angus}}, \citenamefont {{Bhattacharjee}}, \citenamefont {{Braff}}, \citenamefont {{Brethauer}}, \citenamefont {{Burgasser}}, \citenamefont {{Cao}}, \citenamefont {{Carlile}}, \citenamefont {{Chambers}}, \citenamefont {{Coulter}}, \citenamefont {{Dominguez-Ruiz}}, \citenamefont {{Dickinson}}, \citenamefont {{de Boer}}, \citenamefont {{Gagliano}}, \citenamefont {{Gall}}, \citenamefont {{Gao}}, \citenamefont {{Gates}}, \citenamefont {{Gomez}}, \citenamefont {{Guolo}}, \citenamefont {{Halford}}, \citenamefont {{Hjorth}}, \citenamefont {{Huber}}, \citenamefont {{Johnson}}, \citenamefont {{Karpoor}}, \citenamefont {{Laskar}}, \citenamefont {{LeBaron}}, \citenamefont {{Li}}, \citenamefont {{Lin}},
  \citenamefont {{Loch}}, \citenamefont {{Lynam}}, \citenamefont {{Magnier}}, \citenamefont {{Maloney}}, \citenamefont {{Matthews}}, \citenamefont {{McDonald}}, \citenamefont {{Miao}}, \citenamefont {{Milisavljevic}}, \citenamefont {{Pan}}, \citenamefont {{Pradyumna}}, \citenamefont {{Ransome}}, \citenamefont {{Rees}}, \citenamefont {{Rest}}, \citenamefont {{Rojas-Bravo}}, \citenamefont {{Sandford}}, \citenamefont {{Ascencio}}, \citenamefont {{Sanjaripour}}, \citenamefont {{Savino}}, \citenamefont {{Sears}}, \citenamefont {{Sharei}}, \citenamefont {{Smartt}}, \citenamefont {{Softich}}, \citenamefont {{Theissen}}, \citenamefont {{Tinyanont}}, \citenamefont {{Tohfa}}, \citenamefont {{Villar}}, \citenamefont {{Wang}}, \citenamefont {{Wainscoat}}, \citenamefont {{Westerling}}, \citenamefont {{Wiston}}, \citenamefont {{Wozniak}}, \citenamefont {{Yadavalli}},\ and\ \citenamefont {{Zenati}}}]{jacobson-galan:23}%
  \BibitemOpen
  \bibfield  {author} {\bibinfo {author} {\bibfnamefont {W.~V.}\ \bibnamefont {{Jacobson-Gal{\'a}n}}}, \bibinfo {author} {\bibfnamefont {L.}~\bibnamefont {{Dessart}}}, \bibinfo {author} {\bibfnamefont {R.}~\bibnamefont {{Margutti}}}, \bibinfo {author} {\bibfnamefont {R.}~\bibnamefont {{Chornock}}}, \bibinfo {author} {\bibfnamefont {R.~J.}\ \bibnamefont {{Foley}}}, \bibinfo {author} {\bibfnamefont {C.~D.}\ \bibnamefont {{Kilpatrick}}}, \bibinfo {author} {\bibfnamefont {D.~O.}\ \bibnamefont {{Jones}}}, \bibinfo {author} {\bibfnamefont {K.}~\bibnamefont {{Taggart}}}, \bibinfo {author} {\bibfnamefont {C.~R.}\ \bibnamefont {{Angus}}}, \bibinfo {author} {\bibfnamefont {S.}~\bibnamefont {{Bhattacharjee}}}, \bibinfo {author} {\bibfnamefont {L.~A.}\ \bibnamefont {{Braff}}}, \bibinfo {author} {\bibfnamefont {D.}~\bibnamefont {{Brethauer}}}, \bibinfo {author} {\bibfnamefont {A.~J.}\ \bibnamefont {{Burgasser}}}, \bibinfo {author} {\bibfnamefont {F.}~\bibnamefont {{Cao}}}, \bibinfo {author} {\bibfnamefont {C.~M.}\
  \bibnamefont {{Carlile}}}, \bibinfo {author} {\bibfnamefont {K.~C.}\ \bibnamefont {{Chambers}}}, \bibinfo {author} {\bibfnamefont {D.~A.}\ \bibnamefont {{Coulter}}}, \bibinfo {author} {\bibfnamefont {E.}~\bibnamefont {{Dominguez-Ruiz}}}, \bibinfo {author} {\bibfnamefont {C.~B.}\ \bibnamefont {{Dickinson}}}, \bibinfo {author} {\bibfnamefont {T.}~\bibnamefont {{de Boer}}}, \bibinfo {author} {\bibfnamefont {A.}~\bibnamefont {{Gagliano}}}, \bibinfo {author} {\bibfnamefont {C.}~\bibnamefont {{Gall}}}, \bibinfo {author} {\bibfnamefont {H.}~\bibnamefont {{Gao}}}, \bibinfo {author} {\bibfnamefont {E.~L.}\ \bibnamefont {{Gates}}}, \bibinfo {author} {\bibfnamefont {S.}~\bibnamefont {{Gomez}}}, \bibinfo {author} {\bibfnamefont {M.}~\bibnamefont {{Guolo}}}, \bibinfo {author} {\bibfnamefont {M.~R.~J.}\ \bibnamefont {{Halford}}}, \bibinfo {author} {\bibfnamefont {J.}~\bibnamefont {{Hjorth}}}, \bibinfo {author} {\bibfnamefont {M.~E.}\ \bibnamefont {{Huber}}}, \bibinfo {author} {\bibfnamefont {M.~N.}\ \bibnamefont
  {{Johnson}}}, \bibinfo {author} {\bibfnamefont {P.~R.}\ \bibnamefont {{Karpoor}}}, \bibinfo {author} {\bibfnamefont {T.}~\bibnamefont {{Laskar}}}, \bibinfo {author} {\bibfnamefont {N.}~\bibnamefont {{LeBaron}}}, \bibinfo {author} {\bibfnamefont {Z.}~\bibnamefont {{Li}}}, \bibinfo {author} {\bibfnamefont {Y.}~\bibnamefont {{Lin}}}, \bibinfo {author} {\bibfnamefont {S.~D.}\ \bibnamefont {{Loch}}}, \bibinfo {author} {\bibfnamefont {P.~D.}\ \bibnamefont {{Lynam}}}, \bibinfo {author} {\bibfnamefont {E.~A.}\ \bibnamefont {{Magnier}}}, \bibinfo {author} {\bibfnamefont {P.}~\bibnamefont {{Maloney}}}, \bibinfo {author} {\bibfnamefont {D.~J.}\ \bibnamefont {{Matthews}}}, \bibinfo {author} {\bibfnamefont {M.}~\bibnamefont {{McDonald}}}, \bibinfo {author} {\bibfnamefont {H.~Y.}\ \bibnamefont {{Miao}}}, \bibinfo {author} {\bibfnamefont {D.}~\bibnamefont {{Milisavljevic}}}, \bibinfo {author} {\bibfnamefont {Y.~C.}\ \bibnamefont {{Pan}}}, \bibinfo {author} {\bibfnamefont {S.}~\bibnamefont {{Pradyumna}}}, \bibinfo {author}
  {\bibfnamefont {C.~L.}\ \bibnamefont {{Ransome}}}, \bibinfo {author} {\bibfnamefont {J.~M.}\ \bibnamefont {{Rees}}}, \bibinfo {author} {\bibfnamefont {A.}~\bibnamefont {{Rest}}}, \bibinfo {author} {\bibfnamefont {C.}~\bibnamefont {{Rojas-Bravo}}}, \bibinfo {author} {\bibfnamefont {N.~R.}\ \bibnamefont {{Sandford}}}, \bibinfo {author} {\bibfnamefont {L.~S.}\ \bibnamefont {{Ascencio}}}, \bibinfo {author} {\bibfnamefont {S.}~\bibnamefont {{Sanjaripour}}}, \bibinfo {author} {\bibfnamefont {A.}~\bibnamefont {{Savino}}}, \bibinfo {author} {\bibfnamefont {H.}~\bibnamefont {{Sears}}}, \bibinfo {author} {\bibfnamefont {N.}~\bibnamefont {{Sharei}}}, \bibinfo {author} {\bibfnamefont {S.~J.}\ \bibnamefont {{Smartt}}}, \bibinfo {author} {\bibfnamefont {E.~R.}\ \bibnamefont {{Softich}}}, \bibinfo {author} {\bibfnamefont {C.~A.}\ \bibnamefont {{Theissen}}}, \bibinfo {author} {\bibfnamefont {S.}~\bibnamefont {{Tinyanont}}}, \bibinfo {author} {\bibfnamefont {H.}~\bibnamefont {{Tohfa}}}, \bibinfo {author} {\bibfnamefont
  {V.~A.}\ \bibnamefont {{Villar}}}, \bibinfo {author} {\bibfnamefont {Q.}~\bibnamefont {{Wang}}}, \bibinfo {author} {\bibfnamefont {R.~J.}\ \bibnamefont {{Wainscoat}}}, \bibinfo {author} {\bibfnamefont {A.~L.}\ \bibnamefont {{Westerling}}}, \bibinfo {author} {\bibfnamefont {E.}~\bibnamefont {{Wiston}}}, \bibinfo {author} {\bibfnamefont {M.~A.}\ \bibnamefont {{Wozniak}}}, \bibinfo {author} {\bibfnamefont {S.~K.}\ \bibnamefont {{Yadavalli}}}, \ and\ \bibinfo {author} {\bibfnamefont {Y.}~\bibnamefont {{Zenati}}},\ }\href {\doibase 10.3847/2041-8213/acf2ec} {\bibfield  {journal} {\bibinfo  {journal} {\apjl}\ }\textbf {\bibinfo {volume} {954}},\ \bibinfo {eid} {L42} (\bibinfo {year} {2023})},\ \Eprint {http://arxiv.org/abs/2306.04721} {arXiv:2306.04721 [astro-ph.HE]} \BibitemShut {NoStop}%
\bibitem [{\citenamefont {{Hiramatsu}}\ \emph {et~al.}(2023)\citenamefont {{Hiramatsu}}, \citenamefont {{Tsuna}}, \citenamefont {{Berger}}, \citenamefont {{Itagaki}}, \citenamefont {{Goldberg}}, \citenamefont {{Gomez}}, \citenamefont {{Kishalay}}, \citenamefont {{Hosseinzadeh}}, \citenamefont {{Bostroem}}, \citenamefont {{Brown}}, \citenamefont {{Arcavi}}, \citenamefont {{Bieryla}}, \citenamefont {{Blanchard}}, \citenamefont {{Esquerdo}}, \citenamefont {{Farah}}, \citenamefont {{Howell}}, \citenamefont {{Matsumoto}}, \citenamefont {{McCully}}, \citenamefont {{Newsome}}, \citenamefont {{Gonzalez}}, \citenamefont {{Pellegrino}}, \citenamefont {{Rhee}}, \citenamefont {{Terreran}}, \citenamefont {{Vink{\'o}}},\ and\ \citenamefont {{Wheeler}}}]{hiramatsu:23}%
  \BibitemOpen
  \bibfield  {author} {\bibinfo {author} {\bibfnamefont {D.}~\bibnamefont {{Hiramatsu}}}, \bibinfo {author} {\bibfnamefont {D.}~\bibnamefont {{Tsuna}}}, \bibinfo {author} {\bibfnamefont {E.}~\bibnamefont {{Berger}}}, \bibinfo {author} {\bibfnamefont {K.}~\bibnamefont {{Itagaki}}}, \bibinfo {author} {\bibfnamefont {J.~A.}\ \bibnamefont {{Goldberg}}}, \bibinfo {author} {\bibfnamefont {S.}~\bibnamefont {{Gomez}}}, \bibinfo {author} {\bibfnamefont {D.}~\bibnamefont {{Kishalay}}}, \bibinfo {author} {\bibfnamefont {G.}~\bibnamefont {{Hosseinzadeh}}}, \bibinfo {author} {\bibfnamefont {K.~A.}\ \bibnamefont {{Bostroem}}}, \bibinfo {author} {\bibfnamefont {P.~J.}\ \bibnamefont {{Brown}}}, \bibinfo {author} {\bibfnamefont {I.}~\bibnamefont {{Arcavi}}}, \bibinfo {author} {\bibfnamefont {A.}~\bibnamefont {{Bieryla}}}, \bibinfo {author} {\bibfnamefont {P.~K.}\ \bibnamefont {{Blanchard}}}, \bibinfo {author} {\bibfnamefont {G.~A.}\ \bibnamefont {{Esquerdo}}}, \bibinfo {author} {\bibfnamefont {J.}~\bibnamefont {{Farah}}},
  \bibinfo {author} {\bibfnamefont {D.~A.}\ \bibnamefont {{Howell}}}, \bibinfo {author} {\bibfnamefont {T.}~\bibnamefont {{Matsumoto}}}, \bibinfo {author} {\bibfnamefont {C.}~\bibnamefont {{McCully}}}, \bibinfo {author} {\bibfnamefont {M.}~\bibnamefont {{Newsome}}}, \bibinfo {author} {\bibfnamefont {E.~P.}\ \bibnamefont {{Gonzalez}}}, \bibinfo {author} {\bibfnamefont {C.}~\bibnamefont {{Pellegrino}}}, \bibinfo {author} {\bibfnamefont {J.}~\bibnamefont {{Rhee}}}, \bibinfo {author} {\bibfnamefont {G.}~\bibnamefont {{Terreran}}}, \bibinfo {author} {\bibfnamefont {J.}~\bibnamefont {{Vink{\'o}}}}, \ and\ \bibinfo {author} {\bibfnamefont {J.~C.}\ \bibnamefont {{Wheeler}}},\ }\href {\doibase 10.3847/2041-8213/acf299} {\bibfield  {journal} {\bibinfo  {journal} {\apjl}\ }\textbf {\bibinfo {volume} {955}},\ \bibinfo {eid} {L8} (\bibinfo {year} {2023})},\ \Eprint {http://arxiv.org/abs/2307.03165} {arXiv:2307.03165 [astro-ph.HE]} \BibitemShut {NoStop}%
\bibitem [{\citenamefont {{Vasylyev}}\ \emph {et~al.}(2023)\citenamefont {{Vasylyev}}, \citenamefont {{Yang}}, \citenamefont {{Filippenko}}, \citenamefont {{Patra}}, \citenamefont {{Brink}}, \citenamefont {{Wang}}, \citenamefont {{Chornock}}, \citenamefont {{Margutti}}, \citenamefont {{Gates}}, \citenamefont {{Burgasser}}, \citenamefont {{Karpoor}}, \citenamefont {{LeBaron}}, \citenamefont {{Softich}}, \citenamefont {{Theissen}}, \citenamefont {{Wiston}},\ and\ \citenamefont {{Zheng}}}]{vasyleyv:23}%
  \BibitemOpen
  \bibfield  {author} {\bibinfo {author} {\bibfnamefont {S.~S.}\ \bibnamefont {{Vasylyev}}}, \bibinfo {author} {\bibfnamefont {Y.}~\bibnamefont {{Yang}}}, \bibinfo {author} {\bibfnamefont {A.~V.}\ \bibnamefont {{Filippenko}}}, \bibinfo {author} {\bibfnamefont {K.~C.}\ \bibnamefont {{Patra}}}, \bibinfo {author} {\bibfnamefont {T.~G.}\ \bibnamefont {{Brink}}}, \bibinfo {author} {\bibfnamefont {L.}~\bibnamefont {{Wang}}}, \bibinfo {author} {\bibfnamefont {R.}~\bibnamefont {{Chornock}}}, \bibinfo {author} {\bibfnamefont {R.}~\bibnamefont {{Margutti}}}, \bibinfo {author} {\bibfnamefont {E.~L.}\ \bibnamefont {{Gates}}}, \bibinfo {author} {\bibfnamefont {A.~J.}\ \bibnamefont {{Burgasser}}}, \bibinfo {author} {\bibfnamefont {P.~R.}\ \bibnamefont {{Karpoor}}}, \bibinfo {author} {\bibfnamefont {N.}~\bibnamefont {{LeBaron}}}, \bibinfo {author} {\bibfnamefont {E.}~\bibnamefont {{Softich}}}, \bibinfo {author} {\bibfnamefont {C.~A.}\ \bibnamefont {{Theissen}}}, \bibinfo {author} {\bibfnamefont {E.}~\bibnamefont
  {{Wiston}}}, \ and\ \bibinfo {author} {\bibfnamefont {W.}~\bibnamefont {{Zheng}}},\ }\href {\doibase 10.3847/2041-8213/acf1a3} {\bibfield  {journal} {\bibinfo  {journal} {\apjl}\ }\textbf {\bibinfo {volume} {955}},\ \bibinfo {eid} {L37} (\bibinfo {year} {2023})},\ \Eprint {http://arxiv.org/abs/2307.01268} {arXiv:2307.01268 [astro-ph.HE]} \BibitemShut {NoStop}%
\bibitem [{\citenamefont {{Smith}}\ \emph {et~al.}(2023)\citenamefont {{Smith}}, \citenamefont {{Pearson}}, \citenamefont {{Sand}}, \citenamefont {{Ilyin}}, \citenamefont {{Bostroem}}, \citenamefont {{Hosseinzadeh}},\ and\ \citenamefont {{Shrestha}}}]{smith:23}%
  \BibitemOpen
  \bibfield  {author} {\bibinfo {author} {\bibfnamefont {N.}~\bibnamefont {{Smith}}}, \bibinfo {author} {\bibfnamefont {J.}~\bibnamefont {{Pearson}}}, \bibinfo {author} {\bibfnamefont {D.~J.}\ \bibnamefont {{Sand}}}, \bibinfo {author} {\bibfnamefont {I.}~\bibnamefont {{Ilyin}}}, \bibinfo {author} {\bibfnamefont {K.~A.}\ \bibnamefont {{Bostroem}}}, \bibinfo {author} {\bibfnamefont {G.}~\bibnamefont {{Hosseinzadeh}}}, \ and\ \bibinfo {author} {\bibfnamefont {M.}~\bibnamefont {{Shrestha}}},\ }\href {\doibase 10.3847/1538-4357/acf366} {\bibfield  {journal} {\bibinfo  {journal} {\apj}\ }\textbf {\bibinfo {volume} {956}},\ \bibinfo {eid} {46} (\bibinfo {year} {2023})},\ \Eprint {http://arxiv.org/abs/2306.07964} {arXiv:2306.07964 [astro-ph.HE]} \BibitemShut {NoStop}%
\bibitem [{\citenamefont {{Hosseinzadeh}}\ \emph {et~al.}(2023)\citenamefont {{Hosseinzadeh}}, \citenamefont {{Farah}}, \citenamefont {{Shrestha}}, \citenamefont {{Sand}}, \citenamefont {{Dong}}, \citenamefont {{Brown}}, \citenamefont {{Bostroem}}, \citenamefont {{Valenti}}, \citenamefont {{Jha}}, \citenamefont {{Andrews}}, \citenamefont {{Arcavi}}, \citenamefont {{Haislip}}, \citenamefont {{Hiramatsu}}, \citenamefont {{Hoang}}, \citenamefont {{Howell}}, \citenamefont {{Janzen}}, \citenamefont {{Jencson}}, \citenamefont {{Kouprianov}}, \citenamefont {{Lundquist}}, \citenamefont {{McCully}}, \citenamefont {{Meza Retamal}}, \citenamefont {{Modjaz}}, \citenamefont {{Newsome}}, \citenamefont {{Padilla Gonzalez}}, \citenamefont {{Pearson}}, \citenamefont {{Pellegrino}}, \citenamefont {{Ravi}}, \citenamefont {{Reichart}}, \citenamefont {{Smith}}, \citenamefont {{Terreran}},\ and\ \citenamefont {{Vink{\'o}}}}]{hosseinzadeh:23}%
  \BibitemOpen
  \bibfield  {author} {\bibinfo {author} {\bibfnamefont {G.}~\bibnamefont {{Hosseinzadeh}}}, \bibinfo {author} {\bibfnamefont {J.}~\bibnamefont {{Farah}}}, \bibinfo {author} {\bibfnamefont {M.}~\bibnamefont {{Shrestha}}}, \bibinfo {author} {\bibfnamefont {D.~J.}\ \bibnamefont {{Sand}}}, \bibinfo {author} {\bibfnamefont {Y.}~\bibnamefont {{Dong}}}, \bibinfo {author} {\bibfnamefont {P.~J.}\ \bibnamefont {{Brown}}}, \bibinfo {author} {\bibfnamefont {K.~A.}\ \bibnamefont {{Bostroem}}}, \bibinfo {author} {\bibfnamefont {S.}~\bibnamefont {{Valenti}}}, \bibinfo {author} {\bibfnamefont {S.~W.}\ \bibnamefont {{Jha}}}, \bibinfo {author} {\bibfnamefont {J.~E.}\ \bibnamefont {{Andrews}}}, \bibinfo {author} {\bibfnamefont {I.}~\bibnamefont {{Arcavi}}}, \bibinfo {author} {\bibfnamefont {J.}~\bibnamefont {{Haislip}}}, \bibinfo {author} {\bibfnamefont {D.}~\bibnamefont {{Hiramatsu}}}, \bibinfo {author} {\bibfnamefont {E.}~\bibnamefont {{Hoang}}}, \bibinfo {author} {\bibfnamefont {D.~A.}\ \bibnamefont {{Howell}}}, \bibinfo
  {author} {\bibfnamefont {D.}~\bibnamefont {{Janzen}}}, \bibinfo {author} {\bibfnamefont {J.~E.}\ \bibnamefont {{Jencson}}}, \bibinfo {author} {\bibfnamefont {V.}~\bibnamefont {{Kouprianov}}}, \bibinfo {author} {\bibfnamefont {M.}~\bibnamefont {{Lundquist}}}, \bibinfo {author} {\bibfnamefont {C.}~\bibnamefont {{McCully}}}, \bibinfo {author} {\bibfnamefont {N.~E.}\ \bibnamefont {{Meza Retamal}}}, \bibinfo {author} {\bibfnamefont {M.}~\bibnamefont {{Modjaz}}}, \bibinfo {author} {\bibfnamefont {M.}~\bibnamefont {{Newsome}}}, \bibinfo {author} {\bibfnamefont {E.}~\bibnamefont {{Padilla Gonzalez}}}, \bibinfo {author} {\bibfnamefont {J.}~\bibnamefont {{Pearson}}}, \bibinfo {author} {\bibfnamefont {C.}~\bibnamefont {{Pellegrino}}}, \bibinfo {author} {\bibfnamefont {A.~P.}\ \bibnamefont {{Ravi}}}, \bibinfo {author} {\bibfnamefont {D.~E.}\ \bibnamefont {{Reichart}}}, \bibinfo {author} {\bibfnamefont {N.}~\bibnamefont {{Smith}}}, \bibinfo {author} {\bibfnamefont {G.}~\bibnamefont {{Terreran}}}, \ and\ \bibinfo
  {author} {\bibfnamefont {J.}~\bibnamefont {{Vink{\'o}}}},\ }\href {\doibase 10.3847/2041-8213/ace4c4} {\bibfield  {journal} {\bibinfo  {journal} {\apjl}\ }\textbf {\bibinfo {volume} {953}},\ \bibinfo {eid} {L16} (\bibinfo {year} {2023})},\ \Eprint {http://arxiv.org/abs/2306.06097} {arXiv:2306.06097 [astro-ph.HE]} \BibitemShut {NoStop}%
\bibitem [{\citenamefont {{Kilpatrick}}\ \emph {et~al.}(2023)\citenamefont {{Kilpatrick}}, \citenamefont {{Foley}}, \citenamefont {{Jacobson-Gal{\'a}n}}, \citenamefont {{Piro}}, \citenamefont {{Smartt}}, \citenamefont {{Drout}}, \citenamefont {{Gagliano}}, \citenamefont {{Gall}}, \citenamefont {{Hjorth}}, \citenamefont {{Jones}}, \citenamefont {{Mandel}}, \citenamefont {{Margutti}}, \citenamefont {{Ramirez-Ruiz}}, \citenamefont {{Ransome}}, \citenamefont {{Villar}}, \citenamefont {{Coulter}}, \citenamefont {{Gao}}, \citenamefont {{Matthews}}, \citenamefont {{Taggart}},\ and\ \citenamefont {{Zenati}}}]{kilpatrick:23}%
  \BibitemOpen
  \bibfield  {author} {\bibinfo {author} {\bibfnamefont {C.~D.}\ \bibnamefont {{Kilpatrick}}}, \bibinfo {author} {\bibfnamefont {R.~J.}\ \bibnamefont {{Foley}}}, \bibinfo {author} {\bibfnamefont {W.~V.}\ \bibnamefont {{Jacobson-Gal{\'a}n}}}, \bibinfo {author} {\bibfnamefont {A.~L.}\ \bibnamefont {{Piro}}}, \bibinfo {author} {\bibfnamefont {S.~J.}\ \bibnamefont {{Smartt}}}, \bibinfo {author} {\bibfnamefont {M.~R.}\ \bibnamefont {{Drout}}}, \bibinfo {author} {\bibfnamefont {A.}~\bibnamefont {{Gagliano}}}, \bibinfo {author} {\bibfnamefont {C.}~\bibnamefont {{Gall}}}, \bibinfo {author} {\bibfnamefont {J.}~\bibnamefont {{Hjorth}}}, \bibinfo {author} {\bibfnamefont {D.~O.}\ \bibnamefont {{Jones}}}, \bibinfo {author} {\bibfnamefont {K.~S.}\ \bibnamefont {{Mandel}}}, \bibinfo {author} {\bibfnamefont {R.}~\bibnamefont {{Margutti}}}, \bibinfo {author} {\bibfnamefont {E.}~\bibnamefont {{Ramirez-Ruiz}}}, \bibinfo {author} {\bibfnamefont {C.~L.}\ \bibnamefont {{Ransome}}}, \bibinfo {author} {\bibfnamefont {V.~A.}\
  \bibnamefont {{Villar}}}, \bibinfo {author} {\bibfnamefont {D.~A.}\ \bibnamefont {{Coulter}}}, \bibinfo {author} {\bibfnamefont {H.}~\bibnamefont {{Gao}}}, \bibinfo {author} {\bibfnamefont {D.~J.}\ \bibnamefont {{Matthews}}}, \bibinfo {author} {\bibfnamefont {K.}~\bibnamefont {{Taggart}}}, \ and\ \bibinfo {author} {\bibfnamefont {Y.}~\bibnamefont {{Zenati}}},\ }\href {\doibase 10.3847/2041-8213/ace4ca} {\bibfield  {journal} {\bibinfo  {journal} {\apjl}\ }\textbf {\bibinfo {volume} {952}},\ \bibinfo {eid} {L23} (\bibinfo {year} {2023})},\ \Eprint {http://arxiv.org/abs/2306.04722} {arXiv:2306.04722 [astro-ph.SR]} \BibitemShut {NoStop}%
\bibitem [{\citenamefont {{Soraisam}}\ \emph {et~al.}(2018)\citenamefont {{Soraisam}}, \citenamefont {{Bildsten}}, \citenamefont {{Drout}}, \citenamefont {{Bauer}}, \citenamefont {{Gilfanov}}, \citenamefont {{Kupfer}}, \citenamefont {{Laher}}, \citenamefont {{Masci}}, \citenamefont {{Prince}}, \citenamefont {{Kulkarni}}, \citenamefont {{Matheson}},\ and\ \citenamefont {{Saha}}}]{soraisam:18}%
  \BibitemOpen
  \bibfield  {author} {\bibinfo {author} {\bibfnamefont {M.~D.}\ \bibnamefont {{Soraisam}}}, \bibinfo {author} {\bibfnamefont {L.}~\bibnamefont {{Bildsten}}}, \bibinfo {author} {\bibfnamefont {M.~R.}\ \bibnamefont {{Drout}}}, \bibinfo {author} {\bibfnamefont {E.~B.}\ \bibnamefont {{Bauer}}}, \bibinfo {author} {\bibfnamefont {M.}~\bibnamefont {{Gilfanov}}}, \bibinfo {author} {\bibfnamefont {T.}~\bibnamefont {{Kupfer}}}, \bibinfo {author} {\bibfnamefont {R.~R.}\ \bibnamefont {{Laher}}}, \bibinfo {author} {\bibfnamefont {F.}~\bibnamefont {{Masci}}}, \bibinfo {author} {\bibfnamefont {T.~A.}\ \bibnamefont {{Prince}}}, \bibinfo {author} {\bibfnamefont {S.~R.}\ \bibnamefont {{Kulkarni}}}, \bibinfo {author} {\bibfnamefont {T.}~\bibnamefont {{Matheson}}}, \ and\ \bibinfo {author} {\bibfnamefont {A.}~\bibnamefont {{Saha}}},\ }\href {\doibase 10.3847/1538-4357/aabc59} {\bibfield  {journal} {\bibinfo  {journal} {\apj}\ }\textbf {\bibinfo {volume} {859}},\ \bibinfo {eid} {73} (\bibinfo {year} {2018})},\ \Eprint
  {http://arxiv.org/abs/1803.09934} {arXiv:1803.09934 [astro-ph.SR]} \BibitemShut {NoStop}%
\bibitem [{\citenamefont {{Yoon}}\ and\ \citenamefont {{Cantiello}}(2010)}]{yooncantiello:10}%
  \BibitemOpen
  \bibfield  {author} {\bibinfo {author} {\bibfnamefont {S.-C.}\ \bibnamefont {{Yoon}}}\ and\ \bibinfo {author} {\bibfnamefont {M.}~\bibnamefont {{Cantiello}}},\ }\href {\doibase 10.1088/2041-8205/717/1/L62} {\bibfield  {journal} {\bibinfo  {journal} {\apjl}\ }\textbf {\bibinfo {volume} {717}},\ \bibinfo {pages} {L62} (\bibinfo {year} {2010})},\ \Eprint {http://arxiv.org/abs/1005.4925} {arXiv:1005.4925 [astro-ph.SR]} \BibitemShut {NoStop}%
\bibitem [{\citenamefont {{Rabatin}}\ and\ \citenamefont {{Collins}}(2023{\natexlab{a}})}]{rabatin:23}%
  \BibitemOpen
  \bibfield  {author} {\bibinfo {author} {\bibfnamefont {B.}~\bibnamefont {{Rabatin}}}\ and\ \bibinfo {author} {\bibfnamefont {D.~C.}\ \bibnamefont {{Collins}}},\ }\href {\doibase 10.1093/mnrasl/slac123} {\bibfield  {journal} {\bibinfo  {journal} {\mnras}\ }\textbf {\bibinfo {volume} {521}},\ \bibinfo {pages} {L64} (\bibinfo {year} {2023}{\natexlab{a}})},\ \Eprint {http://arxiv.org/abs/2210.03597} {arXiv:2210.03597 [astro-ph.GA]} \BibitemShut {NoStop}%
\bibitem [{\citenamefont {{Rabatin}}\ and\ \citenamefont {{Collins}}(2023{\natexlab{b}})}]{rabatin:23b}%
  \BibitemOpen
  \bibfield  {author} {\bibinfo {author} {\bibfnamefont {B.}~\bibnamefont {{Rabatin}}}\ and\ \bibinfo {author} {\bibfnamefont {D.~C.}\ \bibnamefont {{Collins}}},\ }\href {\doibase 10.1093/mnras/stad2195} {\bibfield  {journal} {\bibinfo  {journal} {\mnras}\ }\textbf {\bibinfo {volume} {525}},\ \bibinfo {pages} {297} (\bibinfo {year} {2023}{\natexlab{b}})},\ \Eprint {http://arxiv.org/abs/2307.04876} {arXiv:2307.04876 [astro-ph.GA]} \BibitemShut {NoStop}%
\bibitem [{\citenamefont {{Hopkins}}(2013)}]{hopkins:13}%
  \BibitemOpen
  \bibfield  {author} {\bibinfo {author} {\bibfnamefont {P.~F.}\ \bibnamefont {{Hopkins}}},\ }\href {\doibase 10.1093/mnras/stt010} {\bibfield  {journal} {\bibinfo  {journal} {\mnras}\ }\textbf {\bibinfo {volume} {430}},\ \bibinfo {pages} {1880} (\bibinfo {year} {2013})},\ \Eprint {http://arxiv.org/abs/1211.3119} {arXiv:1211.3119 [astro-ph.GA]} \BibitemShut {NoStop}%
\bibitem [{\citenamefont {{Plez}}\ \emph {et~al.}(1992)\citenamefont {{Plez}}, \citenamefont {{Brett}},\ and\ \citenamefont {{Nordlund}}}]{plez:92}%
  \BibitemOpen
  \bibfield  {author} {\bibinfo {author} {\bibfnamefont {B.}~\bibnamefont {{Plez}}}, \bibinfo {author} {\bibfnamefont {J.~M.}\ \bibnamefont {{Brett}}}, \ and\ \bibinfo {author} {\bibfnamefont {A.}~\bibnamefont {{Nordlund}}},\ }\href@noop {} {\bibfield  {journal} {\bibinfo  {journal} {\aap}\ }\textbf {\bibinfo {volume} {256}},\ \bibinfo {pages} {551} (\bibinfo {year} {1992})}\BibitemShut {NoStop}%
\bibitem [{\citenamefont {{Ireland}}\ \emph {et~al.}(2008)\citenamefont {{Ireland}}, \citenamefont {{Scholz}},\ and\ \citenamefont {{Wood}}}]{ireland:08}%
  \BibitemOpen
  \bibfield  {author} {\bibinfo {author} {\bibfnamefont {M.~J.}\ \bibnamefont {{Ireland}}}, \bibinfo {author} {\bibfnamefont {M.}~\bibnamefont {{Scholz}}}, \ and\ \bibinfo {author} {\bibfnamefont {P.~R.}\ \bibnamefont {{Wood}}},\ }\href {\doibase 10.1111/j.1365-2966.2008.14037.x} {\bibfield  {journal} {\bibinfo  {journal} {\mnras}\ }\textbf {\bibinfo {volume} {391}},\ \bibinfo {pages} {1994} (\bibinfo {year} {2008})},\ \Eprint {http://arxiv.org/abs/0810.0560} {arXiv:0810.0560 [astro-ph]} \BibitemShut {NoStop}%
\bibitem [{\citenamefont {{Thirumalai}}\ and\ \citenamefont {{Heyl}}(2010)}]{thirumalai:10}%
  \BibitemOpen
  \bibfield  {author} {\bibinfo {author} {\bibfnamefont {A.}~\bibnamefont {{Thirumalai}}}\ and\ \bibinfo {author} {\bibfnamefont {J.~S.}\ \bibnamefont {{Heyl}}},\ }\href {\doibase 10.1111/j.1365-2966.2010.17414.x} {\bibfield  {journal} {\bibinfo  {journal} {\mnras}\ }\textbf {\bibinfo {volume} {409}},\ \bibinfo {pages} {1669} (\bibinfo {year} {2010})},\ \Eprint {http://arxiv.org/abs/1006.2181} {arXiv:1006.2181 [astro-ph.SR]} \BibitemShut {NoStop}%
\bibitem [{\citenamefont {{Thirumalai}}\ and\ \citenamefont {{Heyl}}(2012)}]{thirumalai:12}%
  \BibitemOpen
  \bibfield  {author} {\bibinfo {author} {\bibfnamefont {A.}~\bibnamefont {{Thirumalai}}}\ and\ \bibinfo {author} {\bibfnamefont {J.~S.}\ \bibnamefont {{Heyl}}},\ }\href {\doibase 10.1111/j.1365-2966.2012.20703.x} {\bibfield  {journal} {\bibinfo  {journal} {\mnras}\ }\textbf {\bibinfo {volume} {422}},\ \bibinfo {pages} {1272} (\bibinfo {year} {2012})},\ \Eprint {http://arxiv.org/abs/1109.5148} {arXiv:1109.5148 [astro-ph.SR]} \BibitemShut {NoStop}%
\bibitem [{\citenamefont {{Squire}}\ and\ \citenamefont {{Hopkins}}(2018)}]{hopkins:18}%
  \BibitemOpen
  \bibfield  {author} {\bibinfo {author} {\bibfnamefont {J.}~\bibnamefont {{Squire}}}\ and\ \bibinfo {author} {\bibfnamefont {P.~F.}\ \bibnamefont {{Hopkins}}},\ }\href {\doibase 10.1093/mnras/sty854} {\bibfield  {journal} {\bibinfo  {journal} {\mnras}\ }\textbf {\bibinfo {volume} {477}},\ \bibinfo {pages} {5011} (\bibinfo {year} {2018})},\ \Eprint {http://arxiv.org/abs/1711.03975} {arXiv:1711.03975 [astro-ph.EP]} \BibitemShut {NoStop}%
\bibitem [{\citenamefont {{Moseley}}\ \emph {et~al.}(2019)\citenamefont {{Moseley}}, \citenamefont {{Squire}},\ and\ \citenamefont {{Hopkins}}}]{moseley:19}%
  \BibitemOpen
  \bibfield  {author} {\bibinfo {author} {\bibfnamefont {E.~R.}\ \bibnamefont {{Moseley}}}, \bibinfo {author} {\bibfnamefont {J.}~\bibnamefont {{Squire}}}, \ and\ \bibinfo {author} {\bibfnamefont {P.~F.}\ \bibnamefont {{Hopkins}}},\ }\href {\doibase 10.1093/mnras/stz2128} {\bibfield  {journal} {\bibinfo  {journal} {\mnras}\ }\textbf {\bibinfo {volume} {489}},\ \bibinfo {pages} {325} (\bibinfo {year} {2019})},\ \Eprint {http://arxiv.org/abs/1810.08214} {arXiv:1810.08214 [astro-ph.GA]} \BibitemShut {NoStop}%
\bibitem [{\citenamefont {{Steinwandel}}\ \emph {et~al.}(2022)\citenamefont {{Steinwandel}}, \citenamefont {{Kaurov}}, \citenamefont {{Hopkins}},\ and\ \citenamefont {{Squire}}}]{steinwandel:22}%
  \BibitemOpen
  \bibfield  {author} {\bibinfo {author} {\bibfnamefont {U.~P.}\ \bibnamefont {{Steinwandel}}}, \bibinfo {author} {\bibfnamefont {A.~A.}\ \bibnamefont {{Kaurov}}}, \bibinfo {author} {\bibfnamefont {P.~F.}\ \bibnamefont {{Hopkins}}}, \ and\ \bibinfo {author} {\bibfnamefont {J.}~\bibnamefont {{Squire}}},\ }\href {\doibase 10.1093/mnras/stac2035} {\bibfield  {journal} {\bibinfo  {journal} {\mnras}\ }\textbf {\bibinfo {volume} {515}},\ \bibinfo {pages} {4797} (\bibinfo {year} {2022})},\ \Eprint {http://arxiv.org/abs/2111.09335} {arXiv:2111.09335 [astro-ph.SR]} \BibitemShut {NoStop}%
\end{thebibliography}%

\end{document}